\documentclass[12pt]{article}
\usepackage{graphicx}
\usepackage{amsfonts}
\usepackage{amsmath}
\usepackage{amssymb}
\usepackage{amsthm, xcolor}
\usepackage[mathscr]{eucal} 
\usepackage{hyperref}

\numberwithin{equation}{section}

\def\N{\mathbb{N}}
\def\Z{\mathbb{Z}}
\def\Q{\mathbb{Q}}
\def\R{\mathbb{R}}
\def\C{\mathbb{C}}
\def\P{\mathbb{P}}

\newcommand{\nin}{\mbox{\ooalign{\hfil/\hfil\crcr$\in$}}}

\def\diag{\mathop{\rm diag}\nolimits}

\def\SO{\mathop{\rm SO}}

\def\U{\mathop{\rm U}}

\theoremstyle{definition}
\newtheorem{thm}{Theorem}[section]
\newtheorem{defn}[thm]{Definition}
\newtheorem{notn}[thm]{Notation}

\newtheorem{props}[thm]{Proposition}
\newtheorem{lemma}[thm]{Lemma}
\newtheorem{rmk}[thm]{Remark}
\newtheorem{anythng}[thm]{}
\newtheorem{cor}[thm]{Corollary}
\newtheorem{conj}[thm]{Conjecture}

\begin{document}

\baselineskip 0.6cm

\begin{titlepage}
  
   \vskip 1cm
 \begin{center}
  
  { \Huge Towards Hodge Theoretic Characterizations of\\[0.3cm] 2$d$ Rational SCFTs}
 
 \vskip 1.2cm
  
 Abhiram Kidambi$^{1,2,a}$, Masaki Okada$^{1,b}$ and Taizan Watari$^{1,c}$ \\[0.4cm]
 $^a$\href{mailto:abhiram.kidambi@ipmu.jp}{abhiram.kidambi@ipmu.jp}, ~
$^b$\href{mailto:masaki.okada@ipmu.jp }{masaki.okada@ipmu.jp },~ $^c$\href{mailto:taizan.watari@ipmu.jp}{taizan.watari@ipmu.jp}\\[0.4cm]

 {\it  $^1$Kavli Institute for the Physics and Mathematics of the Universe (WPI), \\
  University of Tokyo, Kashiwa-no-ha 5-1-5, Kashiwa 277-8583, Japan \\[0.3cm] $^2$ Riemann Center for Geometry and Physics, Institut f\"ur Theoretische Physik, \\Leibniz Universit\"at Hannover, Appelstra\ss e 2, 30167 Hannover}
  \end{center}
\vskip 0.4cm

\begin{center}
  \textbf{Abstract}
  \end{center} 
The study of rational conformal field theories in the moduli space of conformal field theories is of particular interest since these theories correspond to points in moduli space where the algebraic and arithmetic structure are usually richer, while also being points where non--trivial physics occurs (such as in the study of attractor black holes and BPS states at rational points). This has led to various attempts to characterize and classify such rational points. In this paper, a conjectured characterization by Gukov--Vafa (Commun. Math. Phys. {\bf 246} (2004) 181) of rational conformal field theories whose target space is a Ricci flat K\"ahler manifold  is analyzed carefully for the case of toroidal compactifications. We refine the conjectured statement as well as making an effort to verify it, using $T^4$ compactification as a test case. Seven common properties in terms of Hodge theory (including complex multiplication) have been identified for $T^4$-target rational conformal field theories. By imposing a subset of the seven properties, however, there remain $\mathcal N = (1,1)$ SCFTs that are not rational. Open questions, implications and future lines of work are discussed.

\end{titlepage}
 
 \tableofcontents

\section{Introduction}

\textbf{Note to the reader}: Although progress reported in this article is in string theory, and not mathematics, we still adopt the mathematical style of presentation using Theorem, Conjecture, Remark, Lemma 
etc. This makes it easier to refer to specific facts and/or arguments.


\subsection{Relevant Background on String Theory and SCFT}

$\mathcal N=(1,1)$ supersymmetric non-linear sigma models in $1+1$ dimensions 
have a non-trivial moduli space when the target space $M$ is a Ricci--flat 
K\"{a}hler manifold. These theories, which are superconformal field 
theories (SCFTs), define a compactification of Type II string theory. 
These SCFTs are \textit{rational} CFTs (RCFT)\footnote{
There are many ways to define a rational CFT. A simple way of explaining what an RCFT is to string theorists is that they have a finite number of primary fields or that their conformal blocks are finite dimensional (See \cite{Moore:1989vd} and references therein). Classifying and characterizing such rational CFTs is an important open problem in string theory. See section \ref{ssec:translate}
for a dictionary between string theory and vertex operator algebra. 
} %
 only at special points in the moduli space. It was observed (and conjectured) 
by Gukov--Vafa (GV) in \cite{Gukov:2002nw} 
(see also \cite{Moore:1998pn,Moore:1998zu} and references therein)
that the points in the moduli space where the SCFTs are rational may be 
characterized 
in terms of the period integrals on $M$ and those on its mirror manifold 
using Hodge theory and number fields (to be more concrete, both $M$ and 
its mirror are CM-type); see section \ref{ssec:GV-conj} for more 
on the observation. 
Our current work presented here is inspired by these observations  
and we elaborate on the problem of characterizing these special 
points in the moduli space where the SCFTs are rational. 

Giving (and establishing) such a characterization is a well-defined 
question in mathematical physics, which may be of interest in its own right.  
Apart from the GV conjecture, and some association with 
enhanced symmetry on worldsheet theory and/or effective field theory 
after compactification, our knowledge of rational SCFTs has been mostly 
construction based. Not much is known beyond the Gepner constructions, and  
lattice vertex operator algebras and their orbifolds. 
If the criteria for the SCFTs are proven to be something close to 
the one by the GV conjecture, that means that there are more 
rational SCFTs than those obtained by these constructional approaches. 
Those rational SCFTs will include the ones in the small-volume limit 
region in the moduli space; rational SCFTs will be an ideal (and a rare) 
tool to study how string theory captures geometry at the short-distance 
(high-energy) region in a situation where classical Einstein gravity 
is not a good approximation.\footnote{
An open problem that is pertinent is here, and for which progress is desirable, is to also understand how close such rational points are to each other in the moduli space of SCFTs.
}

Further progress can be expected in a couple of other directions, 
when the GV \cite{Gukov:2002nw} observation is understood 
better, with more systematically constructed examples of rational SCFTs 
with geometric interpretations. For example, on the side of arithmetic 
geometry, it is known that complex analytic CM-type manifolds $M$ are 
known to admit arithmetic models \cite{taniyama1961complex,shimura2016abelian,piatetski1973arithmetic,rizov2005complex}, 
at least when $M$ is either an abelian variety or a K3 surface, and
some of the $L$-functions defined for these arithmetic models are expected to have modular transformation properties.
It will be an interesting subject of research to explore relations  
between such modular objects in arithmetic geometry and $g=1$ 
chiral correlation functions of the corresponding RCFTs in string 
theory. References \cite{Kondo:2018mha,Kondo:2019jpi} studied that for the case 
where the target space is a CM elliptic curve. To carry out a similar study 
for abelian varieties and K3 surfaces, a better understanding of 
the relation between CM manifolds and RCFTs is necessary 
to get started. 

Also in particle phenomenology, the work of GV \cite{Gukov:2002nw} is of 
significance.  
In Type IIB Calabi--Yau orientifold compactifications, the gravitino 
mass and the cosmological constant are not generically 
much smaller than the Planck scale of the effective theory on $3+1-$dimensions
due to non-zero fluxes \cite{Denef:2008wq}. When the complex structure of 
the Calabi--Yau threefold has period integrals characterized by 
number fields (such as in, \cite{Gukov:2002nw, DeWolfe:2004ns, Aspinwall:2005ad}), 
then the gravitino mass can be much smaller than the Planck scale 
for a much larger fraction of flux configurations \cite{Moore:2004fg, 
DeWolfe:2004ns, Kanno:2017nub}. 
If the GV observation is true, then we may attribute particle phenomenologies 
such as electroweak gaugino dark matter, gauge coupling unification, and 
small gravitino mass, to large chiral algebra on the worldsheet theory, 
and not to a larger symmetry of the spacetime field theory. 


\subsection{Outline and Summary of the Paper}
We begin in section \ref{ssec:GV-conj} with a review of the conjectured 
connection between rational SCFTs and CM-type Hodge 
structures (Conj. \ref{conj:GV-original}) by following Ref. \cite{Gukov:2002nw}. 
We do so while highlighting a few aspects in which the conjecture needs 
to be refined for its applications to various Ricci-flat K\"{a}hler manifolds. 
RCFTs among toroidal compactifications have been completely 
classified \cite{Harvey:1987da,Wendland:2000ye,Hosono:2002yb}, so we use 
the information as test data in refining the GV conjecture. 
Moreover, we can also try to verify/find counter examples within 
toroidal compactifications. 
Results from the work of Meng Chen (MC) \cite[Theorem 2.5 \& Proposition 3.10]{Chen:2005gm} 
are vital to the underlying logic of the analysis performed in this study. 
The refined version of the GV conjecture is presented in Theorems~\ref{thm:RCFT-2-GV-forT4} and \ref{thm:RCFT-2-GV-forT4-genI} with auxiliary comments in section \ref{sec:discussions}. 

Sections \ref{sec:choose-cpx-str} and \ref{ssec:mirror-exist} 
clarify subtleties raised in section \ref{ssec:GV-conj}
by studying $T^4$-target rational SCFTs. We will learn that 
RCFTs with $T^{2n}$ target are associated not with 
complex tori with sufficiently many complex multiplications, 
but with CM-type abelian varieties (in section~\ref{ssec:pol}). 
We also see that we can always find a polarizable complex structure 
while demanding that the Hodge $(2,0)$ component of the $B-$field 
is absent (only for $T^4$, Prop. \ref{props:I-polNalgB}), and that 
there is always a mirror 
SCFT that allows a geometric interpretation (for $T^4$, 
Thm. \ref{thm:mirror-exist}). 
An observation that the K\"{a}hler form for a $T^4$-target 
rational CFT is in the algebraic part (non-transcendental part) 
of $H^2(T^4;\Q)$ (Thm. \ref{thm:Kahler-is-alg}) may also be 
interesting in its own right. We do not have evidence, however, 
that this is true for all the rational CFTs that are not $T^4$-target. 

Sections \ref{ssec:torus-RCFT}--\ref{ssec:CM-abel-surface} are, 
for the most part, review on related materials to be used 
in this article. Section~\ref{ssec:translate} and 
Appendix \ref{sec:app} contain only textbook-level materials.
We include them in this preprint so that 
readers with math background can get a rough sense from 
section \ref{ssec:translate} of what the authors mean by 
jargons such as compactification, CFT and ${\cal N}=(1,1)$ SCFT.   
Appendix \ref{sec:app}, on the other hand, collects some definitions, 
notations and useful facts in number theory and Hodge theory, 
for the convenience of some readers with background in string theory. 
The authors are happy to follow the advice from the editor/referees 
whether to keep Section \ref{ssec:translate} and Appendix \ref{sec:app}
or to drop them from the manuscript. 


\section{Preliminaries}
\label{sec:prelim}
\subsection{A Pertinent String Theory -- VOA Dictionary}
\label{ssec:translate}

This subsection is absolutely not for a string theorist, but may be of 
some use to those who view themselves as members of the vertex operator 
algebra community. String theory may be viewed as machinery producing 
a conformal field theory (CFT) from a set of data associated with a geometry.
This section \ref{ssec:translate} summarizes which data determine what
in the machinery, and also explains the string theoretic terminology in 
this paper in the language of the vertex operator algebra community. 

\vspace{5mm}

\noindent {\bf A bosonic conformal field theory (CFT)} 
(as a countable noun):\footnote{
We use the word ``{\it bosonic conformal field theory}'' in the same sense 
as ``{\it vertex operator algebra}'' of Ref. \cite{Kapustin:2000aa}. 
MC \cite{mengchenphd} uses {\it generalized vertex algebra} or 
{\it OPE algebra} for the same thing. 
} %
it consists of data that include (but are not exhausted by) 
\begin{enumerate}
\item a pair of vertex operator algebras, one for the \textit{left-mover} 
(i.e., one with the holomorphic local coordinates on a Riemann surface) and 
one for the \textit{right-mover} (i.e., one with the anti-holomorphic 
local coordinates), and 
\item a set of representations under the two mutually commuting 
vertex operator algebras. 
\end{enumerate}
The direct sum of all those representation spaces is called the {\it total 
Hilbert space}, ${\cal H}_{\rm tot}$. The states in the left-mover 
(resp. right-mover) vertex operator algebra must correspond to all the states 
of the total Hilbert space whose right-mover (resp. left-mover) conformal 
weights are zero. Such vertex operator algebras are called the left-mover 
(holomorphic) (resp. the right-mover (anti-holomorphic)) 
{\it chiral algebras of the CFT}. 

When a string theorist refers to a bosonic CFT, it is often assumed implicitly 
that the both of the chiral algebras contain an operator called the 
energy-momentum tensor of conformal weight 2, that the central charge 
of the left-mover and the right-mover are not more than 26, that 
${\cal H}_{\rm tot}$ has a positive definite Hermitian inner product, and 
that the  
partition function computed from ${\cal H}_{\rm tot}$ is modular invariant. 
The set of data of a bosonic CFT also includes an  
${\rm End}({\cal H}_{\rm tot})$-valued formal power series for any state in 
${\cal H}_{\rm tot}$, not just for the states in the left-mover and right-mover 
chiral algebras. The authors do not intend to explain all the 
concepts in this paragraph, because not knowning them does not pose 
a problem in following the materials in this article. 
 
\vspace{5mm} \noindent
{\bf A torus compactification} associated with 
the data $(T^m;G, B)$: it is a bosonic CFT corresponding to the data
$(T^m; G, B)$. Here, 
 \begin{enumerate}
 \item $T^m = \R^m/\Z^{\oplus m} = \R^m/H_1(T^m;\Z)$ is a real $m$-dimensional 
torus (a manifold) with a set of real coordinates 
$(X^I) = (X^1,X^2,\cdots, X^m) \in \R^m$.  
\item $G$ is a Riemannian metric on $T^m$
that remains constant under translations in $\R^m$, and
 \item $B$ a closed 2-form on $T^m$, referred to as the $B$-field.
\end{enumerate} 
In the language of string theory, $G$ is the vacuum metric and $B$ the vacuum 
$B$-field configuration, to be precise, but we will omit the word `vacuum' 
for brevity in this article. 

For any bosonic CFT (i.e., torus compactification) for $(T^m;G,B)$, both 
the left-mover and right-mover chiral algebras contain the direct sum 
of $m$ copies of the Heisenberg Lie algebra. Different choices of the data 
$G, B$ as above correspond to different choices of the set of 
representations of the $m+m$ Heisenberg Lie algebras. 
The set of choices of such a set of representatons of the $m+m$ Heisenberg 
Lie algebras (and hence the set of data $(T^m; G, B)$) forms a moduli space. 
For further details, we refer the reader to \cite{Kapustin:2000aa}. 

\vspace{5mm}
\noindent
{\bf A bosonic CFT is rational} when both the left-mover (holomorphic) 
chiral algebra and the right-mover (anti-holomorphic) chiral algebra 
are rational, meaning that both of the two vertex operator algebras have 
a finite number of distinct irreducible modules \cite[p.90]{frenkel2004vertex}. 

\vspace{5mm}
\noindent
{\bf A torus ($T^m$) compactification} for $G$ and $B$ {\bf is rational}
if both of the holomorphic and anti-holomorphic chiral algebras are larger 
than $m$ copies of the Heisenberg Lie algebras to the extent that each 
of the chiral algebras becomes the vertex operator algebra of a rank-$m$ 
even positive definite lattice. 

\vspace{5mm}
\noindent 
{\bf An $\mathcal N=(1,1)$ SCFT} (as a countable noun) {\bf associated 
with the data $(M; G, B)$}, where $M$ is a real manifold, $G$ a Ricci-flat 
Riemannian metric on $M$, and $B$ a closed 2-form on $M$: It is an 
$\mathcal N=(1,1)$ SCFT (a countable noun) determined uniquely in 
string theory by the data $(M; G, B)$. An $\mathcal N=(1,1)$ SCFT is a set 
of information that includes (but is not exhausted by) 
\begin{enumerate} 
\item two vertex operator superalgebras (one 
for left-mover (holomorphic) and the other for right-mover (anti-holomorphic)), 
each one of which contains an $\mathcal N=1$ superconformal algebra, and  
\item a set 
of representations under the mutually (anti-)commuting algebras. 
\end{enumerate}
The states in the left-mover (resp. right-mover) vertex operator superalgebra 
must correspond to all the states in ${\cal H}_{\rm tot}$ whose right-mover 
(resp. left-mover) conformal weight is zero. 
These vertex operator superalgebras are called the left-mover and 
right-mover {\it chiral superalgebra of the SCFT}. This definition 
does not rule out an $\mathcal N=(1,1)$ SCFT whose chiral superalgebra 
contains an $\mathcal N=2$ (or more) superconformal algebra. 

\vspace{5mm}
\noindent
{\bf If there exists a complex structure $I$} for a Riemannian manifold 
$(M; G)$ such that $G$ is compatible with $I$ and $(M; G,;I)$ is 
K\"{a}hler, then the $\mathcal N=(1,1)$ SCFT for the data $(M; G, B)$,  
with $B$ as above, has a special property.  Each one of the 
left-mover and right-mover chiral superalgebras contains an $\mathcal N=2$ 
superconformal algebra. In fact, there is a unique way  
to specify the vertex operator with conformal-weight $1$ from the data $I$ 
such that the $\mathcal N=1$ superconformal algebra is enhanced to 
an $\mathcal N=2$ superconformal algebra. We refer the reader 
to \cite{VanEnckevort:2003qc} for more information. 

\vspace{5mm}
\noindent 
{\bf An $\mathcal N=(1,1)$ SCFT is rational} when both the left-mover and 
right-mover chiral superalgebras have finitely many distinct irreducible 
modules. It is clear that the rationality of an $\mathcal N=(1,1)$ SCFT depends on the entire chiral superalgebra determined by the data $(M;G,B)$, and not by the $\mathcal N=(2,2)$ superconformal algebra determined by the data 
$(M;G,B;I)$. It is also a known fact that the $\mathcal N=(1,1)$ SCFT for $(T^m;G,B)$ is rational 
if and only if a torus compactification for (i.e., bosonic CFT for) 
the same set of data $(T^m;G,B)$ is rational. 

\subsection{The Gukov--Vafa Conjecture}
\label{ssec:GV-conj}

Amongst bosonic CFTs with torus ($T^m$) as target, RCFTs have been 
identified completely \cite{Wendland:2000ye,Hosono:2002yb,Harvey:1987da}. In the case of $m=1$, for example, 
the moduli space of bosonic CFTs is $\R_{>0}$ and is physically parametrized by 
the radius-squared of the target space $S^1$. Amongst such bosonic CFTs, only those CFTs for which the radius-squared (in units of the string length) is a rational number are rational. So, the subset $\Q_{>0} \subset \R_{>0}$ classifies all RCFTs for the case of $m=1$. 

However, not much is known for cases 
other than torus compactifications. Certain explicit constructions such as the 
Gepner models are known, but it is not known how many more rational 
CFTs or SCFTs exist.  

The GV conjecture \cite{Gukov:2002nw} was formulated in an attempt to ascertain where one encounters a rational SCFT in the moduli space of 
2$d$ $\mathcal N=(1,1)$ SCFTs obtained as a non-linear-sigma model of 
a manifold $M$ that admits a Ricci-flat K\"{a}hler metric.  
\begin{conj} \cite[\S 7]{Gukov:2002nw}
\label{conj:GV-original}
Consider a 2$d$ $\mathcal N=(1,1)$ SCFT obtained as a non-linear sigma model 
of $(M; G, B)$; here, $M$ is a real $2n$-dimensional manifold, $B$ 
a closed $2-$form on $M$, and $G$ a Riemannian metric on $M$ that can be 
Ricci-flat and K\"{a}hler under some suitable complex structure $I$. 
The SCFT is rational if and only if the following conditions are satisfied:
\begin{enumerate}
 \item the rational Hodge structure on the cohomology groups of $M$ is of CM-type, 
 \item the rational Hodge structure on the cohomology groups of the 
mirror manifold $W$ of $M$ is also CM-type, 
 \item the CM fields of both are isomorphic. 
\end{enumerate}
Conjecture \ref{conj:GV-original} presented as above is a slight 
modification of the conjecture presented in Ref. \cite{Gukov:2002nw}, 
although its scientific merit is not compromised.  $\bullet$
\end{conj}

We first repeat the justification argument for Conjecture \ref{conj:GV-original} for the convenience of a less initiated reader. We then later discuss 
the refinement of the statement of Conjecture \ref{conj:GV-original}. 
For a physicist-friendly overview of theory of complex multiplication 
covering more than elliptic curves, readers are referred to the appendix 
of \cite{Kanno:2017nub}; in the case of elliptic 
curves to get started, see \cite{Moore:1998pn,Moore:1998zu}. 

The first piece of evidence in support of Conjecture \ref{conj:GV-original} 
above is the case of $T^2$ compactification. Consider a $T^2$ compactification 
associated with the data $(T^2; G, B)$, where $G$ and $B$ are 
a constant Riemannian metric and a constant 2-form on $T^2$, respectively. 
There is a unique complex structure $I$ with which the metric $G$ is 
compatible. 
Let (the ${\rm SL}(2;\Z)$-orbit of) $\tau$ be the complex structure 
parameter of a complex $g=1$ Riemann surface $M=(T^2; I) = \C/(\Z + \tau \Z)$, 
and $\displaystyle \rho := \int_{T^2} (B+i\omega )/[(2\pi)^2\alpha']$ be 
the complexified K\"{a}hler parameter; $\omega = \omega(-,-):= 2^{-1} G(I-,-)$ 
is the K\"{a}hler form. 
It is then known that the CFT is rational if and only if 
both $\Q(\tau)$ and $\Q(\rho)$ are degree-2 extension fields over $\Q$, 
and are isomorphic to each other \cite{Moore:1998pn,Moore:1998zu}.
The condition that $[\Q(\tau):\Q] = 2$ (resp. $[\Q(\rho):\Q] =2$)
is equivalent to the condition that the rational Hodge structure 
on $H^1(M;\Q)$ (resp. $H^1(W;\Q)$, where $W$ is the mirror manifold 
isomorphic to $\C/(\Z+\rho \Z)$) is of CM-type. 
Those observations in this example have been abstracted to become 
the statement of Conjecture \ref{conj:GV-original} above. 

Let us consider another example: the $\Z_5$-orbifold 
of the tensor product of five copies of the 2$d$ $\mathcal N=(2,2)$ 
minimal models with the central charges $c_L=c_R=3k/(k+2)$ and $k=3$.
This SCFT is rational, and is interpreted as a 2$d$ 
non-linear sigma model whose target space is a quintic Calabi--Yau threefold 
with a very special complex structure and a very special complexified 
K\"{a}hler parameter. The complexified K\"{a}hler parameter 
is chosen at the small volume limit within the complex $1-$dimensional 
moduli space, and the complex structure parameter is chosen 
at the Fermat point\footnote{
The $101-$dimensional moduli space of complex structure corresponds to 
choosing an arbitrary homogeneous function $F(\Phi_{i=1,\cdots,5})$ 
of degree-5 on $\P^4$ that defines a threefold $M$ through $M = 
\{ [\Phi_i] \in \P^4 \; | \; F(\Phi)=0 \} \subset \P^4$. 
The Fermat point in the moduli space corresponds to the choice 
$F = \sum_{i=1}^5 (\Phi_i)^5$.
} %
  of the complex $101-$dimensional moduli space.  
The cohomology group $H^3(W;\Q)$ of the mirror manifold $W$ 
is 4-dimensional over $\Q$, and is of CM--type, where the CM--field is 
$\Q(\zeta_5)$, a cyclotomic extension field over $\Q$ generated 
by a primitive 5th root of unity $\zeta_5$. 
The cohomology group $H^3(M;\Q)$ also contains a rational Hodge substructure 
that is $4-$dimensional over $\Q$, level$-3$, and is of CM--type; the 
CM--field is $\Q(\zeta_5)$ on this substructure.  
Various jargons pertaining to Hodge structure are explained in the appendix 
of \cite{Kanno:2017nub}, or any textbook or lecture note on Hodge theory 
by mathematicians. This example indicates that GV's 
Conjecture \ref{conj:GV-original} has been generalized 
from the case of $T^2$ compactification in a proper way. 

We are yet to verify or refute this conjecture, and that is what we do 
in section \ref{sec:analysis-T4}. Such an effort also lets us notice 
that the statement of the conjecture needs to be refined to be 
verified, as we see below in Discussions \ref{statmnt:subtlty-justL=n?}, 
\ref{statmnt:subtlty-offdiag}, \ref{statmnt:subtlty-choose-cpx-str} and 
 \ref{statmnt:subtlty-mirror?}.  

\begin{anythng}
\label{statmnt:subtlty-justL=n?}
Let $M$ be a Ricci-flat K\"{a}hler manifold of complex dimension $n$; the 
cohomology group $H^n(M;\Q)$ is endowed with a rational Hodge structure 
by the complex structure of $M$. The rational Hodge structure on $H^n(M;\Q)$
is not necessarily simple, but has a decomposition into simple Hodge 
substructures
\begin{align}
  H^n(M;\Q) \cong \oplus_{a\in A} [H^n(M;\Q)]_a ~.
\end{align}
Let $D_a := {\rm End}([H^n(M;\Q)]_a)^{\rm Hdg}$ be the algebra of Hodge 
endomorphisms of a simple component $[H^n(M;\Q)]_a$; it is always a 
division algebra.\footnote{
See Definition \ref{defn:divis-alg} and Notation \ref{notn:end-alg}. 
} %
 The endomorphism algebra 
of $H^n(M;\Q)$ is of the form 
\begin{align}
  {\rm End}(H^n(M;\Q))^{\rm Hdg} \cong \oplus_\alpha M_{n_\alpha}(D_\alpha), 
\end{align}
where the simple components $a\in A$ are grouped into those 
with isomorphic $D_a$'s and a common level, and the equivalence classes 
are labeled by $\alpha$'s; $n_\alpha$ is the number of simple 
components ($a$'s) in an equivalence class $\alpha$. 

For example, in the case of the Fermat quintic Calabi--Yau threefold $M$, 
the 204-dimensional vector space $H^3(M;\Q)$ has a decomposition 
into simple rational Hodge substructures \cite[\S3]{shioda1982geometry}, 
\begin{align}
 [H^3(M;\Q)]_{\ell=3} \oplus
     \left( \oplus_{a=1}^{50} [H^3(M;\Q)]_{\ell=1, a} \right), 
\end{align}
and each one of the components is of $4-$dimensional over $\Q$, 
supporting a rational Hodge substructure of level$-1$ (with the 
exception of the first component whose Hodge substructure is of level $3$) 
with the endomorphism field $D_\alpha \cong \Q(\zeta_5)$. 

\vspace{5mm}

Back to the general case, the Hodge structure on $H^n(M;\Q)$ is said 
to be {\it of CM-type} if and only if the Hodge structure on individual 
substructures on $[H^n(M;\Q)]_a$ are of CM-type. All the division algebras 
$D_a$ are then fields, with $[D_a:\Q] = \dim_\Q [H^n(M;\Q)]_a$. 
So, the Hodge structure on $H^n(M;\Q)$ is of CM-type in the case that $M$ 
is the Fermat quintic threefold. There are, however, K\"{a}hler manifolds 
where the Hodge structure is of CM-type on the level-$n$ component 
(the simple component containing the $(n,0)$ Hodge component; 
the notion of {\it level} is explained, e.g., in \cite[App. B]{Kanno:2017nub}), 
but is not of CM-type in other simple components. We should therefore remain 
open minded as to whether the first two conditions in 
Conjecture \ref{conj:GV-original} should be imposed on some of 
the simple components of $H^n(M;\Q)$ and $H^n(W;\Q)$, or on all 
of their simple components. Examples studied in \cite{Gukov:2002nw} are 
not sufficient to resolve this difference in the conditions.\footnote{
\label{fn:BV}
Calabi--Yau threefolds of the form of Borcea--Voisin 
orbifolds \cite{borcea1998calabi,voisin1993miroirs} will be a good 
testing ground in resolving this issue. To work on this class of cases, 
however, we should work on K3 surfaces first. 
} %

The third condition in Conjecture \ref{conj:GV-original} refers 
to the fields of CM-type Hodge structures of $M$ and the mirror 
manifold $W$. An easy way to make sense of this condition is to think 
of them as the endomorphism fields of the unique simple component 
of $H^n(M;\Q)$ and $H^n(W;\Q)$ containing the Hodge $(n,0)$ component, 
the level-$n$ simple component.  
If one should require other simple components to be of CM-type, 
as discussed before, then one may also have to refine the 
the third condition; whether it is read as an isomorphism of the CM fields 
of the level-$n$ components on both sides, or as isomoprhisms of 
some pairs of simple components of $H^*(M;\Q)$ and $H^*(W;\Q)$. 
In general, $\dim_\Q[H^n(M;\Q)]$ is not necessarily equal to 
$\dim_\Q[H^n(W;\Q)]$, so there is no natural choice of pairs of 
simple components besides the pair of the level-$n$ components. 
\end{anythng}

\begin{anythng}
\label{statmnt:subtlty-offdiag}
For a Ricci-flat K\"{a}hler manifold $M$,
its Hodge diamond can be non--zero not only in the 
vertical diagonal terms ($h^{k,k}$ with $k=0,\cdots, n$) 
and horizontal diagonal terms ($h^{p,n-p}$ with $p=0,\cdots, n$),
but also in the off--diagonal terms. Certainly all the off--diagonal 
terms are zero when $M$ is an elliptic curve, K3 surface, or a 
Calabi--Yau threefold. But $h^{q,0}$ with $q \neq 0, n$ can be 
non-zero when $M$ is 
\begin{itemize}
\item a complex torus of $n\geq 2$ dimensions, or
\item a hyper-K\"{a}hler manifold of real 8-dimensions and higher, or 
\item a product of Ricci-flat K\"{a}hler manifolds one of which 
is either a complex torus or a hyper-K\"{a}hler manifold. 
\end{itemize}
Moreover, there are Calabi--Yau fourfolds where all the $h^{q,0}$'s 
are zero for $q=1,\cdots, 3$, but the off-diagonal term $h^{2,1}$ 
is non-zero (see \cite[(90)]{Braun:2014xka} 
for a class of toric hypersurface fourfolds where $h^{2,1}\neq 0$). 

The authors of this paper are not aware of a proof indicating that 
the Hodge structure on $H^k(M;\Q)$ with $k\neq n$ is always of CM-type 
when one just requires that the Hodge structure on $H^n(M;\Q)$ is of 
CM-type (see Rmk. \ref{statmnt:CM-from-H2-to-H1}). 
We then face a question whether we should read the conditions 
in Conjecture \ref{conj:GV-original} as that for $H^n(M;\Q)$ and $H^n(W;\Q)$
(the vertical part of the Hodge diamond of $M$), or that for all the 
cohomology groups including the off-diagonal parts of the cohomology group
of $M$. 
\end{anythng}

\begin{anythng}
\label{statmnt:subtlty-choose-cpx-str}
Consider a case where the target space $(M; G)$ is either a torus $T^{2n}$ 
of real $2n$ dimensions with $n\geq 2$, or a hyper-K\"{a}hler manifold. 
On one hand, for such 
a smooth manifold $M$ and a Riemannian metric $G$ on it, there is a continuous 
freedom in choosing a complex structure $I$ with which the metric $G$ is 
compatible.  

On the other hand, a 2$d$ non-linear sigma model is specified by only the 
data $(M; G)$, without referring to a complex structure on $M$. 
Whether the SCFT is rational or not should therefore be a property 
of $(M; G)$, not of the data $(M; G; I)$. 

Since Conjecture \ref{conj:GV-original} tries to characterize 
rational SCFTs by using a Hodge structure, there is no way of interpreting 
the conditions there without choosing a complex structure. If the 
conjecture is to be applicable for the class of manifolds we are referring 
to here, then we should read the conditions and characterizations in 
Conjecture \ref{conj:GV-original} either as those for arbitrary $I$ with 
which the metric is compatible (this is not a good guess as we will see 
in section \ref{ssec:pol}), or as those for a class of $I$'s that should 
be specified more carefully.  
\end{anythng}

\begin{anythng}
\label{statmnt:subtlty-mirror?}
The statement of Conjecture \ref{conj:GV-original} is written 
by referring to a mirror manifold $W$. It is not always true, however, 
that an $\mathcal N=(2,2)$ SCFT as a non-linear sigma model with a Ricci-flat 
K\"{a}hler $M$ as the target space has a mirror-equivalent $\mathcal N=(2,2)$ 
SCFT that can be interpreted as a non-linear sigma model of another Ricci-flat 
K\"{a}hler manifold $W$. Even when there is, it is not guaranteed that 
there is a unique choice of $W$. 

It is an interesting question whether there is always such a mirror 
manifold when the $M$-target $\mathcal N=(2,2)$ SCFT is rational. 
Would it be an improvement 
if Conjecture \ref{conj:GV-original} is stated without referring 
to a mirror manifold? 
\end{anythng}

In this article, we work on the cases with $M = T^{2n}$, most intensively 
with $M=T^4$. The experimental data collected in this article 
do not help in resolving the issue raised in 
Discussion \ref{statmnt:subtlty-justL=n?}, but will shed 
some light on the issues raised in \ref{statmnt:subtlty-choose-cpx-str} 
and \ref{statmnt:subtlty-mirror?}. The authors admit 
that they have not made all possible efforts imaginable in exploiting 
the experimental data to clarify the issues in \ref{statmnt:subtlty-offdiag}
and \ref{statmnt:subtlty-mirror?}.

\subsection{Rational CFTs with Torus Target}
\label{ssec:torus-RCFT}

Since RCFTs in torus compactifications have been completely classified, 
we may use the established results to refine and test 
Conjecture \ref{conj:GV-original}.  
In this section \ref{ssec:torus-RCFT}, we quote results from 
Ref. \cite{Harvey:1987da,Wendland:2000ye} relevant to our analysis. 

\begin{props} (\cite{Harvey:1987da} and  \cite[Lemma 4.5.1]{Wendland:2000ye})
{\it Let $T^{m} = \R^m/\Z^{\oplus m}$ be a real $m$-dimensional torus
with a smooth structure, $X^I$s with $I=1,\cdots, m$ a 
set of coordinates of $\R^m$ with periodicity $\Delta X$ 
(i.e., $X^I \sim X^I + \Delta X$), let $G = G_{IJ} dX^I \otimes dX^J$
be a constant Riemannian metric on $T^m$ (i.e., $G_{IJ} \in \R$ are 
independent of the coordinates $X^K$'s), and $B = 2^{-1} B_{IJ} dX^I \wedge dX^J$
a 2-form on $T^m$ where $B_{IJ}$ are independent of the coordinates.  
 
The bosonic CFT for the data $(T^m; G, B)$ is rational if and only if }\footnote{
The author of \cite{Wendland:2000ye} adopts the convention 
$\Delta X = 2 \pi R$, 
$R=\sqrt{\alpha'}$ and $\alpha' =2$. We will use the convention 
$\Delta X = 2\pi \sqrt{\alpha'}$ throughout this article. The metric 
and $B$-field satisfying (\ref{eq:cond-GnB-rational}) are therefore 
said to be {\it rational}. 
} %
\begin{align}
 \left( \frac{\Delta X}{(2\pi) \sqrt{\alpha'}} \right)^2 G_{IJ} \in \Q, \qquad
 \left( \frac{\Delta X}{(2\pi) \sqrt{\alpha'}} \right)^2 B_{IJ} \in \Q. 
  \label{eq:cond-GnB-rational}
\end{align}
{\it The condition for the 
$\mathcal {N}=(1,1)$ SCFT associated with the data $(T^m;G,B)$ to be rational 
is also the same as above. } 
\end{props}
 
We are interested in the cases with $m=2n$, when there is a possibility of 
introducing a complex structure on the target space $T^m$. The author of \cite{Wendland:2000ye} has further derived this
\begin{cor}\cite[Thm. 4.5.5]{Wendland:2000ye}
\label{cor:Wend-ExEx}
{\it Let $(T^{2n};G, B)$ be a set of data for which the (S)CFT is rational. 
Then there exists a surjective homomorphism 
$\varphi: T^{2n} \cong \R^{2n}/\Z^{\oplus 2n} \longrightarrow
 \prod_{a=1}^n \C/(\Z + \tau_a \Z)$ with respect to the abelian group 
law on $\R^{2n}$ and $\C^n$, where each one of $\C/(\Z + \tau_a \Z)$ is 
a CM elliptic curve (i.e., $[\Q(\tau_a):\Q] = 2$), 
and there is a metric on $\C/(\Z + \tau_a \Z)$, given by 
$ds^2 = g_a (du^a \otimes d\bar{u}^{\bar{a}} + {\rm h.c.})$ with\footnote{
$u^a$ with $a=1, \cdots, n$ are the complex coordinates of the $a$-th
elliptic curve $\C/(\Z+\tau_a \Z)$, which has the periodicity 
$u^a \sim u^a + 1 \sim u^a + \tau_a$. 
} %
$g_a \in \Q$ so that the pull-back of the metric $ds^2$ by $\varphi$ 
agrees with the metric $G$ on $T^{2n}$. }
\end{cor}

This result lends support towards the justification of the 
GV Conjecture \ref{conj:GV-original} in the following sense. 
Firstly, there is already an implicit choice of complex structure $I_0$ on 
$\prod_a \C/(\Z+\tau_a \Z)$, with which the metric $ds^2$ is compatible. 
The metric $G$ is compatible with the complex structure $I= \varphi^*(I_0)$. 
The complex torus $(T^{2n};I)$ is of CM--type since 
$\prod_a \C/(\Z + \tau_a \Z)$ is of CM--type. The metric $g_a \in \Q$ should be 
split into $g_a = {\rm Im}(\rho_a)/{\rm Im}(\tau_a)$ so that 
${\rm Im}(\rho_a)$ parametrizes the volume of $\C/(\Z + \tau_a \Z)$. 
It also follows that $\Q(i {\rm Im}(\rho_a)) \cong \Q(\tau_a)$. 

This observation alone still falls short of resolving the issue 
raised in \ref{statmnt:subtlty-mirror?}. It also remains to be an 
open question whether the class of complex structures of the form 
$I = \varphi^*(I_0)$ are all those where a GV--like statement 
holds true (this issue was raised in \ref{statmnt:subtlty-choose-cpx-str}).
We will discuss those issues in sections \ref{sec:choose-cpx-str} 
and \ref{sec:metric}, 
before examining whether Conjecture \ref{conj:GV-original}
holds true or not in section \ref{sec:analysis-T4}. 

\subsection{Horizontal and Vertical Generalized Complex Structures}
\label{ssec:gen-HS-mirror}

Generalized complex structure/K\"{a}hler structure and their 
relation to mirror symmetry are reviewed in this 
section \ref{ssec:gen-HS-mirror}. For a reader familiar with 
the work in such references as \cite{Golyshev:1998vzz, Kapustin:2000aa, 
 Hitchin:2003cxu, VanEnckevort:2003qc, Gualtieri}, this 
section \ref{ssec:gen-HS-mirror} does not do anything more than 
preparing notations. 

\begin{anythng}
\label{statmnt:H-GCS}
{\bf Horizontal generalized Hodge structure on $H^*(T^{2n};\Q)$:}
Let $I$ and $B'$ be a complex structure and a $\R$-valued closed 2-form 
on $T^{2n}$, respectively.  Using $I$ and $B'$, a linear operator 
${\cal I}_{B'}$ on the space of sections of $T(T^{2n}) \oplus T^*(T^{2n})$
is introduced:
\begin{align}
 {\cal I}_{B'} : (\partial_{X^I} ,  dX^I) \longmapsto 
   (\partial_{X^L}, dX^L)
  \left( \begin{array}{cc} \delta^L_{\; K} & \\ B'_{LK} & \delta_L^{\; K} 
 \end{array} \right)
 \left( \begin{array}{cc}
        I^K_{\; J} &  \\ & (I^{-1})_{\; K}^{J} \end{array} \right)
   \left( \begin{array}{cc}
      \delta^J_{\; I} & \\ -B'_{JI} & \delta_J^{\; I}
 \end{array} \right),
  \label{eq:def-gen-cpx-str}
\end{align}
where we have chosen a basis $\{ \partial_{X^I} \}$ and $\{ dX^I \}$
for the tangent and cotangent spaces at each point on $T^{2n}$; 
$B' =: 2^{-1} B'_{IJ}dX^I \wedge dX^J$ is the usual convention 
(see \cite[App. B.4]{Polchinski2}).  
In the absence of $B'$, ${\cal I}_{B'}$ multiplies $(+i)$ to holomorphic 
tangent vectors and (0,1)-forms, and multiplies $(-i)$ to 
anti-holomorphic tangent vectors and (1,0)-forms. The operator  
${\cal I}_{B'}$ in (\ref{eq:def-gen-cpx-str}) is an example\footnote{
See Refs. \cite{Hitchin:2003cxu, Gualtieri} for how the notion 
of a generalized complex structure is defined for a general real 
manifold $M$ not necessarily a torus. 
} %
 of a {\it generalized complex structure} on $T^{2n}$ \cite{Golyshev:1998vzz, 
Kapustin:2000aa}.

Let $\Lambda := H_1(T^{2n};\Z) \oplus H^1(T^{2n};\Z)$, and $q$ be the bilinear
form given by $q(\partial_{X^I},\partial_{X^J})=q(dX^I, dX^J)=0$ and 
$q(\partial_{X^I}, dX^J) = \delta_{I}^{\; J}$. 
The integral Hodge structure introduced by $\diag(I, (I^{-1})^T)$ 
on $\Lambda_\R := \Lambda \otimes \R$ has been deformed by the 2-form $B'$
to be ${\cal I}_{B'}$. The deformed version can be expressed by a 
representation of $\U(1) \cong S^1$ given as follows: First, noting that 
\begin{align}
  ({\cal I}_{B'})^T \cdot q \cdot {\cal I}_{B'} = q,
\end{align}
we choose an element $i X_{I,B'}$ of the Lie algebra 
$\mathfrak{so}(\Lambda_\R, q)$ acting on $\Lambda_\R$:
\begin{align}
 {\cal I}_{B'} = \exp \left[ \frac{\pi}{2}i X_{I,B'} \right].
\end{align}
With this (c.f. \ref{statmnt:pHdgStr-by-S1-repr}),  
\begin{align}
  h_{I,B'} : S^1 \ni e^{i \alpha} \longmapsto \exp \left[ i \alpha X_{I, B'} \right]
    \in {\rm GL}(\Lambda_\R).
\end{align}
The vector space $\Lambda \otimes \C$ splits into the 
$h_{I,B'}(e^{i\alpha}) = e^{i\alpha}$ representation space (containing 
(0,1)-forms), and $h_{I,B'}(e^{i\alpha}) = e^{-i\alpha}$ representation 
space (containing (1,0)-forms), but this deformed version of the 
``Hodge'' decomposition is not something we wish to think as 
a pure Hodge structure of some given weight any more.   

The operator ${\cal I}_{B'}$ and $\exp[ i\alpha X_{I,B'}]$ are 
elements of the Lie group $\SO(\Lambda_\R, q)$; the $S^1$ subgroup 
is denoted by $S^1_{I,B'}$. Now, we may think of the spinor representation 
$\rho_{\rm spin}$ of the $\SO(\Lambda_\R, q)$ group and its restriction 
to the $S^1_{I,B'}$ subgroup; the representation $\rho_{\rm spin}|_{S^1_{I,B'}}$ 
of $S^1_{I,B'}$ is denoted by $\rho_{\rm spin}(h_{I,B'})$. 
The $\rho_{\rm spin}$ representation space of $\SO(\Lambda_\R, q)$ 
has an isomorphism\footnote{
\label{fn:conv-spinRpr-coh}
Here is a brief note on the convention. Let $\Lambda_\R$ be a 
$2m$-dimensional vector space, and $q$ its symmetric bilinear form  
of signature $(m,m)$. Suppose that $L_\R \subset \Lambda_\R$ is an 
isotropic subspace of $m$-dimensions, and $\{e_{I=1,\cdots,m} \}$ its basis. 
Then set $L'_\R := [L_\R^\perp \subset \Lambda_R]$, and choose a basis 
$\{ e_{m+I; I=1,\cdots,m} \}$ of $L'_\R$. The representation space of 
$\rho_{\rm spin}$ of $\SO(\Lambda_\R, q)$ is constructed as follows.
The Clifford algebra is given by 
\[
{\rm Cliff}(\Lambda_\R,q) :=  
\R[x_{1,\cdots, m}, x_{m+1,\cdots, m+m}]/(\{x_I, x_J\}, \{x_{m+I},x_{m+J}\}, \{x_I,x_{m+J}\}-2q(e_I, e_{m+J})),
\] 
and the representation space we want is the left-ideal $\mathfrak{a}_L$ 
of ${\rm Cliff}(\Lambda_\R,q)$ generated by the element 
$x_{1\cdots m} := x_1x_2\cdots x_m$. See \cite[\S3.2.1]{Golyshev:1998vzz} for 
more information. 

 When $\Lambda_\R = H_1(T^{m};\R) \oplus H^1(T^m;\R)$, one may set 
the maximal isotropic subspace $L_\R$ to be $H_1(T^m;\R)$. The isomorphism 
$\mathfrak{a}_{H_1(T^m;\R)} \cong H^*(T^m;\R)$ is given by 
assigning $x_{m+I_1}x_{m+I_2}\cdots x_{m+I_k}(x_{1\cdots m})$ to 
$dX^{I_1}\wedge dX^{I_2} \wedge \cdots \wedge dX^{I_k} \in H^k(T^{m};\R)$. 
} %
 with $H^*(T^{2n};\Q) \otimes \R$, so we always use 
this interpretation freely. The spinor representation of the SO group 
splits into the two irreducible representations, one on 
$H^{\rm even}(T^{2n};\Q)\otimes \R$ and the other on 
$H^{\rm odd}(T^{2n};\Q) \otimes \R$.

The representation $\rho_{\rm spin}(h_{I,B'})$ of $S^1_{I,B'}$ introduces 
something similar (but not quite) to the rational mixed Hodge 
structure\footnote{ \label{fn:mixed-HS}
A mixed rational Hodge structure generalizes the pure rational Hodge 
structure on $H^k(M;\Q)$ with a fixed $k$, when $M$ is not necessarily 
a compact smooth K\"{a}hler manifold, but a possibly open and singular 
variety. A {\it mixed rational Hodge structure} \cite{peters2008mixed}
on a vector space $V_\Q$ over $\Q$ consists of one decreasing 
filtration $F^\bullet$ (called {\it Hodge filtration}), 
where $V \otimes \C \supset \cdots \supset F^p \supset F^{p+1}$, and 
one {\it increasing} filtration $W_\bullet$ (called {\it weight filtration}), 
where $W_m \subset W_{m+1} \subset \cdots \subset V_\Q$. 
The component $(F^pW_m\otimes \C/F^pW_{m-1}\otimes \C) \cap \overline{
(F^qW_m\otimes \C/F^qW_{m-1}\otimes \C)}$ is regarded the $(p,m-p)$ component. 

Technically, it is not impossible to think of the generalized 
rational Hodge structure in Def. \ref{defn:gen-HS} as a mixed rational 
Hodge structure. To get started, note that the difference between a 
decreasing filtration $W^\bullet_h$ in 
Def. \ref{defn:gen-HS} and an increacing filtration $W_\bullet$ 
of a mixed rational Hodge structure is relatively minor. 
One may set a weight filtration of a mixed Hodge structure 
by $W_m := W_h^{2n-m}$; we would have to think of $2n$-forms as 
weight $(m=0)$ then, but we could close our eyes to that.  

It is not impossible to use the $S^1$ representation $h$ 
of a generalized Hodge structure to introduce the decreasing filtration 
$F^\bullet$ of a mixed Hodge structure; an idea that comes to the minds 
of the authors is to set $F^\Delta := \oplus [{\rm charge} \geq \Delta]$. 
The range of $(\Delta, m)$ with a non-zero $h^{\Delta, m-\Delta}$ in this 
rational mixed Hodge structure (from $(\rho_{\rm spin}(h_{I,B'}), W_h^\bullet)$)
on $H^{*}(T^{2n};\Q)$ is quite different from the mixed rational Hodge 
structure on $H^n(M;\Q)$ of a non-compact and/or singular complex 
$n$-dimensional variety $M$. The range becomes the conventional one 
when we set a dictionary $p = (m+\Delta)/2$. 
Conversely, the information in the Hodge filtration of a mixed Hodge structure 
on $H^k(M;\Q)$ can also be translated into the language of the $S^1$ 
representation; a differential form 
$(\wedge^{p} dz)(\wedge^{k-p}d\bar{z}) / (\prod_{i=1}^{p+q-k} z_i)$ generating 
the component $[H^k(M;\Q)]^{p,q}$ (not necessarily $p+q = k$) is assigned 
a charge $\#[dz] - \#[d\bar{z}] - \#[{\rm poles}] = p - (k-p)-(p+q-k)
= p-q$.  

Despite the similarity between the mixed and generalized Hodge structures 
at the technical level, both structures are based on completely different 
geometric intuitions. It does not seem possible for a set of data 
$(\rho_{\rm spin}(h_{I,B'}), W_h^\bullet)$ to think of something 
like $F^p \sim [{\rm charge}\geq \Delta = (2p-m)] \subset
 (W_m=W_h^{2n-m})\otimes \C$ consistently with varying choice of $m$.
} %
on $H^{\rm even}(T^{2n};\Q)$ and $H^{\rm odd}(T^{2n};\Q)$. 
The $k$-th cohomology group $H^k(T^{2n};\Q) \otimes \R$ of 
$H^*(T^{2n};\Q) \otimes \R$ alone is not regarded as a representation space 
of the $S^1_{I,B'}$ subgroup when $B' \cdot I - (I^{-1})^T \cdot B' 
 \propto (B')^{(2,0)} - (B')^{(0,2)} \neq 0$. 
But there is still a filtration structure 
\begin{align}
  \{ 0 \} \subset W^{2n}_h & \; \subset W^{2n-2}_h \subset \cdots \subset W^2_h \subset W^0_h = H^{\rm even}(T^{2n};\Q), \label{eq:filtr-on-coh-even} \\
  \{ 0 \} \subset W^{2n-1}_h & \; \subset W^{2n-3}_h \subset \cdots \subset W^3_h \subset W^1_h = H^{\rm odd}(T^{2n};\Q),  \label{eq:filtr-on-coh-odd}
\end{align}
of vector subspaces over $\Q$, where $W^k_h \otimes_\Q \R$ supports 
a sub--representation of $\rho_{\rm spin}(h_{I,B'})$; here, 
\begin{align}
  W^{2n-2\ell}_h := \oplus_{m=0}^{\ell} H^{2n-2m}(T^{2n};\Q), \qquad 
  W^{2n-1-2\ell}_h := \oplus_{m=0}^\ell H^{2n-1-2m}(T^{2n};\Q).  
  \label{eq:subrepr-space-H-GHS}
\end{align}
The induced representation of $\rho_{\rm spin}(h_{I,B'})$ on 
$(W^k_h/W^{k+2}_h)\otimes \R$ agrees with the representation of $S^1_{I,B'}$ 
that describes the pure Hodge structure of weight-$k$ on $H^k(T^{2n};\Q)$
obtained from the complex structure $I$ alone. 
$S^1_{I,B'} \ni e^{i\alpha} \longmapsto e^{-i\Delta \alpha}$ with 
$\Delta = (p-q)$ on the Hodge $(p,q)$ component
 (cf \ref{statmnt:pHdgStr-by-S1-repr}).
\end{anythng}

It is therefore motivated to introduce the following notion.
\begin{defn}
\label{defn:gen-HS}
On a vector space $V_\Q$ over $\Q$, one may introduce a set 
of data $(h, W^\bullet)$ called a {\it generalized rational Hodge structure} whose properties are specified below. $W^\bullet$ is a decreasing filtration, 
a sequence of vector subspaces over $\Q$, 
$\{ 0\} \subset \cdots \subset W^1 \subset W^0 = V_\Q$, and $h$ 
is a representation of $S^1$, 
$h: S^1 \ni e^{i\alpha} \longmapsto h(e^{i\alpha}) \in {\rm GL}(V_\Q \otimes \R)$, 
where each one of the subspaces $W^k\otimes \R$'s supports 
a sub--representaton of $h$. We call the $h(e^{i\alpha}) = e^{-i\alpha \Delta}$ 
subspace of $W^k \otimes \C$ the {\it charge $\Delta$ component of } 
$W^k\otimes \C$. One might be interested in introducing the notion that 
a generalized rational Hodge structure is {\it polarizable}, but the 
authors do not feel fully ready to do so.\footnote{
An idea will be to generalize the notion of a polarizable rational 
Hodge structure of a K\"{a}hler manifold (cf Def. \ref{defn:polrz-HS}). 
When the decreasing filtration 
$W^\bullet$ terminates at $0 \subsetneq W^{2n} \cong \Q$, the set of information 
to be called a polarization of a generalized Hodge structure will include 
a bilinear pairing  $(-,-)_0: V_\Q \times V_\Q \rightarrow V_\Q$ such that 
$(W^k,W^\ell)_0 \subset W^{k+\ell}$, generalizing the notion of 
the wedge product on the middle dimensional cohomology group $H^n(M;\Q)$
when $\dim_\C M=n$. One may also include $(-,-)_{2p}: V_\Q \times V_\Q
 \rightarrow V_\Q$ where $(W^k,W^\ell)_{2p} \subset W^{k+\ell+2p}$. 
A polarization on the Hodge structure on $H^k(M;\Q)$ on an abelian 
variety $M$ with $\dim_\C M = n$ has been generalized to $(-,-)_{2(n-k)}$.
So, a tentative definition may be to demand a set of information 
$(-,-)_{2p}$ for $p=n,n-1,\cdots,0,\cdots, -n$, with a positive 
definiteness condition similar to the one for a pure rational 
Hodge structure (see Def. \ref{defn:polrz-HS}). 
} %
\end{defn}

So, for a closed (constant) 2-form $B'$ and a complex structure $I$ 
on $T^{2n}$, the set of data $(\rho_{\rm spin}(h_{I,B'}), W^\bullet_h)$ on $H^*(T^{2n};\Q)$
introduces a generalized rational Hodge structure. We call it 
the {\it horizontal generalized rational Hodge structure} for $(B'; I)$. 

\begin{lemma}\cite{Golyshev:1998vzz,Hitchin:2003cxu} 
\label{lemma:interpret-twist-byB}
Let us note in passing that {\it the spinor representation of the linear 
transformation }
\[ \left( \begin{array}{cc} {\bf 1} & \\ B' & {\bf 1} \end{array} \right) 
   \in \SO(\Lambda_\R, q) \]
{\it is $\exp \left[ 2^{-1} B' \wedge \right]$ on the representation space 
$H^*(T^{2n};\R)$.  }
\end{lemma}

\begin{anythng}
\label{statmnt:V-GCS}
 {\bf Vertical generalized Hodge structure on $H^*(T^{2n};\Q)$:}
Let $\omega = 2^{-1} \omega_{IJ} dX^I \wedge dX^J$ and 
$B' = 2^{-1} B'_{IJ} dX^I \wedge dX^J$ be a symplectic form and 
a real--valued closed 2-form on $T^{2n}$, respectively. Using $\omega$ 
and $B'$, a linear operator ${\cal J}_{B'}$ on $\Lambda_\R$ is introduced:
\begin{align}
 {\cal J}_{B'}: (\partial_{X^I}, dX^I) \longmapsto 
   (\partial_{X^L}, dX^L)
  \left( \begin{array}{cc} \delta^L_{\; K} & \\ B'_{LK} & \delta_L^{\; K} 
 \end{array} \right)
 \left( \begin{array}{cc}
        &  (\omega^{-1})^{JK} \\ \omega_{KJ} & \end{array} \right)
   \left( \begin{array}{cc}
      \delta^J_{\; I} & \\ -B'_{JI} & \delta_J^{\; I}
 \end{array} \right).
\end{align}
This operator satisfies 
\begin{align}
  ({\cal J}_{B'})^T \cdot q \cdot {\cal J}_{B'} = q, 
\end{align}
so we may find an element $i X_{\omega, B'}$ of the Lie algebra 
$\mathfrak{so}(\Lambda_\R, q)$,
\begin{align}
  {\cal J}_{B'} =: \exp \left[ \frac{\pi}{2}i X_{\omega, B'} \right], 
\end{align}
and define a representation \cite[\S8.4]{Golyshev:1998vzz}
\begin{align}
  h_{\omega, B'} : S^1 \ni e^{i\alpha} \longmapsto \exp \left[ i \alpha X_{\omega, B'}\right] \in {\rm GL}(\Lambda_\R). 
\end{align}
The $S^1$ subgroup of $\SO(\Lambda_\R, q)$ determined this way may be 
denoted by $S^1_{\omega, B'}$. 

We may introduce a representation of the $S^1_{\omega,B'}$ subgroup on 
$H^*(T^{2n};\Q) \otimes \R$ by restricting the spinor representation 
of $\SO(\Lambda_\R, q)$ on $H^*(T^{2n};\R)$. This is denoted by 
$\rho_{\rm spin}(h_{\omega,B'})$. The representation splits into 
the representation on $H^{\rm even}(T^{2n};\Q)\otimes \R$ and on
$H^{\rm odd}(T^{2n};\Q)\otimes \R$. For a general $\omega$ and $B'$, however, 
we have no reason to expect that there is a filtration structure (there 
is a sub--representation space defined over $\Q$) like we have in 
\ref{statmnt:H-GCS} or in Def. \ref{defn:gen-HS}. 

One can still verify by computation that the vector 
subspaces \cite[\S4.1 Ex.2]{Hitchin:2003cxu}
\begin{align}
  \C e^{2^{-1}(B'\pm i \omega)} \subset H^*(T^{2n};\Q)\otimes \C
\end{align}
are where the representation $\rho_{\rm spin}(h_{\omega,B'})$ becomes 
$1-$dimensional, with $\rho_{\rm spin}(h_{\omega,B'}): e^{i\alpha} \mapsto 
e^{ \mp in\alpha}$. 
Following some computation, one also finds that the vector spaces 
\begin{align}
  \C e^{2^{-1}(B' \pm i \omega)} dX^I \subset H^*(T^{2n};\Q)\otimes \C
\end{align}
for any $I=1,\cdots, 2n$ are where the representation 
becomes $1-$dimensional, with $\rho_{\rm spin}(h_{\omega,B'}): e^{i\alpha} \mapsto 
e^{\mp i\alpha(n-1)}$. Although it is possible to write down the 
generators of all the $2^{2n}$ one dimensional representations in a similar 
fashion, these are all that we will use in this article. 
\end{anythng}

\begin{anythng}
The arguments \ref{statmnt:H-GCS} and \ref{statmnt:V-GCS} are purely  mathematical, and are independent of each other. 
In the context of torus compactification, however, we have 
a metric $G$ on $T^{2n}$. When we choose a complex structure $I$ 
with which $G$ is compatible, we have a natural choice of 
a symplectic form, the K\"{a}hler form $\omega = \omega(-,-):= 2^{-1}G(I-,-)$;
when we write $\omega = 2^{-1} \omega_{IJ} dX^I\wedge dX^J$, then 
$\omega_{IJ} = I^K_{\; I}G_{KJ} = (I^T G)_{IJ}$. 

The operators ${\cal I}_{B'}$ and ${\cal J}_{B'}$ on $\Lambda_\R$
commute, and so do $X_{I,B'}$ and $X_{\omega, B'}$ when a common $B'$ is used 
for both. So, the two U(1) subgroups $S^1_{I,B'}$ and $S^1_{\omega, B'}$
in $\SO(\Lambda_\R,q)$ commute. 
\end{anythng}

We now proceed to make contact with the following result from string theory. 
\begin{lemma}
\label{lemma:when-mirror-exists}
 \cite[Prop. 4 and Prop. 8]{VanEnckevort:2003qc}
{\it Consider the $\mathcal N=(1,1)$ SCFT associated with a set of data $(T^{2n}; G, B)$. 
When we specify a pair of rank-$n$ primitive subgroups $\Gamma_f$ and $\Gamma_b$
of $H_1(T^{2n};\Z)$ so that $\Gamma_f \oplus \Gamma_b \cong H_1(T^{2n};\Z)$}, 
there is a fibration $T^{2n} \rightarrow \R^n/\Gamma_b = T^n$. 
The T--duality transformation along the fiber $T^n$ in string theory 
implies that there is a lattice isometry 
\begin{align}
 g: \Lambda = (\Gamma_f \oplus \Gamma_b) \oplus (\Gamma_f^\vee \oplus \Gamma_b^\vee) \rightarrow 
 (\Gamma_f^\circ \oplus \Gamma_b) \oplus 
     ((\Gamma_f^\circ)^\vee \oplus \Gamma_b^\vee) =: \Lambda^\circ 
\end{align}
with $g: \Gamma_f \cong (\Gamma^\circ_f)^\vee$ and 
$g: \Gamma_f^\vee \cong \Gamma_f^\circ$, and there is also an isomorphism 
$f$ from the total Hilbert space of the $\mathcal N=(1,1)$ SCFT 
for $(T^{2n};G,B)$ to that for $(T^{2n}_\circ;G^\circ, B^\circ)$. 

Suppose that $I$ is a complex structure on $T^{2n}$ with which $G$ is 
compatible. Then $I$ specifies one additional holomorphic (left-mover) 
U(1) current $J_L$ and one more in right-mover $J_R$ in the superchiral 
algebra so that the the original $\mathcal N=(1,1)$ superconformal algebra extends 
to an $\mathcal N=(2,2)$ superconformal algebra. 
The T--duality isomorphism $f$ between the Hilbert spaces specifies 
two current operators $J^\circ_L := f J_L f^{-1}$ and 
$J^\circ_R := - f J_R f^{-1}$ in the superchiral algebras of the 
$\mathcal N=(1,1)$ SCFT for $(T^{2n}_\circ; G^\circ,B^\circ)$, but it is not 
guaranteed that there exists some complex structure $I^\circ$ on 
$T^{2n}_\circ$ compatible with $G^\circ$ so that the pair $J^\circ_L$ 
and $J^\circ_R$ is reproduced from $I^\circ$.
We say that $(T^{2n};G, B;I)$ {\it has a geometric SYZ-mirror} when such an 
appropriate complex structure $I^{\circ}$ exists. 

{\it A geometric SYZ--mirror exists for the T--duality along $\Gamma_f \subset 
H_1(T^{2n};\Z)$ if and only if the following conditions are satisfied:}
\begin{align}
   \omega|_{\Gamma_f \otimes \R} = 0, \qquad  B|_{\Gamma_f \otimes \R} = 0. 
  \label{eq:cond-mirror-exists}
\end{align}
$\square$
\end{lemma}

Note that the condition (\ref{eq:cond-mirror-exists}) does not 
ask to find an isotropic $n$-dimensional vector space $\R^n$ 
within $H_1(T^{2n};\R)$, but to find an isotropic $n$-dimensional 
vector space $\Q^n \cong \Gamma_f \otimes \Q$ within $H_1(T^{2n};\Q)$. 
It is clear that a generic choice of $(G, B; I)$ would not have 
a geometric SYZ-mirror \cite[\S9.5]{Golyshev:1998vzz}. 

\begin{figure}[tb]
\begin{center}
\begin{tabular}{ccc}
  \includegraphics[width=0.45\linewidth]{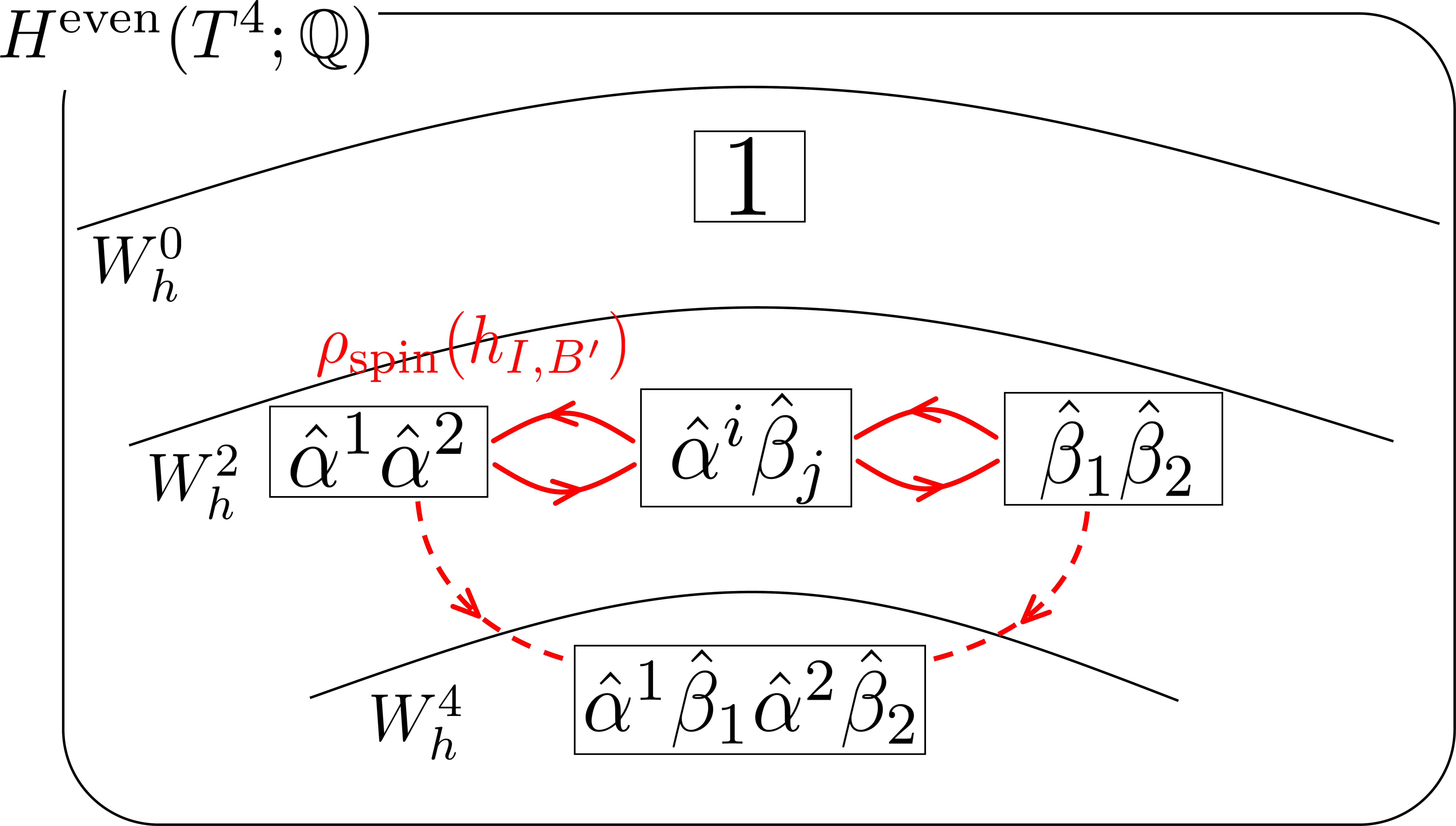} & & 
  \includegraphics[width=0.45\linewidth]{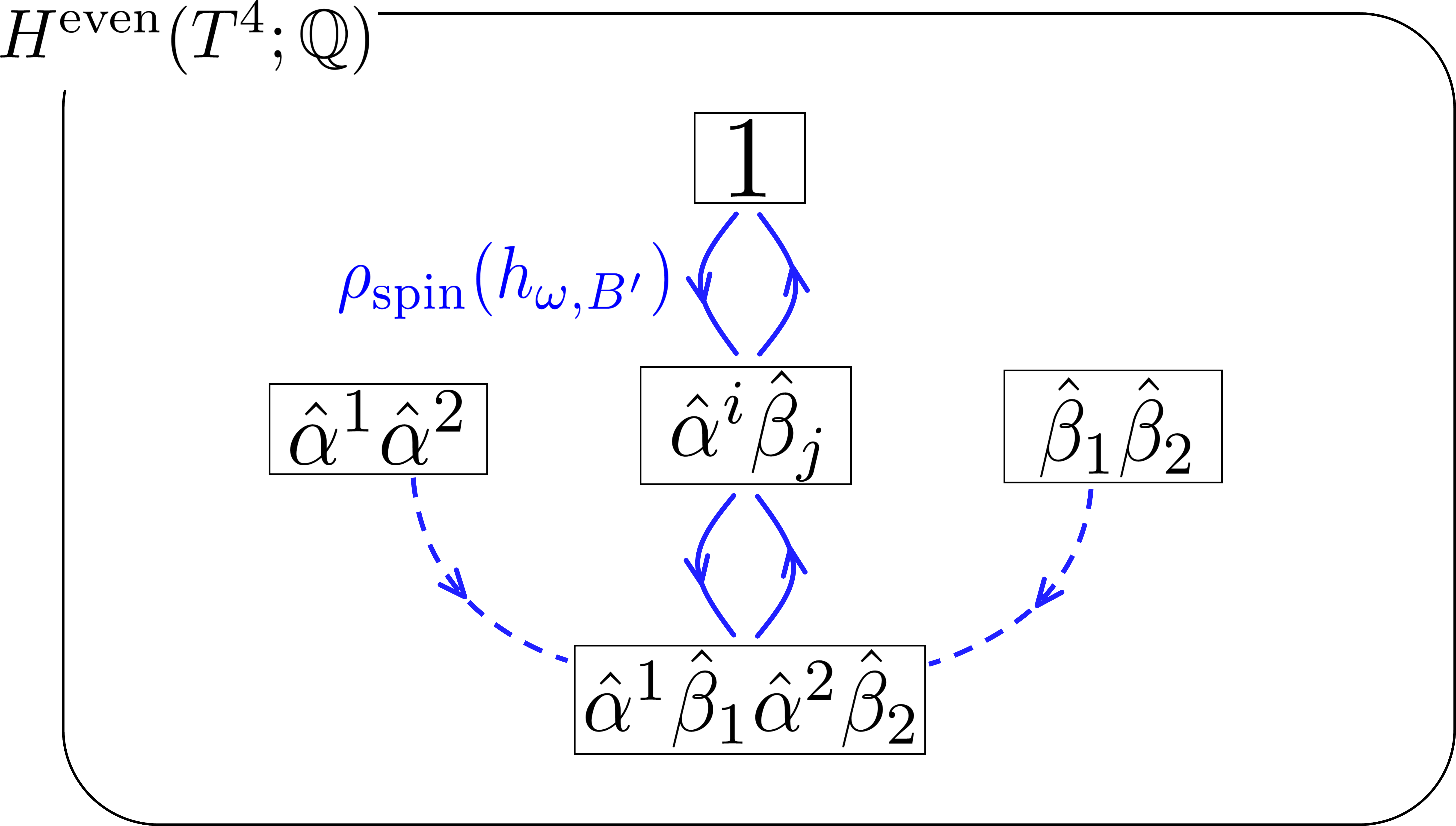}  \\
 (a) & & (b) 
\end{tabular}
\caption{  \label{fig:gen-HS-h+v}
This figure illustrates how the (a) horizontal and (b) vertical $S^1$ 
subgroups act on $H^*(T^4;\R)$, and how the filtration $W^\bullet_h$ 
is introduced on $H^*(T^4;\Q)$. See also Fig. \ref{fig:mult-filtr}
for the filtrations $g^*(W^\bullet_{h\circ})$. 
}
\end{center}
\end{figure}

\begin{anythng}
Suppose that the $\mathcal N=(1,1)$ SCFT for $(T^{2n}; G, B)$ with $(J_L, J_R)$ for 
a complex structure $I$ has a geometric SYZ--mirror for a T--duality along 
a rank-$n$ subgroup $\Gamma_f \subset \Gamma_f \oplus \Gamma_b \cong 
H_1(T^{2n};\Z)$. We use the same notation as 
in Lemma \ref{lemma:when-mirror-exists}. 

It is understood in string theory\footnote{
The isometry $g: (\Lambda, q) \cong (\Lambda^\circ,q^\circ)$
induces the isomorphisms 
$\SO(\Lambda_\R, q) \cong \SO(\Lambda_\R^\circ,q^\circ)$
and ${\rm Cliff}(\Lambda_\R, q) \cong {\rm Cliff}(\Lambda_\R^\circ,q^\circ)$; 
we abuse the notation and denote those two isomorphism as $g$.  
Thus, the left-${\rm Cliff}(\Lambda_\R^\circ,q^\circ)$-module 
$\mathfrak{a}_{\Gamma_f^\circ+\Gamma_b}$ can be regarded as 
a left-${\rm Cliff}(\Lambda_\R, q)$ module as well. 
A linear map $g: \mathfrak{a}_{\Gamma_f+\Gamma_b} \rightarrow
 \mathfrak{a}_{\Gamma_f^\circ+\Gamma_b}$ is determined by demanding 
that it is compatible with the action of ${\rm Cliff}(\Lambda_\R,q)$. 
Combining this isomorphism with the cohomology interpretation 
in footnote \ref{fn:conv-spinRpr-coh}, we obtain $g: H^*(T^{2n};\Q)
 \rightarrow H^*(T^{2n}_\circ;\Q)$ in the main 
text (e.g., \cite[\S3.3, \S3.5 and \S9.3]{Golyshev:1998vzz}). 
} %
 that there is an isomorphism 
$H^*(T^{2n};\Q) \cong H^*(T^{2n}_\circ;\Q)$ given by the map of D--brane 
charges having the same physical properties.
We abuse the notation and use $g$ for this isomorphism, too.
The cohomology group $H^*(T^{2n}_\circ;\Q)$ has a grading 
$\oplus_k H^k(T^{2n}_\circ;\Q)$, as well as a filtration structure 
$W^\bullet_{h\circ}$ (i.e., $W^\bullet_h$ in (\ref{eq:subrepr-space-H-GHS}) 
for the mirror theory), and the data $(B^\circ, I^\circ)$ introduces 
a pure rational Hodge structure (resp. the horizontal 
generalized rational Hodge structure) on $H^k(T^{2n}_\circ;\Q)$ 
(resp. $H^*(T^{2n}_\circ;\Q)$). Those structures can be superimposed 
on $H^*(T^{2n};\Q)$ by pulling them back via the isomorphism $g$. 
Using the isometry $g: (\Lambda,q) \rightarrow (\Lambda^\circ,q^\circ)$, 
$g^{-1} {\cal I}_{B^\circ} g = {\cal J}_{B}$ and 
$g^{-1} S^1_{I^\circ,B^\circ}g = S^1_{\omega,B}$ \cite{Kapustin:2000aa}. 
In the spinor representation, the horizontal generalized Hodge structure 
$(\rho_{\rm spin}(h_{I^\circ,B^\circ}), W^\bullet_{h\circ})$ on $H^*(T^{2n}_\circ;\Q)$ 
is mapped into a generalized rational Hodge structure on $H^*(T^{2n};\Q)$ by 
$(\rho_{\rm spin}(h_{\omega, B}), g^*(W^\bullet_{h\circ}))$. 
We call this generalized rational Hodge structure as 
the {\it vertical generalized rational Hodge structure}.\footnote{
The vertical generalized rational Hodge structure splits into pure rational 
Hodge structures of weights ranging from 0 to $2n$, when the condition 
\begin{align}
  \omega|_{\Gamma_b \otimes \R} = 0 , \qquad 
  B|_{\Gamma_b \otimes \R} = 0
\end{align}
is also satisfied.
} %
See Fig. \ref{fig:gen-HS-h+v} for an illustration. 

The rational Hodge structure on $H^n(T^{2n}_\circ;\Q)$ by $I^\circ$ 
is polarized with respect to the wedge product on $T^{2n}_\circ$:
\begin{align}
 (W^n_{h\circ}/W^{n+2}_{h\circ}) \times (W^n_{h\circ}/W^{n+2}_{h\circ}) 
   \ni (\psi, \chi) \longmapsto \int_{T^{2n}_\circ} \psi \wedge \chi \in \Q.  
\end{align}
When this bilinear form (symmetric if $n$ is even, and anti--symmetric
if $n$ is odd) is pulled back by $g$ to $H^*(T^{2n};\Q)$, it becomes 
\begin{align}
  g^*(W_{h\circ}^n/W^{n+2}_{h\circ}) & \; \times g^*(W_{h\circ}^n/W_{h\circ}^{n+2})
       \ni (g^*(\psi), g^*(\chi))    \label{eq:def-vert-polarizatn} \\
  & \; \longmapsto    (-1)^{\frac{n(n-1)}{2}} \int_{T^{2n}}
   \left(\sum_{k=0}^n (-1)^k \Pi_{2k} g^*(\psi)\right) \wedge  g^*(\psi) \in \Q, 
   \nonumber 
\end{align}
where $\Pi_{2k}$ is the projection $H^*(T^{2n};\Q) \rightarrow H^{2k}(T^{2n};\Q)$. 
\end{anythng}

\begin{rmk}
The most natural choice of the operators ${\cal I}_{B'}$ and ${\cal J}_{B'}$
are for $B'$ equal to the $B$--field in the data $(T^{2n};G, B)$ of 
an $\mathcal N=(1,1)$ SCFT. In this case, ${\cal I}_{B}{\cal J}_{B}$ is 
an operation multiplying $-1$ to the right-moving momentum 
on $\Lambda_\R$ \cite[eq. (2.10)]{VanEnckevort:2003qc}.
It is still possible, mathematically, to define 
those operators with $B'$ chosen differently than $B$ itself; 
then different U(1) subgroups $S^1_{I,B'}$ and $S^1_{\omega, B'}$ are 
specified within $\SO(\Lambda_\R ,q)$.  For a technical reason 
in the presentation, we will also use such subgroups in footnote
\ref{fn:B-alg-vU(1)}. 
\end{rmk}

\begin{anythng}
\label{statmnt:multi-mirror}
Suppose that there are two geometric SYZ-mirrors for the $\mathcal N=(1,1)$ SCFT 
with a set of data $(T^{2n};G,B;I)$. 
Let $\Gamma_{fi} \oplus \Gamma_{bi} \cong H_1(T^{2n};\Z)$ for $i=1,2$ 
be the split of the 1-cycles into those in the fiber (to be taken T-dual)
and those in the base, and $g_{i}$ and $f_{i}$ (with $i=1,2$) 
the corresponding lattice isometries and the Hilbert space isomorphisms, 
respectively. 

Both representations $\rho_{\rm spin}(h_{I^\circ_{(i)},B^\circ_i})$ on 
$H^*(T^{2n}_{\circ(i)};\R)$ for $i=1,2$ are pulled back by $g_i$ to 
one identical representation on $H^*(T^{2n};\R)$, and that is 
the representation $\rho_{\rm spin}(h_{\omega, B})$. It is therefore 
economical to deal with $\rho_{\rm spin}(h_{\omega, B})$ instead of 
$\rho_{\rm spin}(h_{I^\circ_{(i)},B^\circ_i})$. 
The gradings and the filtrations pulled back to 
$H^*(T^{2n};\Q)$---$g_i^*(H^k(T^{2n}_{\circ(i)};\Q))$ and 
$g_i^*(W^k_{h\circ(i)})$---are however not identical\footnote{
The pulled back filtrations, $g_i^*(W^\bullet_{h\circ(i)})$, 
are identical, if $\Gamma_{f1} = \Gamma_{f2}$ (even when 
$\Gamma_{b1} \neq \Gamma_{b2}$).   
} %
 for different geometric SYZ-mirrors $i=1,2$ (cf Fig. \ref{fig:mult-filtr}).  
\end{anythng}


\subsection{Coarse Classification of CM-type Abelian Surfaces}
\label{ssec:CM-abel-surface}

A complex torus $M = \C^n/\Z^{\oplus n} = (T^{2n};I)$ of dimension $n$ 
is regarded as an {\it abelian variety} when there exists a {\it polarization}, 
which means the existence of a $\psi \in H^2(M;\Z) \cap H^{1,1}(M;\R)$ such that 
the bilinear form $\psi(I -, -): (X, Y) \mapsto \psi(I X, Y)$ for 
$X, Y \in H_1(T^{2n};\R)$ is positive definite. It is a non-trivial 
condition on $I$ whether a polarization exists or not. 
Mathematicians tend to favor abelian varieties over general complex tori
because abelian varieties may be treated as algebraic varieties (rather 
than complex analytic manifolds). String theorists, however, do not have any a priori reason to be in favor of a complex structure on $T^{2n}$ that 
allows a polarization over those that do not. So, here, we introduce 
the following definition for a general complex torus that is not 
necessarily an abelian variety. 
\begin{defn}
\label{def:CM-cpx-trs}
Let $M=(T^{2n};I)$ be a complex torus of dimension $n$; then 
a rational Hodge structure is given on $H^1(M;\Q)$. The following 
two conditions are known to be equivalent:\footnote{
The proof of Props. 17.3.4 and 17.3.5 of \cite{birkenhakeabelian} does not 
assume that the Hodge structure in question admits a polarization.
} %
\begin{itemize}
 \item [(i)] The algebra ${\rm End}(H^1(M;\Q))^{\rm Hdg}$ over $\Q$ 
    contains a commutative semi-simple subalgebra\footnote{
\label{fn:comm-ss-alg}
When a finite dimensional semi-simple algebra over $\Q$ is commutative, 
then it is of the form of $\oplus_{\alpha \in {\cal A}} F_\alpha$, the direct 
sum of a finite number of number fields $F_\alpha$. Conversely, an 
algebra of this form is always semi-simple and commutative. 

{\bf An addendum in arXiv ver. 2:} only one of the authors (TW) is held 
responsible for the rest of this footnote. Although the condition (i) 
is a natural generalization of the definition of CM-type from abelian
varieties to complex tori, there is another way to generalize that is 
equally natural. One might also think of defining complex tori with 
{\it sufficiently many complex multiplications} by imposing the following 
condition: (i$'$) The semi-simplification of the algebra $\mathfrak{R}: = 
{\rm End}(H^1(M;\Q))^{\rm Hdg}$ over $\Q$, i.e., the quotient 
$\mathfrak{R}/J$ by the radical $J:= J(\mathfrak{R})$ of 
$\mathfrak{R}$, contains a commutative semi-simple subalgebra of 
dimension $2n$. It turns out (as explained shortly), however, that the 
two conditions (i) and (i') are equivalent; so it does not matter which 
one is used for the definition. 

To see that the condition (i) implies the condition (i$'$), one just has 
to note that a semi-simple subalgebra $F \subset \mathfrak{R}$ is 
injectively mapped into the quotient $\mathfrak{R}/J$. 
To see that the condition (i$'$) implies (i), think of a case where $M$ is 
an indecomposable complex torus of dimension $n$ for simplicity 
(we use terminology of 
\cite{birkenhaketori}). The algebra $\mathfrak{R}$ is a local algebra, 
and the division algebra $\mathfrak{R}/J$ contains 
a commutative algebra $F$ (a field $F$ in this simple case) because 
of the condition (i$'$). Now, there is a sequence of complex subtori 
$M \supset JM \supset J^2 M \supset \cdots \supset J^m M = \{0\}$; 
here, we adapt the idea of Loewy series of a module 
(\cite[p. 346]{MR1245487}, \cite[\S1]{MR998775}) to a complex torus.  
The division algebra $\mathfrak{R}/J$ and its subfield $F$ is realized on 
each of the complex tori $J^iM/J^{i+1}M$ with $i=0,\cdots, m-1$ 
as the endomorphism algebra.  If a non-zero element $\phi+J$ of 
$\mathfrak{R}/J$ or 
$F$ were to be realized trivially in any one of the tori $J^iM/J^{i+1}M$, 
then $\phi \in \mathfrak{R}$ must be nilpotent, which is a contradiction. 
So, a degree $2n$ field $F$ has an embedding into ${\rm End}_\Q(J^iM/J^{i+1}M)$ 
for some $i \in \{0,\cdots, m-1\}$, and we learn that the complex torus 
$J^iM/J^{i+1}M$ is of $n$ dimension, which further implies that 
$J^{i+1}M=0$, and $J^iM = M$, so $m=1$, $i=0$, and $J(\mathfrak{R})=0$ 
in the end. Under the condition (i$'$), the semi-simple commutative 
subalgebra $F \subset \mathfrak{R}/J$ is actually a subalgebra of 
$\mathfrak{R}$ (i.e., the condition (i)), and an indecomposable complex 
torus satisfying the condition (i$'$) is always a simple complex torus. 
$\square$
} %
 of dimension $2n$.
 \item [(ii)] The Hodge group of the Hodge structure ${\rm Hg}(M)$ is 
   commutative. 
\end{itemize}
The definition of a Hodge group is given in Def. \ref{defn:Hdg-grp};
very little intuition on the Hodge group is required, however, in following 
the related arguments in section \ref{ssec:pol}.  

When either one of the above conditions (and hence both) are satisfied, 
we say that the complex torus has/is with {\it sufficiently many complex 
multiplications}, and also that the rational Hodge structure is 
{\it with sufficiently many CM}. In the case a complex 
torus $M$ (and its rational Hodge structure on $H^1(M;\Q)$) with the 
property (i), (ii) admits a polarization, we say that $M$ is 
a {\it CM abelian variety}, and the Hodge structure is of {\it CM-type}.
\end{defn}

\begin{rmk}
\label{rmk:difference-cm-tpri-abelV}
It may seem a little odd to use different jargons for one and 
the same properties, (i) and (ii), depending on whether existence 
of a polarization is guaranteed or not. Such a choice of jargons 
partially reflects the fact that the properties (i) and (ii) mean 
a lot more when they are combined with a polarization. 

Here are a few properties that hold true 
only when a polarization exists 
(see e.g., \cite{birkenhakeabelian,milnecm,shimura2016abelian}):
\begin{itemize} 
 \item The algebra ${\rm End}(H^1(M;\Q))^{\rm Hdg}$ is semi-simple, 
   so this algebra has the structure (\ref{eq:WeddB-alg-str}). 
 \item The semi-simple algebra ${\rm End}(H^1(M;\Q))^{\rm Hdg}$ 
 with the structure (\ref{eq:WeddB-alg-str}) acts faithfully\footnote{
A structure theory on ${\rm End}(H^1(M;\Q))^{\rm Hdg}$ of 
a complex torus, not necessarily with a polarization, is found 
in \cite[\S1.7 and \S 1.8]{birkenhaketori}. 
} %
 on the vector space $H^1(M;\Q)$ (by definition); when this algebra
 contains a $2n$-dimensional commutative subalgebra (so, 
  its structure is of the form in footnote \ref{fn:comm-ss-alg}), 
  that means that the division algebra $D_\alpha$ is commutative, 
  (i.e., $D_\alpha = k_\alpha$ and $q_\alpha = 1$), and 
  the number field $F_\alpha$ is a degree-$n_\alpha$ extension\footnote{
Footnote \ref{fn:construct-nonprim-CM-F} provides a pedagogical 
explanation on how to construct an extension $F_\alpha/k_\alpha$.
} %
  of the number field $k_\alpha$. 
 \item Both $k_\alpha$ and $F_\alpha$ are CM-fields. 
\end{itemize}
\end{rmk}

A little more information is provided in section \ref{ssec:pol} 
on complex tori with sufficiently many complex multiplications. 
Although complex tori with sufficiently many complex multiplications are 
more general 
than CM abelian varieties (making it desirable to have 
a theory relating rational SCFTs with such tori), we will be able to 
confirm such a connection only for CM abelian varieties. This is for 
scientific reasons, not a matter of mathematical taste, preference 
or interests. 

For that reason, it makes sense to prepare ourselves to work 
specifically with CM abelian varieties. Let us quote a result 
of classification of CM-type abelian varieties of $n=2$ dimensions. 
That is essentially done by classifying the CM algebras 
$\oplus_{\alpha \in {\cal A}} F_\alpha$ of dimension $2n=4$. 
\begin{lemma}
\label{lemma:classify-deg4-CM-alg}
\cite[pp.64--65 Ex.8.4.(2)]{shimura2016abelian}
{\it There are four different kinds of CM algebras $\oplus_\alpha F_\alpha$ over 
$\Q$ of dimension 4. }
\begin{itemize}
 \item [(A)] The CM algebra is a CM field (i.e., $|{\cal A}|=1$), 
     and $F=F_\alpha \cong \Q[x,y]/(y^2-d, x^2-p)$ for some square-free 
     integers $d > 1$ and $p<0$. This field $F$ is an extension of 
     the imaginary quadratic field $K^{(2)} \cong \Q[x]/(x^2-p)$. 
     The totally real subfield of $F$ is $\Q[y]/(y^2-d)$. 
 \item [(A')] The CM algebra is of the form $K^{(2)}_1 \oplus K^{(2)}_2$, 
   where $K^{(2)}_1 \cong \Q[x_1]/(x_1^2-p_1)$ and
    $K^{(2)}_2 \cong \Q[x_2]/(x_2^2-p_2)$ are imaginary quadratic fields 
   that are not mutually isomorphic. That is, $p_1,p_2$ are negative 
   square-free integers, and $p_1 \nin p_2 (\Q^\times)^2$. 
   $|{\cal A}|=2$ in this case. 
 \item [(B, C)] The CM algebra is a CM field (i.e., $|{\cal A}|=1$), 
   $F=F_\alpha =: K$, that does not contain a CM subfield. Such a degree-4 
   CM field always has a structure $K \cong \Q[x,y]/(y^2-d, x^2-p-qy)$ 
   for some square-free integer $d>1$ and rational numbers $p,q$ such that 
   $p<0$, $q\neq 0$ and $d' := p^2-q^2d>0$.  
   The two distinct cases $d' \in d (\Q^\times)^2$ and $d' \nin d (\Q^\times)^2$ 
   are called the case (B) and (C), respectively; the field extension 
   $K/\Q$ is Galois and non-Galois, respectively, in the two cases. 
   The totally real subfield is $\Q[y]/(y^2-d)$. 
\end{itemize}
{\it The distinction between 
the cases B and C is not very important in the analysis in this article.  }
\end{lemma}

The case (A) looks as if it is the cases (B, C) with just setting $q=0$. 
There is a clear difference between the case (A) and the cases (B, C), 
however. The difference is seen in the reflex field of the CM field $F$ 
in the case (A) and the field $K$ in the cases (B, C)
(e.g., \cite{shimura2016abelian, milnecm}). To be more explicit:
\begin{anythng}
\label{statmnt:CMtype-n-reflexF}
For the CM field $F$ in {\bf case (A)}, the four embeddings 
$F\hookrightarrow \overline{\Q}$ are denoted by $\tau_{\pm \pm}$, where 
$\tau_{\pm *}: y \mapsto \pm \sqrt{d}$ and 
$\tau_{*\pm}: x \mapsto \pm \sqrt{p} = \pm i \sqrt{-p}$.  
Throughout in this article, we mean by $\sqrt{d}$ for $d \in \R_{>0}$ 
the positive square root, and by $\sqrt{p} = i\sqrt{-p}$ for 
$p \in \R_{<0}$ the square root in the upper half complex plane. 

There seem to be two choices of a CM-type of the CM-field, 
$\{ \tau_{++}, \tau_{-+}\}$ and $\{\tau_{++}, \tau_{--} \}$; in fact,  
we have an alternative presentation $F \cong \Q[x', y]/(y^2-d, (x')^2-pd)$
due to the isomorphism $xy \leftrightarrow x'$, and the set of embeddings 
$\{ \tau_{++}, \tau_{--}\}$ for $F\cong \Q[x,y]/(y^2-d, x^2-p)$ is regarded 
as $\{ \tau_{++}, \tau_{-+}\}$ for $F\cong \Q[x',y]/(y^2-d, (x')^2-pd)$. 
So, we do not lose generality by thinking only of the CM-type 
$\Phi := \{ \tau_{++}, \tau_{-+}\}$. 

For the CM-type $(F, \Phi)$, the reflex field $F^r$ 
is $\Q[\xi^r]/((\xi^r)^2-p)$, 
which is an imaginary quadratic field. The reflex field of the reflex 
field is $F^{rr} = \Q(\sqrt{p})$, and $F^{rr}$ is a proper-subfield of $F$
in the case (A). 
It is also easy to see this directly from the fact that the CM-type $\Phi$ 
is not primitive, but is induced from the CM-type 
$(K^{(2)}, \tau_{*+}:x\mapsto \sqrt{p})$ \cite[\S8]{shimura2016abelian}. 

For the CM field $K$ in the {\bf cases (B, C)}, on the other hand, 
the four embeddings 
$K \hookrightarrow \overline{\Q}$ are denoted by $\tau_{\pm \pm}$, where 
\begin{align}
 \tau_{\pm *}: y \mapsto \pm \sqrt{d}, \qquad 
 \tau_{\pm +}:& \;  x \mapsto \sqrt{p \pm q\sqrt{d}}
                         = i \sqrt{-p\mp q\sqrt{d}}, \\
 \tau_{\pm -}: & \; x \mapsto - \sqrt{p \pm q \sqrt{d}}
                         = -i\sqrt{-p\mp q\sqrt{d}}.
  \label{eq:introduce-sh-notation-below}
\end{align}
We introduce a short-hand notation $\sqrt{+} := \sqrt{p+q\sqrt{d}}$ and 
$\sqrt{-} := \sqrt{p-q\sqrt{d}}$ for the pure imaginary complex numbers 
in the upper half plane and use it for the sake of compactness of notation 
in this article. 

There seem to be two inequivalent choices of a CM-type of the CM-field $K$, 
namely, $\{ \tau_{++}, \tau_{-+} \}$ and $\{\tau_{++}, \tau_{--} \}$. We can 
change the presentation of the field $K$ (with different values of $p, q$) 
so that the choice $\{ \tau_{++}, \tau_{--} \}$ is regarded as 
$\{ \tau_{++}, \tau_{-+}\}$ in the new presentation, as we have done in the 
case (A). So, we do not lose generality by considering only one CM-type 
$\Phi := \{ \tau_{++}, \tau_{-+}\}$. 

The reflex field $K^r$ of $(K, \Phi)$ is not (necessarily) isomorphic to $K$
in the cases (B, C), but is a degree-4 number field  
\begin{align}
 K^r \cong \Q[y', \xi^r]/((y')^2-d', (\xi^r)^2-2p+2y').
\end{align}
The reflex field of $K^r$, denoted by $K^{rr}$, is $K$ itself in the 
cases (B, C). That is the difference between the case at hand and case (A); 
this difference in algebraic notation/terminology is also reflected in 
geometric notation/terminology concerning abelian varieties 
as we quote a statement below (Lemma \ref{lemma:classify-abel-surf}). 

Before moving on, however, let us introduce notations $\tau^r_{\pm \pm}$
for the four embeddings of the reflex field 
$K^r \hookrightarrow \overline{\Q}$. 
\begin{align}
 \tau^r_{\pm *}: y' \mapsto \pm \sqrt{d'}, \qquad 
 \tau^r_{\pm +}: & \; \xi^r \mapsto \sqrt{p+q\sqrt{d}} \pm \sqrt{p-q\sqrt{d}}, \\
 \tau^r_{\pm -}: & \; \xi^r \mapsto
    - \left( \sqrt{p+q\sqrt{d}} \pm \sqrt{p-q\sqrt{d}} \right).
\end{align}
\end{anythng}

\begin{anythng}
\label{statmnt:CMtype-n-reflexF-Aprm}
Let us also write down a little bit of information on the reflex field 
in the {\bf case (A')}, because we use that later in this article. 
The two embeddings of the imaginary quadratic fields $K^{(2)}_i$ are given 
by $\tau_{i\epsilon_i}: x_i \mapsto \epsilon_i \sqrt{p_i} =
 \epsilon_i i\sqrt{-p_i}$ for $\epsilon_i \in \{ \pm \}$. 
A CM-type\footnote{
See \cite[Def. 1.17]{milnecm} for the definition of the reflex field 
of a CM algebra that is not a CM field. 
} %
 of the CM algebra $K^{(2)}_1 \oplus K^{(2)}_2$ 
must be for $\Phi = \{ (\tau_{1\epsilon_1}, \tau_{2\epsilon_2}) \}$
for some choice of $(\epsilon_1,\epsilon_2)$. For any one of them, 
the reflex field is $K^r = \Q[x_1,x_2]/(x_1^2-p_1,x_2^2-p_2)$, 
which is isomorphic to $\Q[\xi^r, y']/((\xi^r)^2 - p_1, (y')^2 - d')$ with 
$d'=p_1p_2 > 0$ through $x_1 \mapsto \xi^r$ and $x_1x_2 \mapsto y'$. 
The four embeddings may be denoted by $\tau^r_{\epsilon',\epsilon^r}$, 
where $\tau^r_{ \pm *}:y'\mapsto \pm \sqrt{d'}$ and 
$\tau^r_{*\pm}: \xi^r \mapsto \pm \sqrt{p_1}$.  $\square$
\end{anythng}

Let us now quote the known result stating how the classification of 
degree-4 CM algebras above is translated to the classification of CM-type 
abelian varieties of complex dimension 2 
(abelian surfaces). 

\begin{lemma}
\label{lemma:classify-abel-surf}
{\it Let $M$ be an abelian surface of CM-type. Then it must be in one of the 
following mutually exclusive cases. }
\begin{itemize}
\item [(A)] There is an isogeny $\varphi: M \longrightarrow E \times E$ 
where $E$ is an elliptic curve of CM-type with 
${\rm End}(H^1(E;\Q))^{\rm Hdg} \cong K^{(2)} \cong \Q[x]/(x^2-p)$. 
In this case, 
\begin{align}
  {\rm End}(H^1(M;\Q))^{\rm Hdg} \cong M_2(K^{(2)}); 
\end{align}
for any square-free integer $d>1$, one can find within\footnote{
\label{fn:construct-nonprim-CM-F}
Such a CM field $F \cong \Q[x,y]/(x^2-p, y^2-d)$ within $M_2(K^{(2)})$
can be constructed as follows. First, think of a map 
$\phi_x: z^1 \mapsto \sqrt{p}z^1$, $z^2 \mapsto \sqrt{p} z^2$. 
The pull-back $\phi_x^*$ generates the center $K^{(2)}{\bf 1}_{2\times 2}$ 
of the algebra $M_2(K^{(2)})$. Next, for a square-free integer $d>1$ and 
a complex multiplication $\xi \in {\rm End}(H_1(E;\Q))^{\rm Hdg}
 \backslash \{ 0\}$, think of a map $\phi_{d,\xi}: z^1 \mapsto d \xi z^2$, 
$z^2 \mapsto \xi^{-1} z^1$. Then $(\phi_{d,\xi}^*) \in 
{\rm End}(H^1(M;\Q))^{\rm Hdg}$ has the property of the generator $y$, 
so we may set $F$ to be the subalgebra of $M_2(K^{(2)})$ generated by 
$x = \phi_x^*$ and $y = \phi_{d,\xi}^*$. 

Such a subfield $F$ within the algebra $M_2(K^{(2)})$ has the following property 
characterized by a polarization ${\cal Q}_{d,\xi} \in \sqrt{p} \left( 
dz^1\wedge d\bar{z}^{\bar{1}} 
+ d{\rm Nm}(\xi) dz^2 \wedge d\bar{z}^{\bar{2}} \right) \Q$.
When we assign to $\phi \in {\rm End}(H_1(M;\Q))^{\rm Hdg} \cong M_2(K^{(2)})$
another endomorphism $\phi' \in {\rm End}(H_1(M;\Q))^{\rm Hdg}$ by 
${\cal Q}(\phi'-,-) = {\cal Q}(-,\phi -)$ (this is called \textit{Rosati involution} 
with respect to ${\cal Q}$), then $\phi'_x = - \phi_x$, 
and $\phi'_{d,\xi} = \phi_{d,\xi}$, so $\phi' \in F$ for any 
$\phi \in F$ (i.e., the subfield $F$ for $d$, $\xi$ is closed 
under the involution by ${\cal Q}_{d,\xi}$). 
See \cite[Prop. 3.6 (b)]{milnecm} for more information.

All above in this footnote are written by mathematicians in many articles 
explaining the theory of complex multiplication of abelian varieties, but 
are often expressed only in abstract and general terms. 
So, we pursued a hands-on style of presentation preferred by string 
theorists here. 

In the context of string theory, it is usually not motivated to fix 
an embedding of a complex analytic manifold $M$ to a projective space 
to see it as an algebraic variety. String theorists are not worried about 
too many automorphisms either. So, there is no particular reason 
to restrict one's attention only to a proper subalgebra $F$ of $M_2(K^{(2)})$. 
It is still good to know that $M_2(K^{(2)})$ contains a subfield $F$ 
that is CM and a degree-2 extension of $K^{(2)}$, because that is enough 
to be able to apply a useful fact written as Lemma \ref{lemma:very-useful}.  
} %
 the algebra $M_2(K^{(2)})$ a subfield $F$ of the property (A) 
in the classification in Lemma \ref{lemma:classify-deg4-CM-alg}.
\item [(A')] There is an isogeny $\varphi: M \longrightarrow E_1 \times E_2$ 
where $E_i$ is an elliptic curve of CM-type with 
  ${\rm End}(H^1(E_i;\Q))^{\rm Hdg} \cong K^{(2)}_i  \cong \Q[x]/(x^2-p_i)$
(for both $i=1,2$). 
\item [(B,C)] $M$ does not contain an abelian subvariety. 
${\rm End}(H^1(M;\Q))^{\rm Hdg} \cong K$. 
\end{itemize}
Abstract elements denoted by $x$, $y$ are then regarded as 
`Hodge structure preserving' endomorphisms on $H^1(M;\Q)$. 
\end{lemma}

\section{Choice of Complex Structure}
\label{sec:choose-cpx-str}

As we have remarked in Discussion \ref{statmnt:subtlty-choose-cpx-str}, 
there is no way not choosing a complex structure on $T^{2n}$
when we wish to establish a Gukov--Vafa-like characterization 
of rational $T^{2n}$-target (S)CFTs. On one hand, it is desirable to find a characterization statement that works well 
for a broader class of complex structures on $T^{2n}$. For example, 
it may be natural for algebraic geometers to pay attention 
only to a complex structure $I$ such that $(T^{2n}; I)$ is an abelian 
variety (e.g. \cite{Chen:2005gm}), i.e., an object in the category of 
algebraic varieties (instead of a general complex torus as an object of 
the category of complex analytic manifolds). 
As string theorists, however, we should give a thought whether such 
a characterization has a chance to work for a complex structure 
not necessarily with a polarization. That is the subject of 
section \ref{ssec:pol}.

On the other hand, if there is a choice of a complex structure $I$
that is motivated well in string theory, there is a chance that we have 
a sharper/clearer characterization statement for rational $T^{2n}$-target 
(S)CFTs. That is what we aim for in section \ref{ssec:B-transc}. 

\subsection{Polarization}
 \label{ssec:pol}

Let us first extract more information from the definition 
of a complex torus $M$ with sufficiently many complex multiplications. 
\begin{anythng}
As a direct consequence of the definition, there must be an algebra of endomorphisms 
of the form $\oplus_{\alpha \in {\cal A}} F_\alpha$ acting faithfully 
on $H^1(M;\Q)$; here, $F_\alpha$ is a number field. 
Since the action is faithful, the comparison of the dimensions implies that the vector space $H^1(M;\Q)$ 
should have a structure 
$H^1(M;\Q) \cong \oplus_{\alpha \in {\cal A}} [H^1(M;\Q)]_\alpha$, where 
$[H^1(M;\Q)]_\alpha$ is a 1-dimensional vector space of $F_\alpha$. 
See footnote \ref{fn:comm-ss-alg} and Lemma \ref{lemma:Wedderburn} 
for more background information. We can apply the following discussion 
to individual pairs $F_\alpha$ and $[H^1(M;\Q)]_\alpha$, so we drop 
the subscript now. 

Any endomorphism in $F \subset {\rm End}(H^1(M;\Q))^{\rm Hdg}$ maps  
the Hodge (1,0) and (0,1) components of $H^1(M;\Q)$ 
to themselves. So, the action of the endomorphisms in $F$ can be 
diagonalized simultaneously on the two Hodge components separately. 
The simultaneous eigenstates of the action of $F$ explained in 
Lemma \ref{lemma:very-useful} should therefore belong either 
to the (1,0) component, or to the (0,1) component. 
The set ${\rm Hom}(F,\C)$ of embeddings of the field $F$ 
is therefore separated into two subsets, $\Phi \subset 
{\rm Hom}(F,\C)$ for the (1,0) components and 
$\overline{\Phi}$ for the (0,1) components. Moreover, 
the set $\overline{\Phi}$ consists of the embeddings 
in $\Phi$ followed by the complex conjugation in $\C$.
This means that a number field $F$ in the context of a complex torus 
with sufficiently many complex multiplications must be a totally 
imaginary field. 

Let $F$ be a totally imaginary field of degree $2n$, 
and $\Phi = \{ \tau_a \} \subset {\rm Hom}(F,\C)$ a set of $n$ 
embeddings, any two of which are not mutually complex conjugate 
of the other.  Then one can construct a complex torus $\C^n/\Z^{\oplus 2n}$
of $n$-dimensions by choosing a basis $\{ \eta_{I=1,\cdots, 2n}\}$ 
of $F/\Q$ and setting $\Z^{\oplus 2n} \hookrightarrow \C^{n}$ to be 
$(n_1,n_2,\cdots, n_{2n}) \mapsto (\tau_1(n_I\eta_I), \; \tau_2(n_I \eta_I), 
\cdots, \tau_{n}(n_I \eta_I)) \in \C^n$. 
The $2n$ vectors $(\tau_{a=1,\cdots,n}(\eta_I)) \in \C^n \cong \R^{2n}$ 
for $I=1,\cdots, 2n$ are automatically linearly independent over $\R$; 
to see this, it is enough to note that the $2n\times 2n$ matrix 
$(\tau_a(\eta_I), \bar{\tau}_a(\eta_I))_{I,a\bar{a}}$ has a non-zero 
determinant (see Lemma \ref{lemma:very-useful}).
 The algebra of endomorphisms of this complex torus contains a subalgebra 
isomorphic to $F$ (use Lemma \ref{lemma:very-useful-2}). 
\end{anythng}
\begin{rmk}
CM fields constitute a special subclass of totally imaginary fields. 
A complex torus with sufficiently many complex multiplications is 
a CM abelian variety if and only if its endomorphism algebra 
$\oplus_\alpha F_\alpha$ is made of totally imaginary fields $F_\alpha$ 
that are all CM fields. The implication $\Rightarrow$ is from the 3rd 
property quoted in Rmk. \ref{rmk:difference-cm-tpri-abelV}. 
The implication $\Leftarrow$ is from Lemma \ref{lemma:gen-Sm}. 
Examples of totally imaginary fields that are not CM fields are 
found\footnote{
Math StackExchange entry ``totally imaginary number field of degree 4'' \\
\url{https://math.stackexchange.com/questions/4372232/}
} %
 in the database LMFDB (\url{www.lmfdb.org}). For example, $F = \Q[x]/(x^4-2x^2+2)$. 
\end{rmk}

Having developed intuitions\footnote{
In the case of CM abelian varieties, the notion of a primitive 
CM-type $(K_\alpha, \Phi_\alpha^{rr})$ is available \cite{shimura2016abelian}, 
so one can work out the embedding of the algebra $\oplus_\alpha F_\alpha$
into the entire endomorphism algebra ${\rm End}(H^1(M;\Q))^{\rm Hdg}$
is determined from the CM type $(F_\alpha, \Phi_\alpha)$. 
For a general complex torus with sufficiently many complex multiplications, 
the authors have not made enough effort to come up with an alternative 
to the primitivity of CM type; so the authors are not ready to write down 
a statement similar to the 2nd item in 
Rmk. \ref{rmk:difference-cm-tpri-abelV} in connection with the theory 
of structure of ${\rm End}(H^1(M;\Q))^{\rm Hdg}$ for a general complex 
tori in \cite[\S1.7 and \S1.8]{birkenhaketori}.
} %
 on complex tori with sufficiently 
many complex multiplications, however, let us see 
\begin{props}\cite[Thm. 2.5]{Chen:2005gm}
\label{props:MChen-thm2.5}
{\it Let $M=(T^{2n};I)$ be an abelian variety, i.e., a complex 
torus that admits a polarization. If there exists a constant 
metric $G$ compatible with $I$ that is rational in the sense 
of (\ref{eq:cond-GnB-rational}), then 
the polarized rational Hodge structure on $H^1(M;\Q)$ is of CM-type. }
\end{props}

For a set of data $(T^{2n}; G, B)$ for which the (S)CFT is rational, 
there is always a complex structure $I$ with which $G$ is compatible 
and which admits a polarization (see Cor. \ref{cor:Wend-ExEx} and 
the discussion that follows). So, Prop. \ref{props:MChen-thm2.5} above 
is not an empty statement for {\it any} $T^{2n}$-target rational (S)CFTs. 

The proof of \cite[Thm. 2.5 + Prop. 2.4]{Chen:2005gm} is just as informative 
as the statement itself. Prop. 2.4 of \cite{Chen:2005gm} proves that 
the properties (i) and (ii) in Def. \ref{def:CM-cpx-trs} are equivalent 
to the property 
\begin{itemize}
\item [(iii)] the Hodge group ${\rm Hg}(M)(\R)$ is compact
\end{itemize}
for an abelian variety $M$; Thm. 2.5 of \cite{Chen:2005gm} proves the 
compactness (iii) of ${\rm Hg}(M)(\R)$ when there exists a rational $G$ 
that is compatible with $I$, and hence the properties (i) and (ii) in 
Def. \ref{def:CM-cpx-trs}. 
In proving the equivalence between the properties (iii) and (i, ii), however, 
Ref. \cite{Chen:2005gm} uses the fact that ${\rm Hg}(M)(\R)^{{\rm Ad}(h(i))}$ 
is compact; to prove the compactness of this group, Thm. 1.3.16 
of \cite{rohde2009cyclic}\footnote{
We refer to the LNM version, not to its arXiv versions. 
} %
uses the positive definiteness of a polarization of the rational Hodge 
structure. To conclude, the equivalence between the properties (iii) 
and (i, ii) breaks down when the rational Hodge structure $H^1(M;\Q)$
does not necessarily have a polarization. 

In our context, even when there is a constant rational metric $G$ 
compatible with the complex structure $I$ of a complex torus $M=(T^{2n};I)$, 
we cannot derive the property (i), the presence of sufficiently many 
complex multiplications (endomorphisms), if $I$ is not polarized. 
We could pay attention to complex tori $M=(T^{2n};I)$ satisfying the 
property (i) in Def. \ref{def:CM-cpx-trs}, but it is not obvious whether 
there exists a constant rational metric compatible with the complex 
structure $I$ (Discussion \ref{statmnt:MCheng-build-constRatMet} 
constructs rational metrics satisfying (\ref{eq:cond-GnB-rational}), 
by exploiting properties of CM fields not available to a general 
totally imaginary field).   
For this reason, we pay attention only to complex structures $I$ that 
admit polarization in the rest of this article. 

There are countably infinitely many such complex structures $I$ 
for a constant rational metric $G$ on $T^{2n}$; complex structures 
compatible with $G$ are parametrized by $S^2$ \cite{Aspinwall:1996mn}, 
and once a 2-form $\psi$ with $\int \psi \wedge \psi >0$ is chosen
from $H^2(T^4;\Q)$, then we should choose the direction of the K\"{a}hler 
form $\omega$ for the metric $G$ in the way $\R \omega$ includes the 
projection of $\psi$ to $\Pi_G$ in the notation to be used in 
section \ref{ssec:B-transc}, so that $\psi$ becomes Hodge (1,1) type
 (cf the discussion in between (\ref{eq:temp-B-1})
 and (\ref{eq:temp-B-2})). 
There are countably infinitely many choices of such $\psi$, and hence 
of a polarizable complex structure $I$.  
Existence of countably infinitely many complex structures 
is also understood naturally from the way Cor. \ref{cor:Wend-ExEx} 
is proven in \cite{Wendland:2000ye}. 

\subsection{Transcendental Part of the B-field}
\label{ssec:B-transc}

We may deal with all the polarizable complex structures $I$ on $T^{2n}$ 
with which a given metric $G$ is compatible and try to characterize 
the rational Hodge structures when $(T^{2n}; G, B)$ yields a rational (S)CFT. 
That is done in section \ref{sssec:result-for-genI}. It is also an option 
to impose further conditions on the choice of $I$ and try to characterize 
the Hodge structures for rational (S)CFTs for such a smaller class 
of complex structures. 
That is what we do in Thms. \ref{thm:MChen-thm3.11-refined}, 
\ref{thm:mirror-(n0)-isIN-alg} and \ref{thm:RCFT-2-GV-forT4}, 
built on Thm. \ref{thm:mirror-exist}. For them to make sense, 
however, we should prove the following 
\begin{props}
\label{props:I-polNalgB}
This is for the case $n=2$. 
{\it Let $(T^{2n=4};G, B)$ be a set of data for which the (S)CFT is rational. 
Then there exists a polarizable complex structure $I$ on $T^{4}$ with which 
$G$ is compatible, and the $B$-field only has the Hodge $(1,1)$ component 
with respect to that $I$. In particular, the $B$-field is in algebraic part 
${\cal H}^2(T^4_I)$.}
\end{props}

Here, 
\begin{defn}
For a general K\"{a}hler manifold $M$ of dimension $n$, 
\begin{align}
  {\cal H}^2(M) := H^{1,1}(M;\R) \cap H^2(M;\Q)
\end{align}
is said to be the {\it algebraic part} of $H^2(M;\Q)$. When $n=2$, 
the orthogonal complement 
\begin{align}
  T_M \otimes \Q := \left[ {\cal H}^2(M;\Q)^\perp \subset H^2(M;\Q) \right]
\end{align}
with respect to the wedge product is said to be the {\it transcendental part}.
The rational Hodge structure on $H^2(M;\Q)$ given by $I$ has a decomposition 
into the substructures on ${\cal H}^2(M)$ and $T_M\otimes \Q$; the 
substructure on ${\cal H}^2(M)$ is of level-0 and that on $T_M\otimes \Q$ 
of level-2. For a 2-form $\psi \in H^2(M;\R)$, its decomposition into 
${\cal H}^2(M)\otimes \R \oplus T_M \otimes \R$ is denoted by 
$\psi^{\rm alg} + \psi^{\rm transc}$, and are called the {\it algebraic 
and transcendental parts/components}. 
\end{defn}

{\bf Proof of Prop. \ref{props:I-polNalgB}:} 
Recall that the metric $G$ determines the real 3-dimensional 
vector subspace $\Pi_G$ of $H^2(T^4;\R)$ that consists of 2-forms that 
are self-dual under the Hodge-* operation with respect to the metric $G$. 
Choice of a complex structure $I$ compatible with $G$ is to specify one 
direction for $\omega = 2^{-1} G(I-,-)$ within $\Pi_G$; so, the choice of $I$ 
comes with a variety $S^2$ \cite[\S2]{Aspinwall:1996mn}; the two directions 
in $\Pi_G$
orthogonal to $\omega$ with respect to the wedge product supports the 
holomorphic (2,0) form $\Omega_M$ on $T^4$. 
Recall also that any 2-form can be decomposed into the self-dual component 
and the anti-self-dual component under the Hodge-* operation; 
let $B=B_\parallel + B_\perp$ be the decomposition of the $B$-field. 

When $B_\parallel=0$, automatically there is no Hodge (2,0) or (0,2) component 
in $B=B_\perp$, regardless of which direction in $\Pi_G$ is chosen (and of how 
complex structure $I$ is chosen). We just have to choose any $I$ in $S^2$
such that a polarization exists (such an $I$ exists; we have already seen 
that at the end of section \ref{ssec:torus-RCFT} for a rational $G$). 

When $B_\parallel \neq 0$, there is virtually no free choice for $I$
after requiring that the Hodge (2,0) component is absent; we have to choose 
$\omega \in \R B_\parallel$. Choosing $\omega \in \R_{<0} B_\parallel$ instead of 
$\omega \in \R_{>0} B_\parallel$ is nothing more than declaring holomorphic 
coordinates on $T^4$ as anti-holomorphic coordinates instead. So, we 
fix $\omega = 2^{-1}G(I-,-)$ by the condition $\omega \in \R_{>0} B_\parallel$, 
and prove that there is a polarization under $I$. 

To this end, note that 
$\int \Omega_M \wedge B_\parallel = 0$ and $\int \Omega_M \wedge B_\perp=0$, 
which is equivalent to 
\begin{align}
\int_{T^4} \Omega_M \wedge B = 0, \qquad \int_{T^4} \Omega_M \wedge (*B)=0. 
  \label{eq:temp-B-1}
\end{align}
So, both $B$ and $*B$ are in $H^{1,1}(T^4_I;\R)$. 
We already know that $B$ is also in $H^2(T^4;\Q)$, 
when $(T^4;G,B)$ is for a rational (S)CFT. 

If $\R (*B) = \R B$, then either $*B=B$ or $*B=-B$. 
In the case $B_\parallel \neq 0$, $*B=B$ is the only option, 
and $B_\parallel = B = * B$. In that situation, either $B$ or $-B$ 
is a polarization. To see this, note first that $\int_{T^4} B\wedge B 
= \int_{T^4}B \wedge (*B) >0$. This means that the Hermitian $2\times 2$ matrix 
$(B_{a\bar{b}})$ in $B=i B_{a\bar{b}}dz^a\wedge d\bar{z}^{\bar{b}}$ has a positive 
determinant, so the product of the two eigenvalues of the matrix is positive.
This proves that either $B$ or $-B$ is positive definite, besides being 
rational.  

If $*B$ and $B$ are linearly independent in $H^2(T^4;\R)$, 
then ${\rm Span}_\R\{ B, *B\} \subset H^{1,1}(T^4;\R)$ is a 
2-dimensional subspace, with signature (1, 1). 
Now, we claim that $\R(*B) \cap H^2(T^4;\Q)$ is not $\{0\}$. 
To see this, it is enough to note that 
\begin{align}
 (*B)_{IJ} = \sqrt{{\rm det}(G)} \; \epsilon_{IJMN}\; G^{MK}G^{NL}B_{KL}\frac{1}{2}, 
  \label{eq:temp-B-2}
\end{align}
where $\epsilon_{IJKL}$ is the $\{ \pm 1\}$-valued totally anti-symmetric 
tensor of rank-4; $\R(*B)$ contains such 2-forms 
as $\sqrt{{\rm deg}(G)}^{\pm 1} (*B)$, which are rational, as promised. 
This means that ${\cal H}^2(T^4_I) = H^2(T^4;\Q) \cap H^{1,1}(T^4_I;\R)$ 
is at least of 2-dimensions over $\Q$ of signature (1, 1). Moreover, 
within the 2-dimensional ${\cal H}^2(T^4_I)$, the there is a line 
$\R \omega$ along the K\"{a}hler form, and there is a rational point 
of ${\cal H}^2(T^4_I)$ arbitrarily close to the line in 
${\cal H}^2(T^4_I)\otimes \R$. Such a rational point is a polarization.  

The last statement in Prop. \ref{props:I-polNalgB} follows from 
Lemma \ref{lemma:no20-noTransc} below. We review it below for the benefit 
of the reader not familiar with it.  $\square$

Such a complex structure in Prop. \ref{props:I-polNalgB} 
is almost unique when $B_\parallel \neq 0$, 
and there will be infinitely many when $B_\parallel =0$ 
(see the discussion at the end of section \ref{ssec:pol}).

\begin{lemma}[well known in math]
\label{lemma:no20-noTransc}
{\it Let $T_M \otimes \Q$ be the transcendental part of a K\"{a}hler 
surface $M$ that has a polarization in $H^2(M;\Q)$. 
When $\psi \in T_M \otimes \Q$ is decomposed into 
$\psi^{(2,0)} + \psi^{(0,2)} + \psi^{(1,1)}$ and $\psi^{(2,0)}=0$, 
then $\psi = 0$.}
\end{lemma}

{\bf Proof:} The input $\psi^{(2,0)}=0$ implies $\psi^{(0,2)}=0$, 
because $\psi \in T_M\otimes \Q$ is real. This means that 
$\psi = \psi^{(1,1)}$ is in ${\cal H}^2(M)$. 

Since we have assumed that $M$ admits a polarization, $M$ is 
algebraic, so the intersection form on ${\cal H}^2(M)$ is 
non-degenerated (Hodge index theorem); 
$H^2(M;\Q) \cong {\cal H}^2(M) \oplus T_M\otimes \Q$ then. 
So, $\psi = \psi^{(1,1)}$ is both in ${\cal H}^2(M)$ and 
$T_M \otimes \Q$, which is possible only if $\psi = 0$. 
$\qquad \qquad \square$

\section{On the Rational Constant K\"{a}hler Metric}
\label{sec:metric}

\subsection{It is in the Algebraic Part}
\label{ssec:Kahler-is-alg}

For a rational constant metric $G$ on $T^{2n}$ and a polarizable 
complex structure $I$ with which $G$ is compatible, there is 
an intriguing property on the K\"{a}hler form $\omega = 2^{-1}G(I-,-)$. 
In section~\ref{ssec:Kahler-is-alg} (and henceforth), any complex structure under consideration 
is always of this kind, even when the authors fail to mention 
that explicitly. 
To prove Thm. \ref{thm:Kahler-is-alg}, we begin with this elementary 
preparation. 

\subsubsection{A Convenient Rational Basis}

We have seen that the rational Hodge structure on $H^1(T^{2n}_I;\Q)$
is of CM type, when $G$ is rational and compatible with a polarizable $I$. 
So, a $2n-$dimensional CM algebra $\oplus_{\alpha \in {\cal A}} F_\alpha$ over $\Q$ 
acts faithfully on the $2n$-dimensional vector space $H^1(T^{2n}_I;\Q)$. 
This can be used to introduce a rational basis of $H^1(T^{2n}_I; \Q)$ with 
which various computations are easier. 

The idea is to use the fact explained in Lemma \ref{lemma:very-useful}; 
we can do so because the individual CM fields $F_\alpha$ act faithfully 
on their corresponding $[F_\alpha:\Q]$-dimensional vector subspaces of 
$[H^1(T^{2n};\Q)]_\alpha$. In the case $F$ is a CM field $K$ with a primitive 
CM type, it often becomes convenient when we choose a basis 
$\{ \eta'_{i=1,\cdots, [K:\Q]} \}$ of $K/\Q$ so that 
$\{ \eta'_{i=1,\cdots, [K:\Q]/2} \}$ forms a basis of the totally real subfield 
$K_0$ of $K$, and use a purely imaginary generator $\xi_*$ 
of the extension $K/K_0$ (i.e., $K=K_0(\xi_*)$ such that $(\xi_*)^2 \in K_0$)
to fill the rest of a basis by $\{ \eta'_{i+[K:\Q]/2} = (\xi_*\eta'_i) \; | \; 
i=1,\cdots, [K:\Q]/2 \}$. 
We apply this prescription to the cases (B, C) for $n=2$; we use  
the basis $\{1, y, x, xy \}$ of $K/\Q$ as the basis 
$\{ \eta'_i \}$ in Lemma \ref{lemma:very-useful}, and then there must be 
an appropriate rational basis $\{ v'_i \}$ of $H^1(T^4;\Q)$ such that 
$v'_i \tau_a(\eta'_i)$ for $a \in \{ \pm \pm \}$ become the simultaneous 
eigenvectors of the action of the endomorphisms in $K \cong 
{\rm End}(H^1(T^4_I;\Q))^{\rm Hdg}$. The rational basis $\{ v'_i \}$ corresponding 
to $\{ \eta'_i \} = \{ 1,y,x,xy\}$ is denoted by 
$\{ \hat{\alpha}^1, \hat{\alpha}^2, \hat{\beta}_1, \hat{\beta}_2\}$ 
in this article. It is further convenient to introduce two complex 
coordinates $z^{a=1,2}$ on $T^4$ so that $dz^1 = v_{a=++}$ and $dz^2 = v_{a=-+}$. 
Namely, 
\begin{align}
  (dz^1, dz^2) = (\hat{\alpha}^1, \hat{\alpha}^2, \hat{\beta}_1, \hat{\beta}_2) \left( \begin{array}{cc} 1 & 1 \\ \sqrt{d} & - \sqrt{d} \\
   \sqrt{p+q\sqrt{d}} & \sqrt{p-q\sqrt{d}} \\
   \sqrt{p+q\sqrt{d}}\sqrt{d} & - \sqrt{p-q\sqrt{d}}\sqrt{d}
  \end{array} \right) =:
 (\hat{\alpha}^i, \hat{\beta}_i ) \left( \begin{array}{c} Z^T \\ \alpha^T
    \end{array} \right);  
  \label{eq:hol-rat-basis-trnsf-BC}
\end{align}
$Z$ is real-valued and $\alpha$ pure-imaginary valued; both are $2\times 2$ 
matrices. 

For the cases (A) and (A'), it is convenient to apply 
Lemma \ref{lemma:very-useful} to the CM elliptic curves referred to 
in Lemma \ref{lemma:classify-abel-surf}. In the case (A'), 
the imaginary quadratic field $K^{(2)}_i \cong \Q[x_i]/(x_i^2-p_i)$ 
acts on the $i$-th elliptic curve of CM-type. We use the basis 
$\{\eta' \} = \{ 1, x_i \}$ of $K^{(2)}_i$, and Lemma \ref{lemma:very-useful}
ensure that there is a rational basis 
$\{ v' \} = \{ \hat{\alpha}^i, \hat{\beta}_i\}$ of $H^1(E_i;\Q)$ so that 
$dz^i = \hat{\alpha}^i + \sqrt{p_i} \hat{\beta}_i$ is the (1,0) form,  
which is also an eigenvector of the action of the endomorphisms in $K^{(2)}_i$. 

In the case (A), we may also choose a rational basis as 
$\{\hat{\alpha}^1,\hat{\beta}_1\} \cup \{ \hat{\alpha}^2, \hat{\beta}_2\}$ 
in $H^1(E\times E;\Q) \cong H^1(E;\Q) \oplus H^1(E;\Q)$, and 
introduce complex coordinates $z^{a=1,2}$ on the two CM elliptic curves $E$
by $dz^1 = \hat{\alpha}^1 + \sqrt{p}\hat{\beta}_1$ and 
$dz^2 = \hat{\alpha}^2+\sqrt{p}\hat{\beta}_2$. 

In both cases (A') and (A), there must be an isogeny $\varphi$ from the 
abelian variety $M=(T^4;I)$ to $E_1\times E_2$ and $E\times E$, respectively. 
We pull back the convenient rational basis $\{\hat{\alpha}^i,\hat{\beta}_i \}$
of $E_1\times E_2$ and $E\times E$ to $H^1(M;\Q)$, respectively, and also 
pull back the complex coordinates $z^{1,2}$ to $M$, and use the same notation, 
$\{\hat{\alpha}^i, \hat{\beta}_i \}$ and $z^{1,2}$. 
In the cases (A') and (A), 
\begin{align}
  (dz^1, dz^2) = (\hat{\alpha}^i, \hat{\beta}_i) \left( \begin{array}{c} Z^T \\ \alpha^T \end{array} \right), \qquad 
   Z=\diag(1,1), \quad \alpha = \diag(\sqrt{p_1}, \sqrt{p_2}); 
  \label{eq:hol-rat-basis-trnsf-A}
\end{align}
in the case (A), $\alpha = \diag(\sqrt{p}, \sqrt{p})$. 

Note that the basis $\{ \hat{\alpha}^1, \hat{\beta}_1, \hat{\alpha}^2, 
\hat{\beta}_2 \}$ of $H^1(T^4;\Q)$ chosen above is generically not 
a set of generators of the entire $H^1(T^4;\Z) \cong \Z^{\oplus 4}$. 
That is not a problem; the observation of GV \cite{Gukov:2002nw}
was that rational CFTs may be characterized by using a rational 
Hodge structure, not an integral Hodge structure, so we just need 
a rational basis. 
Although there are infinitely many mutually non-isomorphic CM-type 
abelian surfaces, they may have one of only three---(A), (A') 
and (B, C)---qualitatively different rational Hodge structures. 
Conveniently, all the analysis in this article needs to be performed for 
just these three cases.

\subsubsection{The Algebraic and Transcendental Parts}

We claim in Thm. \ref{thm:Kahler-is-alg} that the K\"{a}hler form 
$\omega$ is always in the algebraic part ${\cal H}^2(T^4_I)\otimes \R$. 
For this purpose, we need to know ${\cal H}^2(T^4_I)$. 

\begin{lemma}
\label{lemma:gen-Sm} [well known in math literatures 
(e.g., \cite{shimura2016abelian,milnecm,Chen:2005gm})]
{\it Let $M$ be a complex torus of dimension $n$ where 
${\rm End}(H^1(M;\Q))^{\rm Hdg}$ is a CM field $F$. Then 
the algebraic part ${\cal H}^2(M) \subset H^2(M;\Q)$ contains 
an $n$ dimensional subspace ${\cal H}^2(M)_{\rm gen}$ specified below 
($h^{1,1}(M) = n^2$, so that is possible).  } The proof also 
introduces a basis on ${\cal H}^2(M)_{\rm gen}$ and also explains 
how to construct a polarization within ${\cal H}^2(M)_{\rm gen} \subset 
{\cal H}^2(M)$. 
\end{lemma}

{\bf Proof:} Let $F_0$ be the totally real subfield of $F$, 
and $\Phi = \{ \tau_{a=1,\cdots, n}\}$ be the CM type corresponding 
to the Hodge (1, 0) components of $H^1(M;\Q)$. 
There must be a basis $\{e_{I=1,\cdots, 2n} \}$ of $H^1(M;\Q)$
and a basis $\{ \eta_{I=1,\cdots, 2n} \}$ of $F/\Q$ so that 
$dz^a := e_I \tau_a(\eta_I)$ for $a=1,\cdots, n$ are the $n$ 
holomorphic 1-forms (cf Lemma \ref{lemma:very-useful}). 

Now, let $\xi_* \in F$ be a generator of $F/F_0$ (i.e., $F=F_0(\xi_*)$)
 so that $\xi_*^2 \in F_0$. Then for any element $\xi \in \xi_*F_0^\times$, 
\begin{align}
 {\cal Q}^{(\xi)} & \; := 
        \sum_{a=1}^n 2\tau_a(\xi) dz^a \wedge d\bar{z}^{\bar{a}}, \\
   &\; = e_I \wedge e_J \sum_{a=1}^n \left( \tau_a(\xi \eta_I \bar{\eta}_J)
         - \tau_a(\xi \bar{\eta}_I  \eta_J) \right),  \nonumber \\
   & \; = e_I \wedge e_J \sum_{a=1}^n \left( \tau_a(\xi \eta_I \bar{\eta}_J)
       + \tau_a(\bar{\xi}\bar{\eta}_I \eta_J ) \right)
    = e_I \wedge e_J {\rm Tr}_{F/\Q}[\xi \eta_I \bar{\eta}_J] \in {\cal H}^2(M)
\end{align}
(for any field extenstion $E/F$, ${\rm Tr}_{E/F}[x] \in F$ for $x \in E$; 
see \cite[A.1.15]{Kanno:2017nub} or any introductory textbook on field 
theory).
Linearly independent choices of $\xi$ from $\xi_* F_0$ generate 
an $n$-dimensional subspace of ${\cal H}^2(M)$, which is denoted 
by ${\cal H}^2(M)_{\rm gen}$. 

For the (1, 1) form ${\cal Q}^{(\xi)}$ to be a polarization, 
first, choose $\{e_I\}$ to be an integral basis of $H^1(M;\Z)$, 
and restrict to $\xi$ such that ${\rm Tr}_{F/\Q}[\xi \eta_I \bar{\eta}_J]
 \in \Z$ for all the pairs $(I, J)$; the basis $\{ \eta_I \}$ should be 
those that correspond to the integral basis $\{ e_I \}$. 
Second, impose inequalities on $\xi \in \xi_* F_0$ so that it is positive 
definite.  $\qquad \qquad \square$

\begin{lemma}
 \label{lemma:gen-Sm-nonSimpl}
{\it Let $M=(T^{2n};I)$ be an abelian variety of CM-type. Then }
\[
 \dim_\Q {\cal H}^2(M) \geq \dim_\C M.
\] 
\end{lemma}

{\bf Proof:} We can split the vector space $H^1(M;\Q)$ into its 
components $\oplus_{a \in A} [H^1(M;\Q)]_a$ supporting simple Hodge 
substructures; let $K_a$ be the CM field ${\rm End}([H^1(M;\Q)]_a)^{\rm Hdg}$.
Thus, it is enough to prove the statement for a simple abelian variety, and 
that was done in Lemma \ref{lemma:gen-Sm}.    $\square$

%
Lemmas \ref{lemma:gen-Sm}
and \ref{lemma:gen-Sm-nonSimpl} above imply that 
\begin{align}
 T_M \otimes \C & \; \subset T_M^{\rm gen}\otimes \C = 
   {\rm Span}_\C \left\{ (dz^a\wedge dz^b)_{a<b}, \; 
       (d\bar{z}^{\bar{a}}\wedge d\bar{z}^{\bar{b}})_{a<b}, 
    \; (dz^a\wedge d\bar{z}^{\bar{b}})_{a\neq b} \right\}, \\
 {\cal H}^2(M)\otimes \C & \; \supset {\cal H}^2(M)_{\rm gen} \otimes \C
   = {\rm Span}_\C \left\{ dz^a \wedge d\bar{z}^{\bar{a}} \right\}
\end{align}
for a CM abelian surface $M$. In fact, 

\begin{anythng}
\label{statmnt:Tm}
This is for $n=2$. {\bf In the cases (B, C, A')} 
of CM abelian surfaces $M$, ${\cal H}^2(M)_{\rm gen}$ is the entire 
${\cal H}^2(M)$; that is almost evident from the details below, 
but we justify this later in Discussion \ref{statmnt:confirm-Tm-n=2}. 

In the case (B, C), 
\begin{align}
 dz^1 \wedge d\bar{z}^{\bar{1}} & \; = -2\sqrt{p+q\sqrt{d}} \left\{ 
    (\hat{\alpha}^1\hat{\beta}_1) + d(\hat{\alpha}^2\hat{\beta}_2)
     + \sqrt{d}(\hat{\alpha}^1\hat{\beta}_2 + \hat{\alpha}^2\hat{\beta}_1)
    \right\},  \label{eq:dz1-dzbar1-caseBC} \\
 dz^2 \wedge d\bar{z}^{\bar{2}} & \; = -2\sqrt{p-q\sqrt{d}} \left\{ 
    (\hat{\alpha}^1\hat{\beta}_1) + d(\hat{\alpha}^2\hat{\beta}_2)
     - \sqrt{d}(\hat{\alpha}^1\hat{\beta}_2 + \hat{\alpha}^2\hat{\beta}_1)
    \right\},  \label{eq:dz2-dzbar2-caseBC}
\end{align}
so
\begin{align}
  {\cal H}^2(M) &\; = {\rm Span}_\Q \left\{ 
     (\hat{\alpha}^1\hat{\beta}_1) + d(\hat{\alpha}^2\hat{\beta}_2), \; 
     (\hat{\alpha}^1\hat{\beta}_2 + \hat{\alpha}^2\hat{\beta}_1) \right\}, 
     \label{eq:Sm-Qbasis-caseBC} \\
  T_M \otimes \Q & \;  = {\rm Span}_\Q \left\{ 
     (\hat{\alpha}^1\hat{\alpha}^2), \; (\hat{\beta}_1\hat{\beta}_2), \;
     (\hat{\alpha}^1\hat{\beta}_2-\hat{\alpha}^2\hat{\beta}_1), \; 
     (\hat{\alpha}^1\hat{\beta}_1 - d \hat{\alpha}^2\hat{\beta}_2) \right\}. 
  \label{eq:Tm-Qbasis-caseBC}
\end{align}
A generator of the Hodge $(2,0)$ component is in $T_M\otimes \C$, because 
\begin{align}
dz^1 \wedge dz^2 & \; = -2\sqrt{d}\left[ 
    \hat{\alpha}^1\hat{\alpha}^2
 + (\sqrt{+}-\sqrt{-})/(2\sqrt{d}) \; [\hat{\alpha}^1\hat{\beta}_1-d\hat{\alpha}^2\hat{\beta}_2]  \right. \nonumber \\
 & \qquad \qquad \quad \left. 
 + (\sqrt{+}+\sqrt{-})/2 \; [
  \hat{\alpha}^1\hat{\beta}_2- \hat{\alpha}^2\hat{\beta}_1]
 +\sqrt{d'}\; \hat{\beta}_1\hat{\beta}_2 \right].
  \label{eq:dz1-dz2-caseBC}
\end{align}
Here, we used the notation introduced 
below (\ref{eq:introduce-sh-notation-below}). 

In the case (A'), 
\begin{align}
  dz^1 \wedge d\bar{z}^{\bar{1}} = -2\sqrt{p_1}\hat{\alpha}^1 \hat{\beta}_1, 
    \qquad 
  dz^2 \wedge d\bar{z}^{\bar{2}} = -2\sqrt{p_2}\hat{\alpha}^2 \hat{\beta}_2, 
\end{align}
so $dz^1 \wedge dz^2 = (\hat{\alpha}^1\hat{\alpha}^2)
 - \sqrt{p_1p_2} (\hat{\beta}_1\hat{\beta}_2)
 + \sqrt{p_2}(\hat{\alpha}^1\hat{\beta}_2)
 + \sqrt{p_1}(\hat{\beta}_1\hat{\alpha}^2)$ is in $T_M \otimes \C$ below:
\begin{align}
 {\cal H}^2(M) & \; = {\rm Span}_\Q \left\{ 
     (\hat{\alpha}^1\hat{\beta}_1), \; (\hat{\alpha}^2\hat{\beta}_2)
          \right\},
       \label{eq:Sm-Qbasis-caseAprm} \\
  T_M \otimes \Q & \;  = {\rm Span}_\Q \left\{ 
     (\hat{\alpha}^1\hat{\alpha}^2), \; (\hat{\beta}_1\hat{\beta}_2), \;
     (\hat{\alpha}^1\hat{\beta}_2), \; (\hat{\alpha}^2\hat{\beta}_1) 
        \right\}. 
   \label{eq:Tm-Qbasis-caseAprm}    
\end{align}

{\bf In the case (A)},\footnote{
\label{fn:SM}
The case (A), where $M$ is isogenous to a product of two copies 
of a CM elliptic curve, is known \cite{shioda1974singular} 
to be the case of rank-2 $T_M$. 
} %
\begin{align}
  T_M\otimes \Q & \; = {\rm Span}_\Q \left\{
     (\hat{\alpha}^1\hat{\alpha}^2+ p \hat{\beta}_1\hat{\beta}_2), \; 
     (\hat{\alpha}^1\hat{\beta}_2 + \hat{\beta}_1\hat{\alpha}^2) \right\}, 
   \label{eq:Tm-Qbasis-caseA}
\end{align}
generated by the real and imaginary part of $dz^1 \wedge dz^2 = (\hat{\alpha}^1+\sqrt{p}\hat{\beta}_1)(\hat{\alpha}^2+\sqrt{p}\hat{\beta}_2)$.  
\begin{align}
{\cal H}^2(M) & \; = {\rm Span}_\Q \left\{ (\hat{\alpha}^1\hat{\beta}_1), \; 
    (\hat{\alpha}^2\hat{\beta}_2), \; 
    (\hat{\alpha}^1\hat{\alpha}^2-p\hat{\beta}_1\hat{\beta}_2), \;
    (\hat{\alpha}^1\hat{\beta}_2 - \hat{\beta}_1\hat{\alpha}^2) \right\}, 
    \label{eq:Sm-Qbasis-caseA}
\end{align}
generated by $dz^1\wedge d\bar{z}^{\bar{1}}/(-2\sqrt{p})$ and  
$dz^2\wedge d\bar{z}^{\bar{2}}/(-2\sqrt{p})$ in ${\cal H}^2(M)_{\rm gen}$, along 
with the real and imaginary part of $dz^1 \wedge d\bar{z}^{\bar{2}}$. 
$\square$
\end{anythng}

\begin{anythng}
\label{statmnt:MCheng-build-constRatMet}
An idea is explained in \cite[Thm. 2.5]{Chen:2005gm} how to construct 
a rational constant K\"{a}hler metric on CM-type complex abelian variety 
$M$. It is done by decomposing $M$ into its factors $M_\alpha$ where 
the algebra ${\rm End}(H^1(M_\alpha;\Q))^{\rm Hdg}$ contains a CM field $F_\alpha$
with $[F_\alpha:\Q] = 2 \dim_\C M_\alpha$. 
Here, we review the construction, as we will refer to this construction 
(already in section \ref{ssec:pol} and also) later 
in section \ref{sssec:ratMet-workedOut} and discussions in 
\ref{statmnt:GV-2-RCFT-caseBC-even} and \ref{statmnt:GV-2-RCFT-caseAprm-even}. 

Let us use the same notation as in Lemma \ref{lemma:gen-Sm} here, except 
dropping the subscript $\alpha$ above. For any $\beta \in \xi_* F_0$, 
\begin{align}
  \omega^{(\beta)} &\; := \frac{i}{2} \sum_{a=1}^n \tau_a(\xi_*)
     \bar{\tau}_{\bar{a}}(\beta) \left(
     dz^a \otimes d\bar{z}^{\bar{a}} -d\bar{z}^{\bar{a}}\otimes dz^a\right), 
    \label{eq:basis-KahlerF-forRatG}  \\
  G^{(\beta)} & \; := \sum_{a=1}^n \tau_a(\xi_*\bar{\beta}) \left( 
     dz^a \otimes d\bar{z}^{\bar{a}}  + d\bar{z}^{\bar{a}}\otimes dz^a\right), \\
   & \; = (e_I \otimes e_J)  \sum_{a=1}^n  
    \left( \tau_a(\xi_*\bar{\beta}\eta_I \bar{\eta}_J) 
        + \tau_a(\xi_*\bar{\beta}\bar{\eta}_I \eta_J) \right)
   = e_I \otimes e_J \; {\rm Tr}_{F/\Q}[\xi_* \bar{\beta} \eta_I \bar{\eta}_J ]. 
\end{align}
In this construction, all the components $G^{(\beta)}_{IJ}$ are rational.  
One just has to impose inequalities on $\xi_* \bar{\beta} \in F_0
 \hookrightarrow \R^n$ so that the metric is positive definite.  
In the application to the cases (B, C, A), 
we may set $\xi_* = x$ in Lemma \ref{lemma:classify-deg4-CM-alg}; 
$F_0=K_0$ in the case (B, C). In the application to the case (A'), 
we may set $\xi_* = x_i$ for $i=1,2$, and $F_0=\Q$ for both $i=1,2$. 
\end{anythng}

\subsubsection{Analysis}
\label{sssec:ratMet-workedOut}

The first one of the conditions in (\ref{eq:cond-GnB-rational})---one 
for the metric---involves an integral basis of $H^1(T^{2n};\Z)$. 
This condition is still in the same form---the components 
are rational numbers when we use a rational basis of $H^1(T^{2n};\Q)$. 
As a preparation for the analysis in section \ref{sec:analysis-T4}, 
let us translate this condition by using the CM-algebra eigenstate basis 
$\{ dz^a, d\bar{z}^{\bar{a}} \}$ of $H^1(T^{2n}_I;\C)$.  

The metric $G$ is Hermitian under the complex structure $I$, that is, 
\begin{align}
 G = h_{a\bar{b}} dz^a\otimes d\bar{z}^{\bar{b}}
    + h_{\bar{a}b}d\bar{z}^{\bar{a}}\otimes dz^b
\end{align}
for some constant Hermitian $n\times n$ matrix $h = (h_{a\bar{b}})$. 
Using the linear relations such as (\ref{eq:hol-rat-basis-trnsf-BC},
 \ref{eq:hol-rat-basis-trnsf-A}), the rationality of the components $G_{IJ}$
in a rational basis is translated to the rationality of all the components 
of the matrix 
\begin{align}
   \left( \begin{array}{cc} Z^T & \overline{Z}^T \\
       \alpha^T & \overline{\alpha}^T \end{array} \right)
   \left( \begin{array}{cc} & h \\ h^T & \end{array} \right)
   \left( \begin{array}{cc} Z & \alpha \\ \overline{Z} & \overline{\alpha} 
   \end{array} \right) = \left( \begin{array}{cc} 
     Z^T h \overline{Z} + \overline{Z}^T h^T Z & 
      Z^T h \overline{\alpha} + \overline{Z}^T h^T \alpha \\
     \overline{\alpha}^T h^T Z + \alpha^T h \overline{Z}  &
      \alpha^T h \overline{\alpha} + \overline{\alpha}^T h^T \alpha
    \end{array} \right).
\end{align}
That is, 
\begin{align}
  Z^T h \overline{Z} + \overline{Z}^T h^T Z  & \; \in M_n(\Q)^{\rm sym} , 
     \label{eq:cond-met-ratnl-n2-1} \\
  \alpha^T h \overline{\alpha} + \overline{\alpha}^T h^T \alpha & \;
   \in M_n(\Q)^{\rm sym}, 
     \label{eq:cond-met-ratnl-n2-2}  \\
  Z^T h \overline{\alpha} + \overline{Z}h^T \alpha & \; \in M_n(\Q).
    \label{eq:cond-met-ratnl-n2-3}
\end{align}
We wish to solve those conditions in terms of the matrix $h = (h_{a\bar{b}})$; 
this is done by working separately for the cases (B, C), (A'), and (A) 
separately. So, the analysis leading to Thm. \ref{thm:Kahler-is-alg} 
is only for $n=2$, $T^{2n=4}$. 
We will use the following parametrization of the $2\times 2$ matrix $h$: 
\begin{align}
  h = \left( \begin{array}{cc} h_1 & c_1-ic'_2 \\
            c_1+ic'_2 & h_2 \end{array} \right),
      \qquad h_{1,2}, \; c_1, \; c'_2 \in \R.
\end{align}

{\bf Case (B, C): } The condition (\ref{eq:cond-met-ratnl-n2-1}) implies 
that 
\begin{align}
h_1+h_2 \in \Q, \qquad h_1 - h_2 \in \sqrt{d}\Q, \qquad 
c_1 \in \Q, \qquad {}^\forall c'_2 \in \R,  
\end{align}
and the condition (\ref{eq:cond-met-ratnl-n2-2}) on top of this implies 
that 
\begin{align}
 c_1=0. 
\end{align}
On the other hand, the condition (\ref{eq:cond-met-ratnl-n2-3}) 
is equivalent to $c'_2=0$ with arbitrary $h_{1,2}$ and $c_1$. 
So, we have\footnote{
For the metric to be positive definite, $a>0$ and $a^2-b^2d>0$. 
} %
\begin{align}
  h = \diag(a+b\sqrt{d}, a-b\sqrt{d}), \qquad {}^\exists a, \; b\in \Q;
\end{align}
the corresponding K\"{a}hler form $\omega(-,-) = 2^{-1}G(I-,-)$ is 
\begin{align}
 \omega = i (a+b\sqrt{d}) dz^1\wedge d\bar{z}^{\bar{1}}
   + i (a-b\sqrt{d}) dz^2\wedge d\bar{z}^{\bar{2}}.
   \label{eq:Kahler-form-BC-4RCFT}
\end{align}
This family of K\"{a}hler forms parametrized by $a,b\in \Q$ is the 
same as the family (\ref{eq:basis-KahlerF-forRatG}) parametrized by 
$\beta \in \xi_*F_0$ ($[F_0:\Q]=[K_0:\Q]=2$ in the case (B,C)).

{\bf Cases (A') and (A):} 
The condition (\ref{eq:cond-met-ratnl-n2-1}) is translated 
to $h_{1,2}, c_1 \in \Q$, and the condition (\ref{eq:cond-met-ratnl-n2-2}) 
on top of this imposes $c_1 \in \sqrt{p_1p_2} \Q$.
So we should have $c_1 =0$ in the case (A'), while 
$c_1 \in \sqrt{p_1p_2}\Q$ is equivalent to $c_1\in \Q$ in the case (A). 
On the other hand, the condition (\ref{eq:cond-met-ratnl-n2-3}) 
implies $c'_2 \in \sqrt{-p_1}\Q \cap \sqrt{-p_2}\Q$; so we should have 
$c'_2=0$ in the case (A'), while we just have $c'_2 \in \sqrt{-p}\Q$.  
To summarize, we should have 
\begin{align}
 (A'): & \qquad  h = \diag( a_1, a_2), \qquad a_{1,2} \in \Q, 
    \label{eq:parametr-Kahler-caseApr} \\
 (A): & \qquad h = \left( \begin{array}{cc} h_1 & c_1 -c_2 \sqrt{p} \\
      c_1+c_2 \sqrt{p} & h_2 \end{array} \right), \qquad 
       h_{1,2}, \; c_1, \; c_2 \in \Q, 
    \label{eq:parametr-Kahler-caseA}
\end{align}
and the corresponding K\"{a}hler forms are 
\begin{align}
  \omega & \; = i a_1 dz^1\wedge d\bar{z}^{\bar{1}}
     + i a_2 dz^2\wedge d\bar{z}^{\bar{2}},  
    \label{eq:Kahler-form-Apr-4RCFT} \\
  \omega & \; = i (dz^1, dz^2) \wedge \left( \begin{array}{cc} 
      h_1 & c_1-c_2\sqrt{p} \\ c_1+c_2 \sqrt{p} & h_2 \end{array} \right)
    \left( \begin{array}{c} d\bar{z}^{\bar{1}} \\ 
        d\bar{z}^{\bar{2}} \end{array} \right), 
   \label{eq:Kahler-form-A-4RCFT}
\end{align}
respectively. The family of K\"{a}hler forms 
in (\ref{eq:Kahler-form-Apr-4RCFT}) in the case (A') is the 
same as the family in (\ref{eq:basis-KahlerF-forRatG}); for 
the case (A), however, the full family of K\"{a}hler forms 
(\ref{eq:Kahler-form-A-4RCFT}) corresponding to rational metrics 
has four rational parameters, whereas the 
family (\ref{eq:basis-KahlerF-forRatG}) constructed by using one 
CM subfield $F \subset {\rm End}(H^1(M;\Q))^{\rm Hdg}$ has $[F_0:\Q]=2$
rational parameters. 

Having done this analysis, we are ready for this:
\begin{thm}
\label{thm:Kahler-is-alg}
This is for $n=2$. {\it Let $(T^{2n=4}; G, B)$ be a set of data for which the 
(S)CFT is rational. For a polarizable complex structure $I$ on $T^{4}$ 
with which $G$ is compatible, the K\"{a}hler form $\omega = 2^{-1}G(I-,-)$ 
is always in the algebraic part of the 2-forms, ${\cal H}^2(T^4_I)\otimes \R $.

Moreover, the combination $i\omega$ is in ${\cal H}^2(T^4_I)\otimes
  \tau^r_{(20)}(K^r)$, where $K^r$ is the reflex field 
in \ref{statmnt:CMtype-n-reflexF} and \ref{statmnt:CMtype-n-reflexF-Aprm}, 
and $\tau^r_{(20)}$ its embedding for the Hodge (2, 0) component 
in $T_{T^4_I}\otimes \C$. }
\end{thm}

{\bf Proof:} It is just necessary to write down the K\"{a}hler forms 
in (\ref{eq:Kahler-form-BC-4RCFT}, \ref{eq:Kahler-form-Apr-4RCFT}, 
\ref{eq:Kahler-form-A-4RCFT}) in the rational basis in 
Discussion \ref{statmnt:Tm}.  In the case (B, C), 
\begin{align}
 i \omega & \; = 2 \tau^r_{++}(a \xi_r + bqd/\xi_r) e_1 
   + 2\tau^r_{++}(bd\xi_r + aqd/\xi_r) e_2, 
  \label{eq:Kahler-form-BC-4RCFT-2}   \\
   & \qquad e_1 := \hat{\alpha}^1\hat{\beta}_1 +d\hat{\alpha}^2\hat{\beta}_2, 
   \quad e_2 := \hat{\alpha}^1\hat{\beta}_2 + \hat{\alpha}^2\hat{\beta}_1.
   \label{eq:def-ratBasis-H2-caseBC}
\end{align}
In the case (A'), 
\begin{align}
 i \omega & \; = 2a_1 \sqrt{p_1}(\hat{\alpha}^1\hat{\beta}_1)
    + 2a_2 \sqrt{p_2}(\hat{\alpha}^2\hat{\beta}_2),
   \label{eq:Kahler-form-Aprm-4RCFT-2}
\end{align}
while 
\begin{align}
 i\omega & \; = \sqrt{p} \left[
   2h_1 (\hat{\alpha}^1\hat{\beta}_1) + 2h_2 (\hat{\alpha}^2\hat{\beta}_2)
   + 2c_1(\hat{\alpha}^1\hat{\beta}_2 - \hat{\beta}_1\hat{\alpha}^2)
   + 2c_2(\hat{\alpha}^1\hat{\alpha}^2-p\hat{\beta}_1\hat{\beta}_2)\right]
  \label{eq:Kahler-form-A-4RCFT-2}
\end{align}
in the case (A).  $\square$ 

Prop. 4.1 and Cor. 5.11 of Ref. \cite{Chen:2005gm} identifies an example 
that looks like a counter example to the (spirit of the) conjecture 
in section \ref{ssec:GV-conj}. The example in \cite[\S4]{Chen:2005gm} 
chose a K\"{a}hler form within $H^{1,1}(M;\R)$ but not 
in ${\cal H}^2(M)\otimes \R$, so there is no wonder in the light 
of Thm. \ref{thm:Kahler-is-alg} above that the metric is not rational  
in that example. This observation suggests that the K\"{a}hler form being 
in the algebraic part is an important element in characterizing the data 
for rational CFTs. So, this property is now implemented as the 
condition 2(b) of Thms. \ref{thm:RCFT-2-GV-forT4} and 
\ref{thm:RCFT-2-GV-forT4-genI} to be meant as a revised version 
of Conjecture \ref{conj:GV-original}. 

\subsection{A Geometric Mirror Always Exists}
\label{ssec:mirror-exist}

\begin{thm}
\label{thm:mirror-exist}
(This is only for the $n=2$ cases) {\it Let $(T^{2n=4};G, B)$ be 
a set of data for which the $\mathcal N=(1,1)$ SCFT is rational. 
Choose\footnote{
We have seen in section \ref{ssec:B-transc} that such a complex 
structure $I$ exists. So, this Theorem is not empty for 
any set of data $(T^{2n};G,B)$.
} %
 a polarizable complex structure $I$ with which $G$ 
is compatible so that the Hodge (2,0) component of $B$ is absent. 
Then one can always find a rank-$n=2$ primitive subgroup $\Gamma_f$
of $H_1(T^4;\Z)$ so that the T-dual along $\Gamma_f$ has a geometric 
SYZ-mirror. }
\end{thm}

{\bf Proof:} In the cases (B, C, A'), we have stated in \ref{statmnt:Tm}
that $\hat{\alpha}^1 \hat{\alpha}^2$ is in the transcendental part
of $H^2(T^4;\Q)$. So, when we choose $\Gamma_f\otimes \Q = 
{\rm Span}_\Q\{ \alpha_1, \alpha_2\}$, $B|_{\Gamma_f}=0$ because we 
chose the complex structure so that $B$ is purely algebraic. 
We have also seen in Thm. \ref{thm:Kahler-is-alg} that 
the K\"{a}hler form is also in the algebraic part, so $\omega|_{\Gamma_f}=0$.

In the case (A), we have seen in (\ref{eq:parametr-Kahler-caseA}) 
how the K\"{a}hler form is parametrized. A rational $B=B^{\rm alg}$ 
should also have 4 parameters in $\Q$ because $\dim_\Q {\cal H}^2(T^4_I)=4$.
Details in \ref{statmnt:Tm} reveals that 
\begin{align}
 B^{\rm alg} & \; = \sqrt{p} (dz^1, dz^2) \wedge \left( \begin{array}{cc}
     h_1^B & c_1^B-c_2^B \sqrt{p} \\ c_1^B + c_2^B \sqrt{p} & h_2^B 
   \end{array} \right)
   \left( \begin{array}{c} d\bar{z}^{\bar{1}} \\ d\bar{z}^{\bar{2}}
   \end{array} \right), \qquad 
   h_{1,2}^B, c_{1,2}^B \in \Q.
\end{align}

With a straightforward computation, one can show that 
$\Gamma_f = {\rm Span}_\Z \{ \alpha'_1, \alpha'_2\}$ with 
\begin{align}
  \alpha'_1 = \alpha_1, \quad 
  \alpha'_2 = \alpha_2 - \left\{ \frac{c_2}{h_1}- \frac{c_1}{h_1}
     \frac{(h_1^Bc_2-h_1c_2^B)}{(h_1^Bc_1-h_1c_1^B)} \right\} \beta^1
   - \frac{h_1^Bc_2-h_1c_2^B}{h_1^Bc_1-h_1c_1^B} \beta^2 
  \label{eq:SYZ-directn-caseA-gen}
\end{align}
satisfies the condition $\omega|_{\Gamma_f}=0$ and $B|_{\Gamma_f}=0$. 
In the case $h_1^Bc_1-h_1c_1^B=0$, the same conditions are satisfied 
when 
\begin{align}
  \alpha'_1 = \alpha_1 + \frac{1}{pv} \beta^1, \qquad 
  \alpha'_2 = \alpha_2 - \frac{c_1}{h_1}\left( v - \frac{1}{pv} \right) \beta^1
   + v \beta^2, \qquad v \in \Q, \; v \neq 0. 
   \label{eq:SYZ-directn-caseA-spc}
\end{align}
The T-dual in the directions (\ref{eq:SYZ-directn-caseA-gen},
 \ref{eq:SYZ-directn-caseA-spc}) have a geometric SYZ-mirror. 

The choices of $\Gamma_f$ above still come with a variety of choices 
of $\Gamma_b$, and moreover, there will be more choices of $\Gamma_f$ 
other than the one above; they are meant to be only examples. $\square$

\section{Refined Gukov--Vafa Theorem for $T^4$}
\label{sec:analysis-T4}

\subsection{CM Horizontal Hodge Structure}
\label{ssec:horizontal}

Let us quickly go through known results to confirm 
CM-type statements on horizontal rational Hodge structures. 
Whenever we refer to rational Hodge structures in this 
section \ref{ssec:horizontal}, that is a horizontal one. 

Prop. \ref{props:MChen-thm2.5} already implies that the rational 
Hodge structure on $H^1(T^{2n}_I;\Q)$ is of CM-type, when we choose a 
polarizable complex structure $I$ on $T^{2n}$ compatible with 
the rational metric $G$ in a set of data $(T^{2n};G,B)$ for a rational 
SCFT. This may be combined with 
\begin{lemma}
\label{lemma:Borcea}
\cite[Prop. 1.2]{borcea1998calabi}
{\it Let $(h_1, V_1)$ and $(h_2, V_2)$ be both a polarizable 
rational Hodge structure. When both are of CM-type, then 
the polarizable rational Hodge structure 
$(h_2\otimes h_1, V_2\otimes V_1)$ on the vector space 
$V_2 \otimes_\Q V_1$ is also CM type.  } $\square$
\end{lemma}
So, the polarizable rational Hodge structure on $\otimes^k (H^1(T^{2n}_I;\Q)$
is also of CM-type, and its rational Hodge substructure on 
$\wedge^k H^1(T^{2n}_I;\Q) = H^k(T^{2n}_I;\Q)$ is also of CM-type.\footnote{
Any maximally commutative endomorphism (sub)algebra on $\wedge^k H^1(T^{2n};\Q)$ 
and the one on the rest in $\otimes^k (H^1(T^{2n};\Q))$ can have dimensions 
as large as the dimension of the vector spaces they act. So, existence of 
a commutative subalgebra $F \subset {\rm End}(\otimes^k(H^1(T^{2n};\Q)))^{\rm Hdg}$ with $\dim_\Q F = \dim_\Q \otimes^k (H^1(T^{2n};\Q))$
implies that there exists a commutative subalgebra $F_1 \subset 
{\rm End}(\wedge^k (H^1(T^{2n};\Q)))^{\rm Hdg}$ and $F_2$ on the 
rest of $\otimes^k (H^1(T^{2n};\Q))$ such that $\dim_\Q F_1 
= \dim_\Q \wedge^k (H^1(T^{2n};\Q))$ (to be rigorous, the 
structure (\ref{eq:WeddB-alg-str}) and the observation in 
footnote \ref{fn:construct-nonprim-CM-F} need to be used).
} %

Within $H^{k=n}(T^{2n}_I;\Q)$, there is just one simple Hodge substructure 
of level-$n$, denoted by $[H^n(T^{2n}_I;\Q)]_{\ell = n}$, which contains 
the 1-dimensional Hodge $(n,0)$ and $(0,n)$ components. The algebra 
${\rm End}([H^n(T^{2n};\Q)]_{\ell=n})^{\rm Hdg}$ should be a CM field, 
and moreover, the CM field should be\footnote{
The authors could not identify a reference to cite for this statement, but 
an easy way to see this may be to note that the Hecke character associated 
with the holomorphic $n$-form is the product of the $n$ Hecke characters 
associated with $\Phi$. For string theorists, a direct 
computation as in \ref{statmnt:confirm-Tm-n=2} may be easier. 
} %
 the reflex field $K^r$ 
of the CM-type $(F,\Phi)$ of the CM-type rational Hodge structure
$H^1(T^{2n}_I;\Q)$. 

\begin{anythng}
\label{statmnt:confirm-Tm-n=2}
In the case of $n=2$, let us confirm explicitly the general statements above;
that also serves as a preparation for the discussion in 
section \ref{ssec:converse}. 

Let us begin with {\bf the case (B, C)}. We may choose a generator 
of the Hodge $(n,0)=(2,0)$ component as $v_{++} := 
dz^1\wedge dz^2/(-2\sqrt{d})$; see  (\ref{eq:dz1-dz2-caseBC}) 
for the expression. 
This generator $v_{++}$ is within the 4-dimensional space $T_M\otimes \C$
and is indeed in the form of $e_i \tau^r_{++}(\eta_i)$ for 
a rational basis $\{ e_i \}$ given in (\ref{eq:Tm-Qbasis-caseBC})
and a rational basis $\{ \eta_i \} = \{ 1, q/\xi^r, \xi^r/2, y'\}$
of the reflex field $K^r$ (see \ref{statmnt:CMtype-n-reflexF}).
It is also possible to confirm that $(dz^1\wedge d\bar{z}^{\bar{2}})$,
$(d\bar{z}^{\bar{1}}\wedge dz^2)$ and 
$(d\bar{z}^{\bar{1}}\wedge d\bar{z}^{\bar{2}})$ are proportional to 
$e_i \tau^r_{-+}(\eta_i)$, $e_i \tau^r_{--}(\eta_i)$ and $e_i \tau^r_{+-}(\eta_i)$, 
respectively. The CM field $K^r$ acts on the 4-dimensional vector space
$T_M\otimes \Q$ in (\ref{eq:Tm-Qbasis-caseBC}) as explained in 
Lemma \ref{lemma:very-useful-2}, which is in a way preserving the Hodge 
decomposition. So, $K^r \subset {\rm End}(T_M\otimes \Q)^{\rm Hdg}$ indeed. 
The fact that $K^r$ is not a CM algebra but a CM field implies that 
the 4-dimensional $T_M\otimes \Q$ as a whole (not its proper subspace) 
is the transcendental part indeed, and ${\cal H}^2(M)$ is no larger than 
${\cal H}^2(M)_{\rm gen}$. 
We also note that the embedding $\tau^r_{++}$ of the number field $K^r$ 
is the one associated with the Hodge (2,0) component (i.e., $\tau^r_{(20)}$ 
in Thm. \ref{thm:Kahler-is-alg}). 

In the {\bf case A'}, literally the same argument as above can be repeated, 
by using \ref{statmnt:CMtype-n-reflexF-Aprm} instead 
of \ref{statmnt:CMtype-n-reflexF}, and using the rational 
basis (\ref{eq:Tm-Qbasis-caseAprm}) instead of (\ref{eq:Tm-Qbasis-caseBC}).
We may think of the embedding $\tau^r_{++}$ of the number field $K^r$ as 
the one associated with the Hodge (2,0) component. 

In the {\bf case A}, $dz^1 \wedge dz^2 =
 (\hat{\alpha}^1\hat{\alpha}^2-p\hat{\beta}_1\hat{\beta}_2)
 + \sqrt{p}(\hat{\alpha}^1\hat{\beta}_2+\hat{\beta}_1\hat{\alpha}^2)$, 
which is expressed in the form of $e_i \tau^r(\eta_i)$ for the reflex 
field $F^r$ in \ref{statmnt:CMtype-n-reflexF}. The endomorphism algebra 
on the 2-dimensional $T_M\otimes \Q$ in \ref{statmnt:Tm} is a degree-2 
extension field, so $T_M\otimes \Q$ is indeed the transcendental part. 
\end{anythng}

\begin{rmk}
\label{statmnt:CM-from-H2-to-H1}
Given the fact that the cohomology group of a Ricci-flat K\"{a}hler 
manifold $M$ has a unique level-$n$ simple Hodge substructure 
$[H^n(M;\Q)]_{\ell=n}$, it is natural to wonder if this simple 
Hodge substructure plays more important role than other parts of the cohomology 
group. In the particular case of $M=T^{2n}$, for example, we may wonder 
whether or not a CM-type Hodge structure of $[H^2(T^{2n};\Q)]_{\ell=n}$ 
implies that the rational Hodge structure on $H^1(T^{2n};\Q)$ is CM-type. 

In the case of $n=2$, this question is split into four cases. 
\begin{itemize}
\item [($\alpha$)] The transcendental part of $H^2(T^{4}_I;\Q)$ is 
of 2-dimensions; its CM field $K'$ should be $K'\cong \Q[\xi]/(\xi^2-p)$ 
for $p \in \Q_{< 0}$, an imaginary quadratic field.
\item [($\alpha'$)] The transcendental part is of 4-dimension, and its 
CM field $K'$ is that of case (A); $K' \cong \Q[\xi,y']/(\xi^2-p_1, (y')^2-d)
\cong \Q[\xi, (\xi y')]/(\xi^2-p_1, (\xi y')^2 - d p_1)$; we may write 
$dp_1=:p_2 \nin p_1 (\Q^\times)^2$ because $d$ is a square-free integer 
and $d\neq 1$. 
\item [($\beta, \gamma$)] The transcendental part is of 4-dimension, 
and its CM field $K'$ is that of cases (B, C). 
\end{itemize}
The case ($\alpha$) has been completely understood, and the question 
is answered affirmative \cite{shioda1974singular} (as mentioned already in 
footnote \ref{fn:SM}). For other cases, certainly 
the CM abelian surfaces of case A', and B, C are examples of 
the cases $\alpha'$, and $\beta, \gamma$ here. It is not obvious, however, 
whether or not all the abelian surfaces with the property $\alpha'$, 
$\beta, \gamma$ are such CM abelian surfaces. 
\end{rmk}

\subsection{The Vertical Hodge Structure is of CM-type}

\subsubsection{Mirror Isogeny and Hodge Isomorphism}

The following result by \cite{Chen:2005gm} exploits various properties 
very special to torus target SCFTs. For example, torus-target SCFTs
forms a self-mirror moduli space of string vacua; all the data 
of constant metric $G$, closed 2-form $B$, complex structure $I$ 
can be treated by constant-valued matrices than a field configuration, 
and all the information on cohomology and Hodge structure follows 
from that on $H^1(T^{2n};\Q)$. Nevertheless, we use the result 
quoted as Prop. \ref{props:MChen-Prop3.10} and reach a nice-looking 
result (Thms. \ref{thm:RCFT-2-GV-forT4} and \ref{thm:RCFT-2-GV-forT4-genI}) 
that can be stated in a language that also (almost) makes sense 
for other families of target spaces. Now, let us begin with 
\begin{props}
\label{props:MChen-Prop3.10}
\cite[Prop. 3.10]{Chen:2005gm} 
{\it Let $(T^{2n}; G, B)$ be a set of data for which the $\mathcal N=(1,1)$ SCFT 
is rational, and $I$ a complex structure with which $G$ is compatible;
here, $I$ may or may not be polarizable. Suppose further that 
this SCFT with $I$ has a geometric SYZ-mirror, with a set of data 
$(T^{2n}; G^\circ, B^\circ;I^\circ)$. Then the complex tori $T^{2n}_I$
and $T^{2n}_{I^\circ}$ are isogenous. }
\end{props}

Now, we can combine this Prop. \ref{props:MChen-Prop3.10} 
with Thm. \ref{thm:mirror-exist} (verified for $n=2$, and 
for a polarizable $I$). 
For any $(T^{2n=4};G,B)$ that yields a rational $\mathcal N=(1,1)$ SCFT, 
there is a geometric SYZ-mirror, and there is an isomorphism 
of rational Hodge structures\footnote{
\label{fn:mirror-is-abelianV}
This also means that the rational Hodge structures on $H^k(T^{2n=4}_\circ;\Q)$ 
by $I^\circ$ are polarizable, because we chose $I$ so that 
the rational Hodge structure on $H^1(T^{2n};\Q)$ is polarizable
\cite[Prop. 9.4.3]{Golyshev:1998vzz}. 
} %
\begin{align}
  (W^k_h/W^{k+2}_h) \cong_{/\Q} g^*(W^k_{h\circ}/W^{k+2}_{h\circ})
\end{align}
for all $k=0,1,\cdots, 2n=4$; the Hodge structure 
is given by $I$ (equivalently, by $\rho_{\rm spin}(h_{I,B})$) 
on the left-hand side, and by $I^\circ$ (equivalently, by 
$\rho_{\rm spin}(h_{\omega,B})$) on the right-hand side. 
By using Prop. \ref{props:MChen-thm2.5} and the discussion 
right after Lemma \ref{lemma:Borcea}, we arrive at the following theorem.
\begin{thm}
\label{thm:MChen-thm3.11-refined}
This is\footnote{
 Since Thm. \ref{thm:mirror-exist} has been 
confirmed only for $n=2$.
} %
 for $n=2$. {\it For a general $(T^{2n=4}; G, B)$ whose corresponding 
$\mathcal N=(1,1)$ SCFT is rational, choose a polarizable complex structure $I$ 
with which $G$ is compatible, and $B^{(2,0)} = 0$; such 
a complex structure $I$ exists because of Prop. \ref{props:I-polNalgB}. 
For any geometric SYZ-mirror of $(T^{2n=4};G,B;I)$ 
(which is known to exist because of Thm. \ref{thm:mirror-exist}), with the 
T-dual taken along $\Gamma_f \subset H_1(T^{2n};\Z)$, the 
vertical rational Hodge structure on $g^*(W^k_{h\circ}/W^{k+2}_{h\circ})$
by $\rho_{\rm spin}(h_{\omega, B})$ is of CM-type, and is Hodge isomorphic 
to the CM-type rational Hodge structure $(W^k_h/W^{k+2}_h)$ by 
$\rho_{\rm spin}(h_{I,B})$ for all $k=0, 1,\cdots, 2n=4$.  

When there are multiple geometric SYZ-mirrors, there are multiple 
different filtrations $g^*(W^\bullet_{h\circ})$ installed on $H^*(T^{2n=4};\Q)$. 
The statement here is meant to apply for one common $\rho_{\rm spin}(h_{\omega, B})$
and all different $g^*(W^\bullet_{h\circ})$. That is not surprising given the 
discussion in \ref{statmnt:multi-mirror}, however. }
\end{thm}

Thm. \ref{thm:MChen-thm3.11-refined} is meant to be a refined 
version of \cite[Thm. 3.11]{Chen:2005gm}. In the present version, 
we make it clear that this Thm. \ref{thm:MChen-thm3.11-refined} 
is applicable and is not an empty statement for any set of data 
$(T^{2n};G,B;I)$ for a rational SCFT, albeit only for $n=2$ at this moment.  

\begin{rmk}
The CM-ness of the vertical rational Hodge structures on 
$g^*(W^k_{h\circ}/W^{k+2}_{h\circ})$ follows immediately from the rational 
nature of the mirror $\mathcal N=(1,1)$ SCFT,\footnote{
The $\mathcal N=(1,1)$ SCFT for $(T^{2n};G,B)$ and $\mathcal N=(1,1)$ SCFT for 
$(T^{2n};G^\circ, B^\circ)$ are isomorphic. So the latter is rational 
when the former is. 
} %
combined with Props. \ref{props:MChen-thm2.5} (and the discussions 
after Lemma \ref{lemma:Borcea}). We need Prop. \ref{props:MChen-Prop3.10}, 
however, for the existence of a horizontal--vertical Hodge isomorphism 
$(W^k_h/W^{k+2}_h) \cong g^*(W^k_{h\circ}/W^{k+2}_{h\circ})$.
\end{rmk}

\subsubsection{The Simple Level-$n$ Vertical Hodge Substructure}

The states $e^{2^{-1}(B \pm i\omega)}$ in $H^*(T^{2n};\Q) \otimes \C$ are the 
generators of the unique $h_{\omega,B}(e^{i\alpha}) = e^{\mp i n \alpha}$ 
eigenstates (see Discussion \ref{statmnt:V-GCS}). These U(1) eigenstates 
must be in $g^*(W^n_{h\circ} \otimes \C)$ of any mirror description; 
we cannot claim that these states must be purely in $g^*(H^n(T^{2n}_\circ;\C))$, 
because it is not guaranteed whether the Hodge $(2,0)$ component 
(with respect to $I^\circ$) of $B^\circ$ vanishes. 

\begin{figure}[tb]
 \begin{center}
  \includegraphics[width=0.6\linewidth]{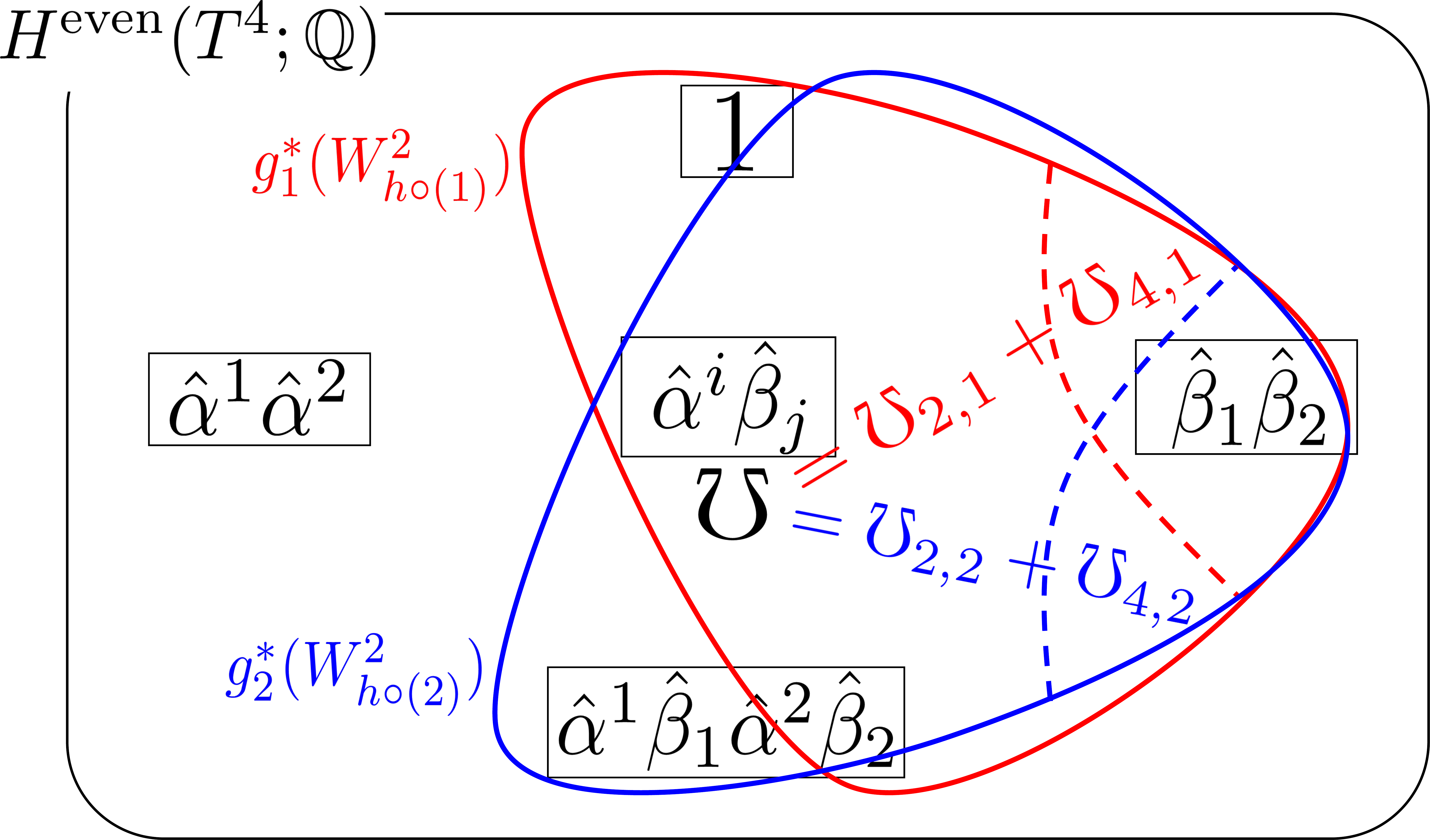} 
  \caption{\label{fig:mult-filtr} 
The grading structures pulled back from multiple different SYZ-mirrors
are not necessarily identical. The decomposition of $\mho \in H^*(T^4;\C)$
into $\mho_4 e_4^\circ$ and $\mho_2$ therefore depends on the SYZ-mirrors.
The combination $\mho$ is still within $g^*(W^2_{h\circ} \otimes \C)$ for 
any geometric SYZ-mirror. }
\end{center}
\end{figure}

Choose one mirror description for definiteness for the moment. Then 
$\mho := e^{2^{-1}(B+i\omega)} \in g^*(W^2_{h\circ})$ has a decomposition 
\begin{align}
  \mho = \mho_4 e_4^\circ + \mho_2 \in g^*(H^4(T^4_\circ;\C)) \oplus
    g^*(H^2(T^4_\circ;\C)), 
\end{align}
where $\Q e_4^\circ = g^*(H^4(T^4_\circ;\Q)) \subset H^{*}(T^4;\Q)$. 
The decomposition is possible in fact within 
\begin{align}
  \mho = \mho_4 e_4^\circ + \mho_2 \in
   \tau^r_{(20)}(K^r) \otimes_\Q g^*(H^4(T^4_\circ;\Q)) \oplus
   \tau^r_{(20)}(K^r) \otimes_\Q g^*(H^2(T^4_\circ;\Q)), 
\end{align}
because $B=B^{\rm alg}$ is rational, and $i\omega \in H^2(T^4;\Q)
\otimes \tau^r_{(20)}(K^r)$ as we have seen in Thm. \ref{thm:Kahler-is-alg}. 
The vertical rational Hodge structure on 
$g^*(H^{n=2}(T^{2n=4}_{I^\circ};\Q))$ is of CM-type 
(Thm. \ref{thm:MChen-thm3.11-refined}), so there must be 
a CM-type simple Hodge substructure of level-$(n=2)$,
$g^*([H^2(T^4_\circ;\Q)]_{\ell=2}) \subset g^*(H^2(T^4_\circ;\Q))$; 
the state $\mho_2$ must be in this level-2 component. 
The CM field is the reflex field $K^r$ because 
${\rm End}(T_M \otimes \Q)^{\rm Hdg} \cong K^r$.  
As a general property (Lemma \ref{lemma:very-useful}), 
the $\dim_\Q(T_M\otimes \Q)$-dimensional level-$n$ simple Hodge 
substructure is generated by the Galois conjugates on the linear 
combination coefficients of the state\footnote{
In applying Lemma \ref{lemma:very-useful}, we should keep in mind 
that we should rescale the state $\mho_2$ to $\mho'_2 \in \C \mho_2$
in general so that $\mho'_2$ is identified with a state of the form 
$e_I \tau^r_a(\eta_I)$ for some rational basis $\{ e_I \}$ of the 
$[K^r:\Q]$-dimensional vector space $g^*([H^2(T^4_\circ;\Q)]_{\ell=2})$ 
and some 
basis $\{ \eta_I \}$ of $K^r/\Q$. In the application here, however, 
we already know that $\mho_2 \in \tau^r_{(20)}(K^r) \otimes H^*(T^4;\Q)$, 
so we should use it as it is for the state of the form $e_I \tau^r(\eta_I)$ 
in Lemma \ref{lemma:very-useful}. 
} %
 $\mho_2$ relatively to a rational basis of the 
$\dim_\Q(T_M\otimes \Q)$-dimensional vector space; in fact, 
it does not matter any one of rational basis of the larger space 
$H^*(T^4;\Q)$ is used for the expansion. The $[K^r:\Q]$ states 
$\{ \mho_2^\sigma \; | \; \sigma \in {\rm Gal}(\overline{\Q}/\Q)\}$
generate the vector space $g^*([H^2(T^4_\circ;\Q)]_{\ell=2})\otimes \C$. 
See Fig. \ref{fig:mult-filtr}. 
The argument here may sound a little abstract; it is still a straightforward 
exercise to work out the decomposition $\mho = \mho_4 e_4^\circ+ \mho_2$
for each example of geometric SYZ-mirrors in section \ref{ssec:mirror-exist}.

It follows from the discussion above that the $[K^r:\Q]$ states 
are in this subspace, 
\begin{align}
  \mho^\sigma = \mho_4^\sigma e^\circ_4 + \mho_2^\sigma
    \in g^*(H^4(T^4_\circ;\C)) \oplus g^*([H^2(T^4_\circ;\C)]_{\ell = 2})
    \subset \C \otimes g^*(W^2_{h\circ}), 
\end{align}
while they also generate a $[K^r:\Q]$-dimensional subspace---denoted 
by $g^*(T^\circ_M)\otimes \C$---of the algebraic part 
$A(T^4_I)\otimes \C \subset H^*(T^4;\C)$, where 
\begin{align}
 A(M ) \otimes \Q :=
    \oplus_{m=0}^{n} \left( H^{2m}(M;\Q) \cap H^{m,m}(M; \R) \right)
\end{align}
for a K\"{a}hler manifold $M$. The composition 
\begin{align}
g^*(T^\circ_M \otimes \Q) \hookrightarrow g^*(W^2_{h\circ}) \rightarrow
  g^*([H^2(T^4_\circ;\Q)]_{\ell=2}), \qquad 
  \mho^\sigma \longmapsto \mho_2^\sigma
\end{align}
is an isomorphism 
of rational Hodge structures. This isomorphism is in fact that of a polarized 
Hodge structure, because the pairing (\ref{eq:def-vert-polarizatn})
is not sensitive to the $W^4_{h\circ}$ component. Therefore, reminding 
ourselves that the discussion above holds true for any mirror description, 
we have 
\begin{thm}
\label{thm:mirror-(n0)-isIN-alg}
{\it Let $(T^{2n=4};G,B)$ be a set of data for a rational $\mathcal N=(1,1)$ SCFT, 
and $I$ a polarizable complex structure with which $G$ is compatible and 
$B^{(2,0)}=0$. Then there exists a $[K^r:\Q]$-dimensional vector subspace
$g^*(T^\circ_M)\otimes \Q \subset A(T^4_I)\otimes \Q$ determined uniquely, 
which admits a CM-type weight-2 polarized rational Hodge structure 
with the endomorphism field $K^r$; its Hodge $(2,0)$ and $(0,2)$ components 
are generated by $e^{2^{-1}(B+i\omega)}$, and the polarization is given 
by (\ref{eq:def-vert-polarizatn}). 

For any geometric SYZ-mirror, $g^*(T^\circ_M)\otimes \Q \subset 
g^*(W^{n=2}_{h\circ})$, and there is an isomorphism of polarized rational 
Hodge structures between $g^*(T^\circ_M)\otimes \Q$ and 
$g^*([H^2(T^4_\circ;\Q)]_{\ell=2})$. } $\square$
\end{thm}

For a general set of data $(T^{2n};G,B)$ for which the $\mathcal N=(1,1)$ SCFT 
is not necessarily rational, and for a general complex structure $I$ 
with which $G$ is compatible, $\dim_\Q(A(T^{2n}_I))$ can be as small as 2.
On the other hand, generically the level-$n$ simple Hodge substructure 
of $H^n(T^{2n}_{I^\circ};\Q)$, if a geometric SYZ-mirror exists, is of 
$b_n = {}_{2n}C_n$ dimensions. Obviously the property stated in 
Thm. \ref{thm:mirror-(n0)-isIN-alg} no longer holds true in such 
a general set up.  

Let us now summarize what we have found so far for $T^4$ in the style 
as close as possible to Conjecture \ref{conj:GV-original}. 
\begin{thm}
\label{thm:RCFT-2-GV-forT4}
Let $(M; G, B)$ a set of data of a real $2n$-dimensional manifold $M$, 
a Ricci-flat metric $G$ and a closed 2-form $B$ on $M$; we assume that 
there exists a complex structure $I$ so that $(M, G, I)$ can be regarded 
as a K\"{a}hler manifold. We have so far verified the following statements 
in the case $M=\R^4/\Z^{\oplus 4} = T^4$. 

{\it Suppose that the $\mathcal N=(1,1)$ SCFT for the set of data $(M;G,B)$ 
is rational. Then }
\begin{enumerate}
  \item there exists a polarizable complex structure $I$ on $M$, 
    with which $G$ is compatible and $(M,G,I)$ becomes K\"{a}hler, 
    and $B^{\rm transc}=0$.
\end{enumerate}
{\it For such a complex structure $I$ ($M_I$ is meant to be the complex 
manifold $(M, I)$), there are}
\begin{enumerate}
\setcounter{enumi}{1}
\item properties on the horizontal and vertical simple level-$n$ rational 
Hodge substructures: 
\begin{enumerate}
   \item The level-$n$ simple Hodge substructure on $[H^n(M;\Q)]_{\ell=n}$ 
  by $I$ is of CM-type, where the CM field 
  ${\rm End}([H^n(M;\Q)]_{\ell =n})^{\rm Hdg}$ is denoted by $K'$. 
  \item There exists a $[K':\Q]$-dimensional vector subspace 
  $A(M_I)\otimes \Q$ denoted by $T_M^v\otimes \Q$ on which  
  a simple level-$n$ rational Hodge structure of weight-$n$
  can be introduced, with the polarization (\ref{eq:def-vert-polarizatn});
  its Hodge $(n,0)$ component is generated by $\mho := e^{2^{-1}(B+i\omega)}$, 
   where $\omega = 2^{-1}G(I-,-)$, and this polarized rational Hodge 
  structure is of CM-type, with the endomorphism field $K'$.
  \item There is an isomorphism of polarized rational Hodge 
       structure of weight-$n$ between the vertical and horizontal 
       simple level-$n$ components $T^v_M\otimes \Q$ and $T_M\otimes \Q$. 
\end{enumerate}
\item There are also properties on the rational Hodge substructures 
other than the level-$n$ components: 
 \begin{enumerate}
   \item All other rational polarizable Hodge structures on 
     $H^k(M;\Q)$ by $I$ are also of CM-type. 
  \item There is a filtration $W^\bullet_v$ on $H^*(M;\Q)$ so that 
     the data $(\rho_{\rm spin}(h_{\omega,B}), W^\bullet_v)$ introduces 
     a generalized rational Hodge structure on $H^*(M;\Q)$, so that 
   \begin{enumerate}
     \item $T^v_M\otimes \Q \subset W^n_v$, and 
     \item the rational Hodge structures on $W^k_v/W^{k+2}_v$ is 
      of CM-type for all $k$, and the one for $k=n$ is polarized
        by the pairing (\ref{eq:def-vert-polarizatn}).
   \end{enumerate}
 \end{enumerate}
\end{enumerate}
{\it Furthermore, }
\begin{enumerate}
\setcounter{enumi}{3}
 \item there is a geometric SYZ-mirror to the $\mathcal N=(2,2)$ SCFT 
     for the data $(M; G, B; I)$, and 
 \item the filtration $W^\bullet_v$ referred to above (and no.6 below) 
    can be interpreted 
    as that of $g^*(W^\bullet_{h\circ})$, the one on the geometric SYZ-mirror.
\end{enumerate}
{\it Finally, there is one more property that makes sense only for 
a family of $(M;G,B)$ that is self-SYZ-mirror (as in the case of 
$M=T^{2n}$ and K3):}
\begin{enumerate}
 \setcounter{enumi}{5}
\item there is a one-to-one correspondence between the simple rational 
    horizontal Hodge substructures on $(W^k_h/W^{k+2}_h)$ and vertical 
    Hodge substructures on $(W^k_v/W^{k+2}_v)$ so that there are Hodge 
    isomorphism.
\end{enumerate}
{\it Furthermore, here is one more property whose generalization 
to $M$ other than a torus is not obvious:}

\begin{enumerate}
 \setcounter{enumi}{6}
 \item the isomorphisms between the horizontal and vertical rational 
   Hodge structures
    can be interpreted 
   as a combination of an isogeny and a mirror map of D-brane charges. 
\end{enumerate}
At this moment, the authors do not have hard evidence to believe 
that the 3rd property follows as a consequence of the 2nd one, because 
an issue remains in Rmk. \ref{statmnt:CM-from-H2-to-H1}. So, 
the properties no.2 and no.3 are listed independently here. 
\end{thm}

{\bf Proof:} This is just a brief note on the origin of the properties 
above. The property no.1 is from Prop. \ref{props:I-polNalgB}. 
The property no.2(a) is essentially due to \cite[Thm. 2.5]{Chen:2005gm}
(quoted as Prop. \ref{props:MChen-thm2.5} in this article), while the fact that 
$B+i\omega \in {\cal H}^2(T^4)\otimes \tau^r_{(n0)}(K^r)$---2.(b)---is 
from Prop. \ref{props:I-polNalgB} and Thm. \ref{thm:Kahler-is-alg}. 
The combination of the properties 3.(b).i and 5, $T_M^v\otimes \Q \subset
 g^*(W^n_{h\circ})$ for any geometric SYZ-mirror, is from the discussion 
leading to Thm. \ref{thm:mirror-(n0)-isIN-alg}. 

For a target space $M$ other than tori and K3 surfaces, the charge $n$ state 
$\mho$ under $\rho_{\rm spin}(h_{\omega, B})$ is no longer $e^{2^{-1}(B+i\omega)}$, 
but is likely to be the one including the worldsheet instanton corrections. 
That is indicated at least by the example of the Gepner construction of 
a Fermat quintic Calabi--Yau threefold (reviewed in section \ref{ssec:GV-conj}). $\square$ 

\subsubsection{Trial Statements for a Complex Structure with $B^{(2,0)} \neq 0$}
\label{sssec:result-for-genI}

The results in Thms. \ref{thm:MChen-thm3.11-refined}, 
\ref{thm:mirror-(n0)-isIN-alg} and \ref{thm:RCFT-2-GV-forT4} 
are valid only for a polarizable complex structure $I$ where 
$B^{(2,0)} = 0$, because Thm. \ref{thm:mirror-exist} 
guaranteeing a geometric SYZ-mirror has been proven only for such complex 
structures. For a general polarizable complex structure $I$, 
not necessarily $B^{(2,0)}=0$, one may still be interested in 
characterizing the data $(T^{2n=4};G,B; I)$ for a rational SCFT
in terms of Hodge structures. Thm. \ref{thm:RCFT-2-GV-forT4-genI}
below is meant to be for that broader class of complex structures. 

Suppose that $(T^4;G,B)$ is for a rational $\mathcal N=(1,1)$ SCFT, 
and $I$ a polarizable complex structure with which $G$ is 
compatible. Suppose further that a geometric SYZ-mirror exists. 
Then Prop. \ref{props:MChen-thm2.5}, 
Thms. \ref{thm:Kahler-is-alg} and \ref{thm:MChen-thm3.11-refined} 
still hold true. Discussion leading to 
Thm. \ref{thm:mirror-(n0)-isIN-alg}, however, needs to be modified 
at one point. The elements $\{\mho^\sigma \}$ do not 
generate a $[K^r:\Q]$-dimensional subspace of $A(M)\otimes \C$, 
but a $[K^r:\Q]$-dimensional subspace of $(e^{B^{\rm transc}/2} A(M)\otimes \C)$; 
let us still use the same notation $g^*(T^\circ_M)\otimes \Q$ for 
the $[K^r:\Q]$-dimensional space in $e^{B^{\rm transc}/2}A(M) \otimes \Q$.
So, we have an analogue of Thm. \ref{thm:RCFT-2-GV-forT4}:
\begin{thm}
\label{thm:RCFT-2-GV-forT4-genI}
Let $(M; G, B)$ a set of data of a real $2n$-dimensional manifold $M$, 
a Ricci-flat metric $G$ and a closed 2-form $B$ on $M$; we assume that 
there exists a complex structure $I$ so that $(M, G, I)$ can be regarded 
as a K\"{a}hler manifold. We have so far verified the following statements 
in the case $M=\R^4/\Z^{\oplus 4} \cong T^4$. In the following, we avoid repeating 
literally the same sentences as in Thm. \ref{thm:RCFT-2-GV-forT4} 
and abbreviate them by ``.......''. 

{\it Suppose that the $\mathcal N=(1,1)$ SCFT for the set of data $(M;G,B)$ 
is rational. Choose a polarizable complex structure $I$ on $M$ 
with which $G$ is compatible and $(M,G,I)$ becomes K\"{a}hler, and 
assume that there is a geometric SYZ-mirror for $(M; G, B; I)$. Then }
\begin{enumerate}
  \item $B^{\rm transc} \in T_M \otimes \Q \subset T_M\otimes \R$.
\end{enumerate}
{\it For the complex structure $I$ ($M_I$ is meant to be the complex 
manifold $(M, I)$), thre are }
\begin{enumerate}
\setcounter{enumi}{1}
\item properties on the horizontal and vertical simple level-$n$ rational 
Hodge substructures: 
\begin{enumerate}
   \item The level-$n$ simple Hodge substructure on $[H^n(M;\Q)]_{\ell=n}$ 
  by $I$ is of CM-type, where ........ 
  \item There exists a $[K':\Q]$-dimensional vector subspace of 
  $e^{B^{\rm transc}/2}A(M_I)\otimes \Q$ denoted by $T_M^v\otimes \Q$ on which  
  a simple level-$n$ rational ....... \footnote{
\label{fn:B-alg-vU(1)}
The state $\mho = e^{2^{-1}(B+i\omega)}$ is in 
$e^{2^{-1}B^{\rm transc}} A(M_I)\otimes \C$, and the state  
$e^{2^{-1}(B^{\rm alg}+i\omega)}$ is in $A(M_I)\otimes \C$. The latter 
state is a U(1) eigenstate of $S^1_{\omega, B'}$ for $B'=B^{\rm alg}$. 
We are not sure if there is any importance in this observation, 
but let us just note this down here. 
} %
 \item There is an isomorphism of .........
\end{enumerate}
\item There are also properties on the rational Hodge substructures 
other than the level-$n$ components:
 \begin{enumerate}
   \item All other rational polarizable Hodge ........ 
  \item There is a filtration $W^\bullet_v$ on $H^*(M;\Q)$ so that 
     ........ 
 \end{enumerate}
\end{enumerate}
{\it Furthermore, }
\begin{enumerate}
\setcounter{enumi}{3}
 \item (ignore this, because we did not prove Thm. \ref{thm:mirror-exist}
   for $I$ with $B^{(2,0)}\neq 0$)
 \item the filtration $W^\bullet_v$ referred to above (and also at no.6 below) 
  can be interpreted as ......... 
\end{enumerate}
{\it Finally, there is one more property that makes sense only for 
a family of $(M;G,B)$ that is self-SYZ-mirror (as in the case of 
$M=T^{2n}$ and K3): }
\begin{enumerate}
 \setcounter{enumi}{5}
\item there is a one-to-one correspondence between ........
\end{enumerate}
{\it Furthermore, here is one more property whose generalization 
to $M$ other than ........:}

\begin{enumerate}
 \setcounter{enumi}{6}
 \item the isomorphisms between the horizontal and vertical ... 
\end{enumerate}
The remark at the end of Thm. \ref{thm:RCFT-2-GV-forT4} also applies 
here. $\square$ 
\end{thm}

\begin{rmk}
Although the rationality of $B^{\rm transc}$ is listed (as the no.1 property) 
separately in the statement of Thm. \ref{thm:RCFT-2-GV-forT4-genI}, there 
may be a way to encode this property together with others.\footnote{
The property 2(b), saying that there exists a vector subspace 
$(T^v_M\otimes \Q) \subset H^*(M;\Q)$ even within 
$e^{B^{\rm transc}/2} A(M_I)\otimes \Q$, makes sense only when 
$B^{\rm transc}$ is rational, though. 
} %

A na\"ive idea (which fails in the following) is to introduce the algebra 
of endomorphisms of a generalized rational Hodge structure, 
\begin{align}
  {\rm End}(H^*(T^{2n};\Q))^{(\rho_{\rm spin}(h_{I,B}), W^\bullet_h)},
  \label{eq:def-EndHdg-alg-twistB}
\end{align}
those that preserve the filtration, and commute with the 
$\rho_{\rm spin}(h1_{I,B})$ action.
When $B^{\rm transc}$ is rational, the algebra above contains 
\begin{align}
  e^{B^{\rm transc}/2} \left( \oplus_k {\rm End}(H^k(T^{2n};\Q))^{\rm Hdg}
     \right) e^{-B^{\rm transc}/2}.  
\end{align}
So, when the rational Hodge structures on $H^k(M;\Q)$ by $I$ are all CM-type, 
and $B^{\rm transc}$ is rational, the algebra (\ref{eq:def-EndHdg-alg-twistB})
contains a commutative semi-simple algebra $F$ over $\Q$ of 
dimension equal to $\dim_\Q(H^*(T^{2n};\Q))$ whose quotient representation 
on $W^k_h/W^{k+2}_h$ is a $\Q$-algebra of dimension $b_{k}(T^{2n})$. 

Conversely, however, it is possible for the case where $B^{\rm transc}$ differs 
from a rational $B'_{\rm ratnl} \in T_M \otimes \Q$ by a (1, 1) form, that 
the algebra (\ref{eq:def-EndHdg-alg-twistB}) contains a commutative 
subalgebra $F$ with $\dim_\Q F = \dim_\Q(H^*(T^{2n};\Q))$, and 
$\dim_\Q (F|_{W^k_h/W^{k+2}_h}) = b_k$, if $(T^{2n}; I)$ is a complex torus 
with sufficiently many complex multiplications. 
\end{rmk}

\subsection{The Converse}
\label{ssec:converse}

Let us now study whether the converse of 
Thm. \ref{thm:RCFT-2-GV-forT4} is true. 
By imposing the 2nd and 3rd properties\footnote{
Since it is not guaranteed whether the 
third property follows from the second, we impose both. 
} %
in Thm. \ref{thm:RCFT-2-GV-forT4} on a set of 
data $(T^{2n=4};G,B)$ and a polarizable $I$ with $B^{\rm{transc}}=0$ 
(implicitly, the 1st property is imposed as well), 
we will see that there is a significant likelihood that the resulting $\mathcal N=(1,1)$ SCFT is rational. However, there are data $(T^4;G,B;I)$ 
satisfying the two properties, and yet the corresponding CFTs are not rational. 
The following analysis is carried out separately for the cases 
(B, C), (A'), and (A). 

\subsubsection{Case (B, C)}

Because of the 2nd and 3rd properties in the statement 
of Thm. \ref{thm:RCFT-2-GV-forT4}, we have an abelian variety 
$M=(T^4, I)$ of CM-type; a CM field $K$ of case (B, C) 
acts on $H^1(M;\Q)$, also on $H^3(M;\Q)$, and its reflex field 
$K^r$ acts on the $[K^r:\Q]=4$-dimensional transcendental part 
$T_M\otimes \Q \subset H^2(M;\Q)$. There is also $[K^r:\Q]$-dimensional 
vector subspace $T^v_M\otimes \Q \subset A(M)\otimes \Q$ on which 
$\rho_{\rm spin}(h_{\omega, B})$ introduces a rational Hodge structure 
with CM by $K^r$, polarized under the 
pairing (\ref{eq:def-vert-polarizatn}). The combination $(B+i\omega)$ 
is in $[(T^v_M\otimes \Q)\cap H^2(T^4;\Q)]\otimes \C$, and 
$e^{2^{-1}(B\pm i\omega)}$ with the sign $+$ and $-$ should generate the 
Hodge (2,0) and (0,2) components, respectively. Let us exploit all 
those information and see whether one can claim that the $B$-field and 
the metric $G$ are rational (the answer is no).

\begin{anythng}
\label{statmnt:GV-2-RCFT-caseBC-even}
The fact that $\mho := e^{(B+i\omega)/2}$ is the only generator 
of the Hodge $(2,0)$ component of the CM-type Hodge structure 
on $T^v_M\otimes \Q$, with the CM field $K^r$ and embedding 
$\tau^r_{(20)}=\tau^r_{++}$, implies that there must be a basis 
$\{ 1, \eta_1, \eta_2, \eta_4\}$ of $K^r/\Q$, so that 
\begin{align}
 \mho & \; = \tau^r_{++} \left[ 1 + e_1 \eta_1 + e_2 \eta_2
     + (\hat{\alpha}^1\hat{\beta}_1\hat{\alpha}^2\hat{\beta}_2) \eta_4
    \right], 
\end{align}
where a rational basis $\{ e_1, e_2 \}$ of ${\cal H}^2(M)$
is the one introduced in (\ref{eq:def-ratBasis-H2-caseBC});  
we must set $\eta_4 = (d\eta_1^2-\eta_2^2) \in K^r$ so that 
$(\mho,\mho)=0$. 
So, there are eight rational parameters for $\eta_1, \eta_2 \in K^r$, 
for the moment. The Hodge (0,2) component should be given by 
\begin{align}
  \overline{\mho} = \tau^r_{+-}\left[ 1 + e_1 \eta_1 + e_2 \eta_2
  + (\hat{\alpha}^1\hat{\beta}_1\hat{\alpha}^2\hat{\beta}_2) \eta_4 \right], 
\end{align}
and the (1,1) components by the two vectors
\begin{align}
 \Sigma & \; =  \tau^r_{-+}\left[ 1 + e_1 \eta_1 + e_2 \eta_2
  + (\hat{\alpha}^1\hat{\beta}_1\hat{\alpha}^2\hat{\beta}_2) \eta_4 \right], \\
 \overline{\Sigma} & \; = 
    \tau^r_{--}\left[ 1 + e_1 \eta_1 + e_2 \eta_2
  + (\hat{\alpha}^1\hat{\beta}_1\hat{\alpha}^2\hat{\beta}_2) \eta_4 \right].  
\end{align}

This Hodge decomposition must be polarized with respect to 
(\ref{eq:def-vert-polarizatn}). The condition $(\mho, \mho)=0$
is built in by construction, $\mho = e^{2^{-1}(B+i\omega)}$. The 
remaining non-trivial information from the polarization is that 
$(\mho, \Sigma)=0$ and $(\mho, \overline{\Sigma})=0$. The two conditions 
are equivalent to 
\begin{align}
  -2^{-1} \left( \tau^r_{++}(X) - \tau^r_{-\pm}(X),
    \tau^r_{++}(X) - \tau^r_{-\pm}(X) \right)_{{\cal H}^2} = 0
\end{align}
for $X=e_1\eta_1 + e_2 \eta_2$, using just the pairing in ${\cal H}^2(M)$; 
those conditions are further rewritten as 
\begin{align}
  d \left( \tau^r_{++}(\eta_1)-\tau^r_{-\pm}(\eta_1) \right)^2
 - \left( \tau^r_{++}(\eta_2)-\tau^r_{-\pm}(\eta_2) \right)^2 = 0
   \label{eq:cond-vert-Pol-caseBC}
\end{align}
in the normal closure of the number field $K^r$. 

The eight rational parameters for $\eta_{1,2} \in K^r$, that is, 
$A,B,C,D,\widetilde{A}, \widetilde{B}, \widetilde{C}, \widetilde{D} \in \Q$ in 
\begin{align}
 \eta_1 =: A + B y' + C\xi^r + D \xi^r y', \qquad 
 \eta_2 =: \widetilde{A} + \widetilde{B} y'
     + \widetilde{C} \xi^r + \widetilde{D} \xi^ry', 
\end{align}
should satisfy the conditions (\ref{eq:cond-vert-Pol-caseBC}). 
Straightforward computation translates the conditions to 
\begin{align}
 d BC = \widetilde{B}\widetilde{C}, \quad 
 d BD = \widetilde{B}\widetilde{D}, \quad 
 d(D^2d'-C^2) = (\widetilde{D}^2d' -\widetilde{C}^2), 
\end{align}
along with 
\begin{align}
 d\left[ d'(B^2-2CD)+p(C^2+d'D^2)\right] = 
 \left[d'(\tilde{B}^2-2\widetilde{C}\widetilde{D})
        + p(\widetilde{C}^2+d'\widetilde{D}^2)\right].
\end{align}
There are four conditions on the eight parameters. 

First, one can immediately see that the rational parameters $A$ 
and $\widetilde{A}$ dropped out. So any $A,\widetilde{A} \in \Q$
has no conflict with the condition (\ref{eq:cond-vert-Pol-caseBC})
for the consistency of the Hodge structure (of $B+i\omega$) with 
the polarization (\ref{eq:def-vert-polarizatn}). 

Second, we prove that $\widetilde{B}=0$ by contradiction. 
If $\widetilde{B}\neq 0$, then $\widetilde{C}$ and $\widetilde{D}$ 
can be solved in terms of $C$, $D$ and $B/\widetilde{B}$. Then
\begin{align}
   (D^2d'-C^2)\left( \frac{B^2}{\widetilde{B}^2}d-1\right)=0.
\end{align}
This is a contradiction\footnote{
If $D=C=\widetilde{D}=\widetilde{C}=0$, then $[T^v_M\otimes \C]^{(2,0)}
=[T^v_M \otimes \C]^{(0,2)} = \C \subset T^v_M\otimes \C$. 
This is not appropriate as a Hodge decomposition. In physics terminology,
this corresponds to $\omega = 0$, and ${\rm volume}(T^4)=0$.  
} %
because neither $d$ nor $d'$ is a square of a rational number. 

Thirdly, $\widetilde{B}=0$ implies that either $B=0$ or $C=D=0$ holds
true. The latter is not possible, however, 
because $\widetilde{D}^2d'-\widetilde{C}^2=0$ would follow, although $d'$ 
is not a square of a rational number. So, $B=0$. We have now proved 
that the $B$-field is 
\begin{align}
  \tau^r_{++}\left[ e_1 (A + By')
    + e_2 (\widetilde{A}+\widetilde{B}y')\right] = A e_1 + \widetilde{A}e_2
\end{align}
for free $A,\widetilde{A} \in \Q$. This is the same as saying that the 
$B$-field is in ${\cal H}^2(M)$.
The rationality condition of the $B$-field (\ref{eq:cond-GnB-rational}) 
follows from the 2nd and 3rd (and 1st) properties in the statements 
of Thm. \ref{thm:RCFT-2-GV-forT4}. 

Next, change the parametrization as follows.
\begin{align}
  C = \frac{1}{2}\left( C' + \frac{p D'}{qd}\right), \quad 
  D = \frac{D'}{2qd}, \quad 
 \widetilde{C} = \frac{1}{2}\left( \widetilde{C}'
      + \frac{p}{q} \widetilde{D}'\right), \quad \widetilde{D}= \frac{\widetilde{D}'}{2q}, 
\end{align}
or equivalently, 
\begin{align}
  \xi^r(C+Dy') = D' \frac{q}{\xi^r} + \frac{C'}{2}\xi^r, \qquad 
 \xi^r(\widetilde{C} + \widetilde{D} y') =
       \frac{\widetilde{C}'}{2}\xi^r+\widetilde{D}'\frac{qd}{\xi^r} .
\end{align}
Then the remaining two conditions on $C, D, \widetilde{C}, \widetilde{D}$ 
are rewritten as 
\begin{align}
  d(C')^2 + \frac{2p}{q}(C'D') + (D')^2 & \; = 
        d (\widetilde{D}')^2+\frac{2p}{q}\widetilde{C}'\widetilde{D}'
           +(\widetilde{C}')^2 , \\
 (C')^2 pd + (D'C')2qd + (D')^2 p & \; = 
   (\widetilde{D}')^2 dp + (\widetilde{D}'\widetilde{C}') 2qd
      + (\widetilde{C}')^2 p . 
\end{align}
So, this is equivalent to 
\begin{align}
  D' C' = \widetilde{C}' \widetilde{D}', \qquad 
  d (C')^2 + (D')^2 = d (\widetilde{D}')^2 + (\widetilde{C}')^2. 
\end{align}
This coupled quadratic equations seem to allow two possibilities, 
\begin{align}
 \frac{\widetilde{C'}}{\widetilde{D}'} = d \frac{C'}{D'}, \qquad 
 \frac{\widetilde{C}'}{\widetilde{D}'} = \frac{D'}{C'}, 
\end{align}
including $D'=\widetilde{D}'=0$ and 
$\widetilde{D}'=C'=0$, respectively. 
The first case is impossible, because 
$(\widetilde{C}')^2 = d (C')^2$ is a 
contradiction for the parameters $C', \widetilde{C}' \in \Q$ for 
$d$ that is not a square. The only option is 
\begin{align}
  (C', D') = (\widetilde{D}', \widetilde{C}'), 
  \label{eq:cond-polHstr-CM-to-kahler-OK}
\end{align}
and 
\begin{align}
     (C', D') = - (\widetilde{D}', \widetilde{C}').
   \label{eq:cond-polHstr-CM-to-kahler-opp}
\end{align}
There are two kinds of solutions, (\ref{eq:cond-polHstr-CM-to-kahler-OK}) 
and (\ref{eq:cond-polHstr-CM-to-kahler-opp}), for the Hodge structure 
on $[T^v_M\otimes \Q]$ to be compatible with the polarization 
(\ref{eq:def-vert-polarizatn}); for solutions of both kinds, 
there are two free rational parameters $C', D' \in \Q$ for $\omega$
(besides the two free parameters $A,\widetilde{A} \in \Q$ for the $B$-field). 

In the first kind of solutions, (\ref{eq:cond-polHstr-CM-to-kahler-OK}), 
we have 
\begin{align}
 \frac{i}{2}\omega & \; = \tau^r_{++}\left( \frac{C'}{2}\xi^r+ \frac{D'q}{\xi^r}\right)
   e_1 + \tau^r_{++}\left( \frac{C'qd}{\xi^r}
                  +\frac{D'}{2} \xi^r \right) e_2, \\
  & \; = \frac{C'}{2} \left( e_1 (\sqrt{+}+\sqrt{-})
                  + e_2 (\sqrt{+}-\sqrt{-})\sqrt{d} \right)  \\
 & \qquad
     + \frac{D'}{2} \left( e_2 (\sqrt{+}+\sqrt{-})
                 +e_1(\sqrt{+}-\sqrt{-})/\sqrt{d} \right),  \nonumber \\
 & \; = i(p\omega^{(x)}-q\omega^{(xy)})\left(-\frac{C'}{4d'}\right)
  + i(-qd\omega^{(x)}+p\omega^{(xy)}) \left(-\frac{D'}{4dd'}\right).
\end{align}
The last expression is a rational linear combination 
of the basis $\omega^{(\beta)}$ of K\"{a}hler 
forms (\ref{eq:basis-KahlerF-forRatG}) corresponding to a rational metric;
the expression in the middle can easily be identified with $i\omega$ 
in (\ref{eq:Kahler-form-BC-4RCFT-2}) for a rational metric with the 
dictionary $2a = C'$ and $2b = D'/d$. 

In the other kind of solutions, (\ref{eq:cond-polHstr-CM-to-kahler-opp}),
on the other hand, 
\begin{align}
 \frac{i}{2}\omega & \; = \tau^r_{++}\left( \frac{C'}{2}\xi^r+ \frac{D'q}{\xi^r}\right)
   e_1 - \tau^r_{++}\left( \frac{C'qd}{\xi^r}
                  +\frac{D'}{2} \xi^r \right) e_2, \\
  & \; = \frac{C'}{2} \left( e_1 (\sqrt{+}+\sqrt{-})
                  - e_2 (\sqrt{+}-\sqrt{-})\sqrt{d} \right)  \\
 & \qquad
     + \frac{D'}{2} \left( - e_2 (\sqrt{+}+\sqrt{-}) +e_1(\sqrt{+}-\sqrt{-})/\sqrt{d} \right),  \nonumber 
\end{align}
Rewriting this in terms of $dz^1\wedge d\bar{z}^{\bar{1}}$ and $dz^2\wedge d\bar{z}^{\bar{2}}$ according to (\ref{eq:dz1-dzbar1-caseBC}) and (\ref{eq:dz2-dzbar2-caseBC}),
\begin{align}
  \frac{i}{2}\omega  = \frac{p-q\sqrt{d}}{4\sqrt{d'}}\left(C'-\frac{D'}{\sqrt{d}}\right)dz^1\wedge d\bar{z}^{\bar{1}}+\frac{p+q\sqrt{d}}{4\sqrt{d'}}\left(C'+\frac{D'}{\sqrt{d}}\right)dz^2\wedge d\bar{z}^{\bar{2}}.
\end{align}
For this K\"{a}hler form to be fitted by the 
expression (\ref{eq:Kahler-form-BC-4RCFT-2}), we have
\begin{align}
  a = -\frac{1}{2\sqrt{d'}}(pC'+qD'),\qquad
  b = \frac{1}{2d\sqrt{d'}}(qdC'+pD').
\end{align}
The fitted parameters $a,b$ are not rational when $C', D' \in \Q$. 
The metric corresponding to this K\"{a}hler form does not satisfy 
the condition (\ref{eq:cond-GnB-rational}).  The resulting metric 
is positive definite for some region in $(C', D') \in \Q^2$, so 
the second kind of solutions include physically sensible ${\cal N}=(1,1)$ 
SCFTs that are not rational.   $\square$ 
\end{anythng}

\begin{anythng}
\label{statmnt:GV-2-RCFT-caseBC-odd}
We have restricted $(B+i\omega)$ by demanding that the vertical 
Hodge structure on $T^v_M\otimes \Q$ is of CM-type, 
with CM by the reflex field $K^r$ of a CM field $K$.
For such a $(B+i\omega)$, whether a solution 
of (\ref{eq:cond-polHstr-CM-to-kahler-OK}) 
or (\ref{eq:cond-polHstr-CM-to-kahler-opp}), the vertical Hodge 
structure on $H^1(T^4;\Q)\oplus H^3(T^4;\Q)$ is already determined. 
Demanding that this rational Hodge structure is also of CM-type
and that  their endomorphism field is $K$, we find in the following that 
no extra condition is found on the parameters $A,\widetilde{A}, C',D' \in \Q$.

The charge $p-q=+1$ components in $H^1(T^4;\Q)\oplus H^3(T^4;\Q)$ can be 
generated by $e^{2^{-1}(B+i\omega)}\hat{\alpha}^{i}$ and 
$e^{2^{-1}(B+i\omega)}\hat{\beta}_i$ with $i=1,2$; the charge $p-q=-1$ components 
are generated by the ones with $(B+i\omega)$ replaced by $(B-i\omega)$, 
as stated in \ref{statmnt:V-GCS}. Now, we use 
\begin{align}
 \frac{1}{2} (B + i\omega) = \tau^r_{++} \left( A +\frac{C'}{2}\xi^r+\frac{D'}{2}\frac{2q}{\xi^r}\right) e_1 + \tau^r_{++}\left(\tilde{A}\pm\frac{D'}{2}\xi^r\pm\frac{C'}{2}\frac{2qd}{\xi^r}\right)e_2 , 
\end{align}
where the $+$ and $-$ choices of $\pm$ correspond to the solution 
(\ref{eq:cond-polHstr-CM-to-kahler-OK}) 
and (\ref{eq:cond-polHstr-CM-to-kahler-opp}), respectively. 

Within the vector space $V_1 :=\mathrm{Span}_\Q\{\hat{\alpha}^{1}, \;
 \hat{\alpha}^{2}\}
\oplus\mathrm{Span}_{\Q}\{\hat{\alpha}^1\hat{\alpha}^2\hat{\beta}_{1}, 
\; \hat{\alpha}^1\hat{\alpha}^2\hat{\beta}_{2}\}$, 
\begin{align}
 ( e^{2^{-1}(B+i\omega)} \hat{\alpha}^1, e^{2^{-1}(B+i\omega)}\hat{\alpha}^2) = 
 \left( \hat{\alpha}^1, \; \hat{\alpha}^2, \;
         \hat{\alpha}^1\hat{\alpha}^2\hat{\beta}_1, \; 
         \hat{\alpha}^1\hat{\alpha}^2\hat{\beta}_2 \right)
 \left(\begin{array}{cc}
    1 & 0 \\
    0 & 1 \\
    Z_2 & - Z_1 \\ dZ_1 & -Z_2 \end{array}\right), 
\end{align}
where
\begin{align}
  Z_1 := \tau^r_{++}
      \left( A +\frac{C'}{2}\xi^r+\frac{D'}{2}\frac{2q}{\xi^r}\right), \qquad 
  Z_2 := \tau^r_{++}
     \left(\tilde{A}\pm\frac{D'}{2}\xi^r\pm\frac{C'}{2}\frac{2qd}{\xi^r}\right).
\end{align}
One finds the following structure when the two generators in 
$[V_1\otimes \C]^{p-q=+1}$ are rearranged as follows:
\begin{align}
  (e^{2^{-1}(B+i\omega)} \hat{\alpha}^1, \; e^{2^{-1}(B+i\omega)} \hat{\alpha}^2) & \; 
    \left(\begin{array}{cc}
    1 & 1 \\
    \mp \sqrt{d} & \pm\sqrt{d}
  \end{array}\right) \\
  & \; = 
 \left( \hat{\alpha}^1, \; \hat{\alpha}^2, \;
         \hat{\alpha}^1\hat{\alpha}^2\hat{\beta}_1, \; 
         \hat{\alpha}^1\hat{\alpha}^2\hat{\beta}_2 \right)
\left(\begin{array}{cc}
    1 & 1 \\
   \tau_{++}( \mp y) & \tau_{-+}(\mp y) \\
   \tau_{++}(\Xi_\pm ) & \tau_{-+}(\Xi_\pm ) \\
   \tau_{++}( \pm \Xi_\pm y) & \tau_{-+}( \pm \Xi_\pm y) 
 \end{array} \right),   \nonumber 
\end{align}
where
\begin{align}
  \Xi_{\pm} = \widetilde{A} \pm A y \pm D' x \pm C' x y \in K. 
\end{align}
Unless $C'=D'=0$ (which we are not interested because the volume of $T^4$ 
is precisely zero), $\{ 1, y, \Xi_\pm, \Xi_\pm y \}$ forms a 
basis of $K/\Q$. So, with Lemma \ref{lemma:very-useful-2}, we see 
that an algebra isomorphic to $K$ acts on the vector space $V_1$ while 
preserving this Hodge decomposition. 

A similar calculation can be carried out for the rest of the vector space,  
$V_3:=\mathrm{Span}_\Q\{\hat{\beta}_{1}, \hat{\beta}_{2} \} 
 \oplus
\mathrm{Span}_\Q\{\hat{\alpha}^{1}\hat{\beta}_1\hat{\beta}_2, \;
 \hat{\alpha}^{2}\hat{\beta}_1\hat{\beta}_2\}$ in 
$H^1(T^4;\Q)\oplus H^3(T^4;\Q)$. The charge $p-q=+1$ components are generated by 
\begin{align}
  \left( \hat{\beta}_1e^{2^{-1}(B+i\omega)}, \; 
   \hat{\beta}_2 e^{2^{-1}(B+i\omega)} \right) & \; 
  \left( \begin{array}{cc}
      1 & 1 \\ \mp \sqrt{d} & \pm \sqrt{d} \end{array} \right)    \\
 & \; = \left( \hat{\beta}_1, \; \hat{\beta}_2, \;
        \hat{\alpha}^1\hat{\beta}_1\hat{\beta}_2, \; 
        \hat{\alpha}^2\hat{\beta}^1\hat{\beta}_2  \right)
  \left(\begin{array}{cc}
     1 & 1 \\
     \tau_{++}(\mp y) & \tau_{-+}( \mp y) \\
     \tau_{++}(-\Xi_\pm ) & \tau_{-+}(-\Xi_\pm) \\ 
     \tau_{++}(\mp \Xi_\pm y) & \tau_{-+}(\mp \Xi_\pm y) 
  \end{array}\right).   \nonumber 
\end{align}
The endomorphism algebra of $V_3$ contains $K$, so this Hodge structure 
is also of CM-type (the property 3.(b)ii). 

The 2nd and 3rd properties in the statement of 
Thm. \ref{thm:RCFT-2-GV-forT4} allows one to choose/find 
a filtration $W^\bullet_v$ (the 5th property demands more, though).  
So, we choose $W^3_v := V_3$. Then the horizontal rational 
Hodge structure on $H^3(T^4;\Q)$ (resp. on $H^1(T^4;\Q)$) is 
isomorphic to the vertical rational Hodge structure 
on $W^3_v$ (resp. $W^1_v/W^3_v$) (the property 6), 
as claimed at the beginning 
of \ref{statmnt:GV-2-RCFT-caseBC-odd}.  $\square$
\end{anythng}

\subsubsection{Case (A')}

Let us work on the case the endomorphism algebra of 
$H^1(M;\Q)$ is the one in case (A'). Let us exploit 
just the 1st, 2nd and 3rd properties of a pair of 
horizontal and vertical Hodge structures in the statement 
of Thm. \ref{thm:RCFT-2-GV-forT4} and see whether 
one can claim that $B$ and $G$ are rational (the answer is no). 
The logic and the procedure of the analysis are precisely 
the same as for the case (B, C). So, we will focus on 
small difference in the following presentation, and often 
avoid repeating the same logic.  

\begin{anythng}
\label{statmnt:GV-2-RCFT-caseAprm-even}
Let us impose on $(B+i\omega)$ the conditions that the vertical 
Hodge structure on $T^v_M\otimes \Q$ is of CM type, with 
the endomorphism field $K^r$ in \ref{statmnt:CMtype-n-reflexF-Aprm}. 
The generator $\mho := e^{2^{-1}(B+i\omega)}$ of the Hodge (2, 0) component of 
$T^\circ_M\otimes \C$ must be in the form of
\begin{align}
 \mho = \tau^r_{++} \left[ 1 + (\hat{\alpha}^1\hat{\beta}_1) \eta_1
      + (\hat{\alpha}^2\hat{\beta}_2) \eta_2
      + (\hat{\alpha}^1\hat{\beta}_1\hat{\alpha}^2\hat{\beta}_2)  \eta_4 \right]
\end{align}
for some basis $\{ 1,\eta_1, \eta_2, \eta_4\}$ of $K^r/\Q$. The property 
$(\mho, \mho) = 0$ implies that $\eta_4 = \eta_1 \eta_2$, so 
there are eight rational parameters for $\eta_1$ and $\eta_2$ at this 
moment. 

Let us parametrize the freedom by $A,B,C,D,\widetilde{A},
 \widetilde{B},\widetilde{C}, \widetilde{D} \in \Q$, where 
\begin{align}
 \eta_1 = A+By'+C\xi^r+D\xi^ry', \qquad 
 \eta_2 = \widetilde{A} +\widetilde{B}y'
     +\widetilde{C}\xi^r+\widetilde{D}\xi^ry'.
\end{align}
For the Hodge decomposition to be compatible with its polarization 
(\ref{eq:def-vert-polarizatn}), we impose $(\mho, \Sigma)=0$ and 
$(\mho, \overline{\Sigma})=0$. As a result, we obtain 
\begin{align}
  B\widetilde{B} + p_1 D \widetilde{D} = 0, \quad
  B\widetilde{D} + \widetilde{B}D=0, \\
  B\widetilde{B} p_2 + C\widetilde{C}=0, \quad 
  B\widetilde{C} + \widetilde{B}C = 0. 
\end{align}
Now, we have four conditions on the eight rational parameters. 

One can prove that $B=\widetilde{B}=0$ (or otherwise we should accept 
an unphysical zero-volume situation (such as $C=D=0$)); the proof is
similar to the case (B, C), so we omit the detail. The four 
conditions above are reduced to $D\widetilde{D}=0$ and $C\widetilde{C}=0$. 
So, there are two kinds of solutions (apart from the zero-volume situations):
\begin{align}
  D=0, \quad \widetilde{C}=0, & \; {\rm so} \; 
   2^{-1}(B+i\omega) & \; = \hat{\alpha}^1\hat{\beta}_1 (A+C\sqrt{p_1})
    + \hat{\alpha}^2\hat{\beta}_2
               (\widetilde{A} + \widetilde{D}\sqrt{p_1}\sqrt{d'}), 
    \label{eq:cond-vHst-B+iw-OK}  \\
  C=0, \quad \widetilde{D}=0, & \; {\rm so} \; 
   2^{-1}(B+i\omega) & \; = \hat{\alpha}^1\hat{\beta}_1 (A+D\sqrt{p_1}\sqrt{d'})
     + \hat{\alpha}^2\hat{\beta}_2 (\widetilde{A} + \widetilde{C}\sqrt{p_1}).
    \label{eq:cond-vHst-B+iw-opp}  
\end{align}
Therefore, the $B$ field has to be rational ($B=\widetilde{B}=0$ and 
${}^\forall A,\widetilde{A} \in \Q$) for both kinds 
of the solutions (\ref{eq:cond-vHst-B+iw-OK}, \ref{eq:cond-vHst-B+iw-opp}).

The first kind of solutions (\ref{eq:cond-vHst-B+iw-OK}) reproduces 
all the rational metric $G$ in (\ref{eq:parametr-Kahler-caseApr},
 \ref{eq:Kahler-form-Aprm-4RCFT-2}); 
$a_1 \sim C$ and $a_2 \sim \widetilde{D}(-p_1)$. In the second kind 
of solutions (\ref{eq:cond-vHst-B+iw-opp}), the metric is not 
rational; $a_1 \sim D\sqrt{d'} \nin \Q$, 
and $a_2 \sim \widetilde{C}\sqrt{p_1/p_2} \nin \Q$. There is a region 
with a positive volume interpretation in the $(a_1,a_2)$ space. 
 $\square$
\end{anythng}

\begin{anythng}
The combination $(B+i\omega)$ is parametrized by four rational parameters.
For such a $(B+i\omega)$, the vertical Hodge structure is also given 
to $H^1(T^4;\Q) \oplus H^3(T^4;\Q)$. For both kinds of solutions,
(\ref{eq:cond-vHst-B+iw-OK}) and (\ref{eq:cond-vHst-B+iw-opp}), 
the vertical Hodge structure is also of CM-type, and there exists 
Hodge isomorphism with the horizontal Hodge structure 
on $H^1(T^4;\Q)\oplus H^3(T^4;\Q)$, as we see below. 

The vector space $V_1 := {\rm Span}_\Q\{ \hat{\alpha}^1, \hat{\alpha}^2\}
\oplus {\rm Span}_\Q\{ \hat{\alpha}^1\hat{\alpha}^2\hat{\beta}_1, \;
  \hat{\alpha}^1\hat{\alpha}^2\hat{\beta}_2 \}$ can be split into 
$V_{11}:= {\rm Span}_\Q\{ \hat{\alpha}^1, \;
 \hat{\alpha}^1\hat{\alpha}^2\hat{\beta}_2\}$ and 
$V_{12}: = {\rm Span}_\Q\{ \hat{\alpha}^2, \; 
\hat{\alpha}^2\hat{\alpha}^1\hat{\beta}_1\}$, and 
\begin{align}
 ({\rm solution}\; (\ref{eq:cond-vHst-B+iw-OK})): \qquad 
 {\rm End}_\Q ( V_{11} )^{\rm Hdg} \cong K^{(2)}_2, \qquad 
 {\rm End}_\Q ( V_{12} )^{\rm Hdg} \cong K^{(2)}_1, \\
 ({\rm solution}\; (\ref{eq:cond-vHst-B+iw-opp})): \qquad 
 {\rm End}_\Q ( V_{11} )^{\rm Hdg} \cong K^{(2)}_1, \qquad 
 {\rm End}_\Q ( V_{12} )^{\rm Hdg} \cong K^{(2)}_2. 
\end{align}
So, as a whole, 
\begin{align}
  {\rm End}_\Q (V_1)^{\rm v.Hdg} \cong K^{(2)}_1 \oplus K^{(2)}_2 \cong 
  {\rm End}_\Q (H^1(T^4_I;\Q))^{\rm Hdg}
\end{align}
for both solutions. The vertical rational Hodge structure 
on $V_1$ is of CM-type, with the CM field $K^{(2)}_1 \oplus K^{(2)}_2$.

Similarly, $V_3 := {\rm Span}_\Q\{\hat{\beta}_1,\hat{\beta}_2\} \oplus 
{\rm Span}_\Q \{ \hat{\alpha}^2\hat{\beta}_2\hat{\beta}_1, \; 
 \hat{\alpha}^1\hat{\beta}_1\hat{\beta}_2 \}$ can also be split 
into $V_{31} := {\rm Span}_\Q \{ \hat{\beta}_1, \; 
\hat{\beta}_1\hat{\alpha}^2\hat{\beta}_2\}$ and $V_{32} := {\rm Span}_\Q\{
\hat{\beta}_2,\; \hat{\beta}_2\hat{\alpha}^1\hat{\beta}_1\}$, and 
\begin{align}
 ({\rm solution}\; (\ref{eq:cond-vHst-B+iw-OK})): \qquad 
 {\rm End}_\Q ( V_{31} )^{\rm Hdg} \cong K^{(2)}_2, \qquad 
 {\rm End}_\Q ( V_{32} )^{\rm Hdg} \cong K^{(2)}_1, \\
 ({\rm solution}\; (\ref{eq:cond-vHst-B+iw-opp})): \qquad 
 {\rm End}_\Q ( V_{31} )^{\rm Hdg} \cong K^{(2)}_1, \qquad 
 {\rm End}_\Q ( V_{32} )^{\rm Hdg} \cong K^{(2)}_2  
\end{align}
for both of the solutions. 
So, the vertical rational Hodge structure on $V_3$ is also 
of CM-type, with the CM field $K^{(2)}_1 \oplus K^{(2)}_2$. 

The 2nd and 3rd properties of Thm. \ref{thm:RCFT-2-GV-forT4}
does not restrict how one introduces a filtration $W^\bullet_v$,
so we may still choose $W^3_v=V_{31}\oplus V_{32}$. Then there is a Hodge 
isomorphism between the horizontal $H^3(T^4;\Q)$ (resp. 
$H^1(T^4;\Q)$) and the vertical $W^3_v$ (resp. $W^1_v/W^3_v$)
(the 6th property in Thm. \ref{thm:RCFT-2-GV-forT4}) for 
both kinds of the solutions.  $\square$
\end{anythng} 

\subsubsection{Case (A)}

Finally, let us work on the case (A). In this case, 
we find that $(B+i\omega)$ that satisfies the 1st, 2nd and 3rd 
properties in the statement of Thm. \ref{thm:RCFT-2-GV-forT4}
correspond to rational $B$ and $G$, and hence for a rational 
$\mathcal N=(1,1)$ SCFT. 

\begin{anythng}
In this case, the reflex field $K^r$ is a degree-2 extension 
field $\Q(\sqrt{p})$. So, the 2-dimensional subspace 
$T^v_M\otimes \Q \subset A(M)\otimes \Q$ should be such that 
both $e^{2^{-1}(B+i\omega)}$ and $e^{2^{-1}(B-i\omega)}$ are contained in 
$T^v_M\otimes \C$. For the vertical Hodge structure on this 
space to be of CM by the degree-2 field $\Q(\sqrt{p})$,  
the generator $\mho$ of the charge $p-q= 2$ component should be 
of the form 
\begin{align}
 \mho = e^{2^{-1}(B+i\omega)} & \;
   = \left( 
  1 + B/2 + E (\hat{\alpha}^1\hat{\beta}_1\hat{\alpha}^2\hat{\beta}_2) \right)
      \tau^r_+(1)
    + \left( \omega'/2 + E'
     (\hat{\alpha}^1\hat{\beta}_1\hat{\alpha}^2\hat{\beta}_2) \right)
   \tau^r_+(\xi^r)  
\end{align}
for some $E,E' \in \Q$ and $B, \omega'$ in the image of 
$T^v_M\otimes\Q \subset A(M)\otimes \Q$ projected into ${\cal H}^2(T^4_I)$. 
The rational constants $E, E'$ are determined 
by $B$ and $\omega'$ by the condition $(\mho, \mho)=0$ in 
$T^v_M\otimes \Q \subset A(M)\otimes \Q$ with respect to the pairing 
(\ref{eq:def-vert-polarizatn}). Free choice of $B \in {\cal H}^2(M)$
corresponds to a rational $B$-field in (\ref{eq:cond-GnB-rational}), 
and a free choice of $\sqrt{p}\omega' \in \sqrt{p}{\cal H}^2(M)$ 
corresponds to $i\omega$ for $\omega$ given 
in (\ref{eq:Kahler-form-A-4RCFT-2}), 
which is for a rational metric. So, the first two properties 
in Thm. \ref{thm:RCFT-2-GV-forT4} are strong enough in the case (A) 
to allow only the data $(G, B)$ for a rational CFT. $\square$
\end{anythng}

\section{Discussions}
\label{sec:discussions}

In this article, we have made an attempt at refining Gukov--Vafa's 
conjecture, Conj. \ref{conj:GV-original}, and verifying it for 
the simple case where the target space is $T^4$. 
Rational CFTs in this case have been completely classified, 
so we used that to refine the conditions to be imposed 
(criteria) for rational SCFTs in the language of the horizontal 
and vertical rational Hodge structures. 

As a result, we arrived at Thms. \ref{thm:RCFT-2-GV-forT4} and 
\ref{thm:RCFT-2-GV-forT4-genI} stated in the language that is 
applicable, for their most part, to SCFTs with a general 
Ricci-flat K\"{a}hler manifold as the target space. 
Thm. \ref{thm:RCFT-2-GV-forT4} extracts properties that 
{\it all} the $T^4$-target RCFTs satisfy. 
The property 2(b) in Thm. \ref{thm:RCFT-2-GV-forT4} rules out 
an example of a non-rational $T^4$-target CFT in \cite[\S4]{Chen:2005gm} 
that looks as if it were a counter example to the version 
Conjecture \ref{conj:GV-original}. 

We have also found by imposing the properties 1, 2, 3, and 6 in 
Thm. \ref{thm:RCFT-2-GV-forT4} on a data $(T^4;G,B;I)$, 
that there are still $T^4$-target $\mathcal N=(1,1)$ SCFTs that 
are not rational. Obviously it is one of the next steps 
to see whether those counter examples can be eliminated 
by implementing the 5th and 7th properties of 
Thm. \ref{thm:RCFT-2-GV-forT4}.  

Although the statements of Thm. \ref{thm:RCFT-2-GV-forT4} are 
phrased (as much as possible) in a way applicable to Ricci-flat 
K\"{a}hler manifolds, we do not make a clear stance on which 
subset of the itemized properties in Thm. \ref{thm:RCFT-2-GV-forT4}
should be imposed as criteria for rationalness of the SCFTs. 
Some of the properties might be derived from others 
(cf Rmk. \ref{statmnt:CM-from-H2-to-H1}). Some of the properties 
may not hold true in some examples of rational SCFTs 
(such as 3.(a) and 3.(b)ii; 
cf discussions \ref{statmnt:subtlty-justL=n?}, \ref{statmnt:subtlty-offdiag}
and footnote \ref{fn:BV}).
Since there is still room for experimental study as 
in this article, the authors do not feel obliged to decide 
now which subset of the properties are necessary conditions 
for the rationalness.
 
There is a chance of having the 6th property as a part of 
necessary criteria for rational SCFTs, only for a family 
of Ricci-flat K\"{a}hler target SCFTs that are self-SYZ-mirror. 
We need to make an effort in properly formulating and 
generalizing the 7th property of Thm. \ref{thm:RCFT-2-GV-forT4} 
to Ricci-flat K\"{a}hler manifolds other than tori. 

\vspace{5mm}

One can also enjoy a moment of speculation. 
Consider K3-target $\mathcal N=(1,1)$ SCFTs. The most na\"ive way to apply 
the refined version Thm. \ref{thm:RCFT-2-GV-forT4} for this class of target space 
is to take the 1st, 2nd, 4th and 5th properties in Thm. \ref{thm:RCFT-2-GV-forT4} 
as the necessary and sufficient conditions for the SCFT to be rational. 
The 3rd and 6th properties do not contain additional information in this case.
An immediate consequence of this is that there exists a complex structure 
on K3 such that there is a polarization, and at the same time $B^{(2,0)}=0$
when a K3-target SCFT is rational. This speculation/conjecture is already 
non-trivial. 
Moreover, the Picard number $\rho$ should be no less than 10, and 
the K\"ahler form must be within the image of $T_M^v\otimes \R \subset 
A(M)\otimes \R$ projected on to the Neron--Severi space ${\cal H}^2\otimes \R$  
for any K3 target rational SCFT.\footnote{
Reference \cite{mengchenphd} discusses how to find an appropriate B-field 
and a symplectic form for a complex CM-type K3 surface $X$ 
with $\rho(X) \geq 10$ such that $X$ has a mirror that is also of CM-type, 
motivated by the GV conjecture \cite{Gukov:2002nw}. 
} 

\vspace{5mm}

There has been a question of how densely rational SCFTs 
populate the moduli space of Ricci-flat K\"{a}hler 
target SCFTs. 
Reference \cite{Gukov:2002nw} conjectured that 
rational SCFTs might have something to do with CM-type 
rational Hodge structures (Conj. \ref{conj:GV-original}), 
and further combined the observation with Andr\'e--Oort 
conjecture in math \cite{ andre1989g, oort1997canonical, 
andre1997distribution},\footnote{
cf also \cite{tsimerman2015proof} and \cite{moonen2011torelli}. 
} %
 which says that CM points are not 
very dense in the moduli space of such manifolds in general 
(except for the moduli space of abelian varieties and K3 surfaces). 
So, it has been hinted that rational SCFTs do not populate 
densely within the whole moduli space of $\mathcal N=(1,1)$ SCFTs with 
a Calabi--Yau threefold target space.  
The experimental study carried out in this article 
concluded that all the rational ($T^4$-target) SCFTs satisfy
the properties of Thm. \ref{thm:RCFT-2-GV-forT4} refined 
from Conj. \ref{conj:GV-original}. So, the inference on scarcity 
of rational SCFTs {\it from} the scarcity of CM-type Calabi--Yau 
manifolds does not have to be questioned at this moment. 
Finer understanding on the remaining issues listed above (e.g., 
footnote \ref{fn:BV}), however, might also change this perspective 
in the future.

The question above may have a consequence beyond mathematical 
physics. Suppose one day that mankind discovers that 
Type IIB flux compactification is theoretically consistent 
only when the SCFT is rational; it is not bad to enjoy 
such a speculation sometimes \cite{Gukov:2002nw}. That may indicate
that the vacuum complex structure of the internal Calabi--Yau 
threefold is something captured by a special subvariety 
of Calabi--Yau moduli space interpreted as a Shimura variety, 
if we speculate along the lines of Gukov--Vafa and Andr\`e--Oort.  
When the moduli space has a group action, discrete and/or continuous, 
its isotropy subgroup at the vacuum point may remain in the 
low-energy effective field theory of the moduli fields as 
gauged and/or accidental symmetry. 

\subsection*{Acknowledgments}   

We thank H. Lange for useful communications. 
We thank M. Ashwinkumar and M. Yamazaki for collaboration 
during early stages of this research project. 
AK thanks DESY (Hamburg) for hospitality during the final 
stage of this project.
This work was supported in part by WPI Initiative (all the authors), 
the Riemann Fellowship (Riemann Center for Geometry and Physics) (AK), 
FoPM, WINGS Program, the University of Tokyo (MO), and 
a Grant-in-Aid for Scientific Research on 
Innovative Areas 6003 (MO and TW), MEXT, Japan. 

\appendix
\section{Appendix: Additional notation and background}
\label{sec:app}
{\bf Some definitions not included (and notations not explained) 
in the main text:}

String theorists do not necessarily have lot of experience with 
number fields or theory of complex multiplication. 
For those readers, the appendices\footnote{
Some materials in the appendix of the preprint version are 
placed within the main text of the journal version.
}  %
 of \cite{Kanno:2017nub} will be useful. Materials there include 
basics about number fields, and the definitions of {\it totally 
imaginary field, totally real field, CM field, the totally 
real subfield of a CM field, CM-type of a 
weight-1 rational Hodge structure, reflex field of a CM-type, 
primitivity of a CM-type}. The notations such as $\Q[x]$, $[E:F]$ and 
${\rm Tr}_{E/F}$ of a field extension $E/F$ as well as its properties 
are also explained there.  So, we do not include those materials 
in this article. A {\it CM algebra} is the direct sum of a finite 
number of CM fields. 

\begin{notn}
$M_n(A)$ for an algebra $A$ is the algebra of $A$-valued $n\times n$ matrices.
\end{notn}

\begin{defn}
\label{defn:divis-alg}
An algebra $D$ over a field $F$ is a {\it division algebra} 
if any non-zero element $x \in D$ has an inverse $x^{-1}$ 
with respect to the multiplication law of the algebra $D$. 

Then $x^{-1} \cdot x = 1 = x \cdot x^{-1}$. A division algebra 
$D$ is regarded as a field if the multiplication law of the 
algebra $D$ is commutative (abelian).
\end{defn}

\begin{defn}
A finite dimensional algebra $A$ over a field $F$ is {\it semi-simple}
if there is no non-zero nilpotent ideal.  
%
\end{defn}

\
\begin{anythng}
Although a minimum explanation on rational (pure) Hodge structure 
is given already in Appendix B of \cite{Kanno:2017nub}, we also repeat 
some of it here without worrying about overlap. That is partly because 
we should have 
an eye on something beyond the most conventional pure rational Hodge 
structure in this article, and also because not much emphasis was 
given to the role played by a polarization of a rational Hodge 
structure in \cite{Kanno:2017nub}. So, let us start from the basics.
\end{anythng}

\begin{defn}
Let $V_\Q$ be a vector space over $\Q$. A {\it pure rational Hodge 
structure on} $V_\Q$ {\it of weight-$m$} is a decomposition of 
a vector space over $\C$, 
\begin{align}
  V_\Q \otimes \C \cong \oplus_{p,q}^{(p+q=m)} [V_\Q \otimes \C]^{p,q}
\end{align}
satisfying $([V_\Q\otimes \C]^{p,q})^{\rm c.c.} = [V_\Q\otimes \C]^{q,p}$. 
The word ``pure'' is often omitted; it is retained only when 
there is a high chance of confusion with a mixed rational Hodge 
structure (mentioned in footnote \ref{fn:mixed-HS}) 
or with a generalized 
Hodge structure we introduce in Def. \ref{defn:gen-HS}. 
\end{defn}

See \cite[App. B]{Kanno:2017nub} or any math literatures for the definition 
of a {\it Hodge substructure, simple rational Hodge structure}, and 
{\it the level of a pure rational Hodge structure}. 

\begin{anythng}
\label{statmnt:pHdgStr-by-S1-repr}
For a rational pure Hodge structure of weight-$m$ on a vector 
space $V_\Q$, the added data on top of the vector field $V_\Q$, i.e., 
the decomposition, can also be encoded by giving a representation 
\begin{align}
  h: S^1 \longrightarrow {\rm GL}(V_\Q \otimes \R), \qquad 
    h(e^{i\alpha})|_{[V_\Q \otimes \C]^{p,q}} = e^{-i\alpha(p-q)} 
\end{align}
(the representation $h$ cannot reproduce the information $m=p+q$, 
so the weight-$m$ needs to be retained along with $h$).
\end{anythng}

\begin{notn}
\label{notn:end-alg}
Let $(V_\Q,h)$ be a pure rational Hodge structure. Then 
\begin{align}
 {\rm End}(V_\Q)^{\rm Hdg} & \; := \left\{ \phi \in {\rm Hom}_\Q(V_\Q,V_\Q)
     \; | \; \phi([V \otimes \C]^{p,q}) \subset [V\otimes \C]^{p,q} \right\}, \\
  & \; = \left\{ \phi \in {\rm Hom}_\Q(V_\Q,V_\Q) \; | \; \phi \circ h(e^{i\alpha}) = h(e^{i\alpha}) \circ \phi \right\}.
\end{align}
We call it {\it the endomorphism algebra}, and its elements {\it endomorphisms}
in this article. We should refer to those elements as 
Hodge-structure-preserving endomorphisms of the vector space $V_\Q$ for 
a general $(V_\Q,h)$; when we deal with the 1st cohomology groups of an abelian 
variety, however, such Hodge-structure-preserving endomorphisms originate 
from the group-law preserving morphisms of an abelian variety to itself. 
So, for this reason, it is not too bad to use the word that does not 
sound right (certainly not right especially for various simple components 
of $H^{k>1}(T^{2n};\Q)$). 
\end{notn}

\begin{defn}
\label{defn:polrz-HS}
A bilinear form ${\cal Q}: V_\Q \times V_\Q \rightarrow \Q$ 
on a vector space $V_\Q$, either symmetric (for even $m$) or 
anti-symmetric (for odd $m$), is 
said to be a {\it polarization of a pure rational Hodge structure 
$(V_\Q, h)$ of weight-$m$}, if ${\cal Q}(x,y) \in \C$ can be non-zero 
for $x \in [V \otimes \C]^{(p_1,q_1)}$ and $y \in [V\otimes \C]^{(p_2,q_2)}$ 
only when $p_1=q_2$ and $q_1=p_2$, and $i^{p-q}{\cal Q}(x,x^{\rm cc}) >0$
for $x_{\neq 0} \in [V\otimes \C]^{p,q}$. 
A pure rational Hodge structure $(V_\Q,h)$ of weight-$m$ is said to  
{\it be polarizable} when $(V_\Q,h)$ admits a polarization. 
\end{defn}

Let $\psi$ be a polarization\footnote{
Its definition is found at the beginning of 
section \ref{ssec:CM-abel-surface}. 
} %
of an abelian variety $X$ of complex dimension $n$. 
Then the rational Hodge structures of weight $m$ on 
$H^m(X;\Q)$ admits a polarization ${\cal Q}_\psi$ given by 
${\cal Q}_\psi(x,y) = \int_X \psi^{n-m} \wedge x \wedge y$. 

\begin{defn}
\label{defn:Hdg-grp}
Let $(V_\Q,h)$ be a pure rational Hodge structure of weight-$m$.
Its {\it Hodge group}, denoted by ${\rm Hg}((V_\Q,h))$ or ${\rm Hg}(h)$, 
is the minimal algebraic variety of ${\rm GL}(V_\Q)$ with the group law 
from ${\rm GL}(V_\Q)$ given by defining equations that involve only 
rational coefficients (i.e., in $\Q$, not in $\C$), so that all 
the points $h(e^{i\alpha})$ for $e^{i\alpha} \in S^1$ satisfy those 
defining equations. 

Most of literatures referring to a Hodge group is for a polarizable 
pure rational Hodge structure. But it is possible to define such 
a notion for a rational Hodge structure that is not necessarily 
polarizable; whether such a ${\rm Hg}(h)$ still has a nice property 
is a separate question.   
\end{defn}

{\bf Representations of semi-simple algebras:} we record a few basic 
known facts about representations of a finite-dimensional semi-simple 
algebras over $\Q$ for convenience of readers. Those facts are used 
in the main text.  

\begin{lemma}
\label{lemma:Wedderburn}
This is known as Wedderburn's theorem. 
A finite dimensional semi-simple algebra $\mathfrak{R}$ over $\Q$
has a structure 
\begin{align}
 \mathfrak{R} \cong \oplus_{\alpha \in {\cal A}} M_{n_\alpha}(D_\alpha)
   \label{eq:WeddB-alg-str}
\end{align}
for some finite set ${\cal A}$, $n_\alpha \in \N$, and 
a division algebra $D_\alpha$.

When $\mathfrak{R}$ has a faithful representation on a vector space $V_\Q$ 
over $\Q$, $\dim_\Q (V_\Q) \geq \sum_{\alpha} n_\alpha q_\alpha^2 [k_\alpha:\Q]$, 
where $k_\alpha$ is the center of $D_\alpha$, and $q_\alpha$ is the positive 
integer such that $[D_\alpha:k_\alpha] = q_\alpha^2$. 
\end{lemma}

The following facts (Lemmas \ref{lemma:very-useful}, \ref{lemma:very-useful-2}) 
are regarded so trivial by mathematicians that we have to 
read that out between the lines in textbooks on semi-simple algebras. 
The authors are unable to refer to a specific text for this reason. For 
the reader with a background in string theory, it will still be 
better that they are written down explicitly.\footnote{
The appendix B.2 of the preprint version of \cite{Kanno:2017nub}
(main text II.B.3 of the journal version) has a little more 
pedagogical explanation on the first half of Lemma \ref{lemma:very-useful}.
The statement here is slightly polished up from the version there, however. 
} %
\begin{lemma}
\label{lemma:very-useful}
Let $F$ be a number field (with $[F:\Q] < \infty$), and 
$\{\tau_{a=1,\cdots, [F:\Q]} \}$ its embeddings to $\overline{\Q} \subset \C$. 
Let $V_\Q$ be a vector space over $\Q$. 
We think of only the cases that $[F:\Q] = \dim_\Q V_\Q$ here. 

Suppose that $F$ acts non-trivially on $V_\Q$ (with $\Q \subset F$ acting 
as the scalar multiplication on the vector space $V_\Q$). The action of $F$ 
on $V_\Q$ can be diagonalized simultaneously; the $\dim_\Q V$ eigenvectors 
can be chosen in a following way. First, choose any non-zero element 
$v_* \in V_\Q$, and arbitrary basis $\{ \omega_{i=1,\cdots, [F:\Q]} \}$ of the 
vector space $F/\Q$. Then $\{ \omega_i \cdot v_* \}_{i=1,\cdots, [F:\Q]}$ can 
be used as a basis of the vector space $V_\Q$ over $\Q$. Now, we can 
choose the eigenvectors to be\footnote{ 
So, the action of $F$ on $V_\Q$ splits into 1-dimensions on $V_\Q \otimes_\Q k$ 
whenever $k \subset \overline{\Q}$ contains the normal closure of $F$ in 
$\overline{\Q}$. 
} %
\begin{align}
  v_a := \sum_i (\omega_i \cdot v_*) \tau_a(\eta_i), \qquad a=1,\cdots, [F:\Q], 
\end{align}
where $\{ \eta_{j=1,\cdots, [F:\Q]} \}$ is the basis of $F/\Q$ dual to 
$\{ \omega_i \}$ with respect to the bilinear form ${\rm Tr}_{F/\Q}[x y]$.  
That is, ${\rm Tr}_{F/\Q}[\omega_i \eta_j] = \delta_{ij}$.
The matrix $(\tau_a(\eta_i))_{ai}$ is used as the inverse matrix 
of $(\tau_a(\omega_j))_{ja}$. 
For any element $x \in F$, its action on $V_\Q \otimes \C$ is 
given by $x \cdot v_a = v_a \tau_a(x)$, i.e., the eigenvalue of $x\cdot$ 
is $\tau_a(x)$ for the eigenvector $v_a$.   
All those eigenvectors $v_a$ ($a=1,\cdots, [F:\Q]$) are obtained from 
one of them, say, $v_{a*}$, by applying Galois transformations on the 
coefficients $\tau_{a*}(\eta_i)$ of the expansion of $v_{a*}$ with respect 
to the rational basis $\{ (\omega_i\cdot v_*)_{i=1,\cdots, [F:\Q]} \}$, because 
$\tau_a = \sigma_a \cdot \tau_{a*}$ for some 
$\sigma_a \in {\rm Gal}(\overline{\Q}/\Q)$. We may express this
in the form of $v_a = v_{a*}^{\sigma_a}$.

For any basis $\{ \eta'_{i=1,\cdots, [F:\Q]} \}$ of $F/\Q$, there exists 
a basis $\{ v'_i \}$ of $V_\Q$ where the simultaneous eigenvectors 
are in the form of $v_a = v'_i \tau_a(\eta'_i)$. To see this, just find 
the rational coefficient matrix $\eta_i = C_{ij} \eta'_j$ and set 
$v'_j := (\omega_i \cdot v_*)C_{ij}$. 
\end{lemma}

\begin{lemma}
\label{lemma:very-useful-2}
Conversely, for any basis $\{ \eta_j \}$ of $F/\Q$ and $\{v_i \}$ of $V$, 
one may construct a non-trivial action of $F$ on the vector space $V$ 
over $\Q$ so that $v_a := \sum_i v_i \tau_a(\eta_i)$ for $a=1,\cdots, [F:\Q]$
are all eigenvectors of the action of $F$. The action of $x \in F$ on $V$ 
claimed here is given as follows. First, write down the multiplication law 
in $F$ as follows: 
\begin{align}
  (x \cdot ): \quad \omega_i \longmapsto \omega_k [A(x)]_{ki}, 
\end{align}
where $\{ \omega_i \}$ is the basis of $[F:\Q]$ dual to $\{ \eta_j \}$, 
and $[A(x)]$ is a $\Q$-valued $[F:\Q] \times [F:\Q]$ matrix. 
Using this matrix, the action of $x$ is 
\begin{align}
  x \cdot : \quad v_i \longmapsto v_k [A(x)]_{ki}. 
\end{align}
$\square$ 
\end{lemma}

The facts above in both ways (Lemmas \ref{lemma:very-useful} and \ref{lemma:very-useful-2}) hold for a general number field $F$ not 
necessarily a CM field; the eigenspace decomposition does not have 
to be relevant to Hodge components. 

\bibliography{rcft}
\bibliographystyle{utphys}

\end{document}